\documentclass[11pt,preprint]{aastex}

\newcommand{\kms}{km~s$^{-1}$}
\newcommand{\subsun}{\mbox{$_{\odot}$}}
\newcommand{\etal}{{\it et al.\/}}
\newcommand{\teff}{$T_{\mbox{\scriptsize eff}}$}
\newcommand{\grav}{log($g$)}
\newcommand{\fe}{[Fe/H]}
\newcommand{\mg}{[Mg/Fe]}
\newcommand{\al}{[Al/Fe]}
\newcommand{\si}{[Si/Fe]}
\newcommand{\ca}{[Ca/Fe]}
\newcommand{\scme}{[Sc/Fe]}
\newcommand{\ti}{[Ti/Fe]}
\newcommand{\crme}{[Cr/Fe]}
\newcommand{\mn}{[Mn/Fe]}
\newcommand{\co}{[Co/Fe]}
\newcommand{\nime}{[Ni/Fe]}
\newcommand{\ba}{[Ba/Fe]}
\newcommand{\mystar}{HE~2148--1247}
\newcommand{\sarastar}{HE~0024--2523}

\newcommand{\eqw}{$W_{\lambda}$}
\newcommand{\ciso}{$^{12}$C/$^{13}$C}

\begin{document}

\title{Abundances In Very Metal Poor Dwarf Stars\altaffilmark{1}}

\author{Judith G. Cohen\altaffilmark{2}, 
Norbert Christlieb\altaffilmark{3},
Andrew McWilliam\altaffilmark{4}, Steve Shectman\altaffilmark{4},
Ian Thompson\altaffilmark{4},
G.J. Wasserburg\altaffilmark{5}, Inese Ivans\altaffilmark{2,7},
Matthias Dehn\altaffilmark{3}, 
Torgny Karlsson\altaffilmark{8}
\& J.Melendez\altaffilmark{2} }

\altaffiltext{1}{Based on observations obtained at the
W.M. Keck Observatory, which is operated jointly by the California 
Institute of Technology, the University of California, and the
National Aeronautics and Space Administration.}
\altaffiltext{2}{Palomar Observatory, Mail Stop 105-24,
California Institute of Technology, Pasadena, Ca., 91125}
\altaffiltext{3}{Hamburger Sternwarte, Universit\"at
Hamburg, Gojenbergsweg 112, D-21029 Hamburg, Germany}
\altaffiltext{4}{Carnegie Observatories of Washington, 813 Santa
Barbara Street, Pasadena, Ca. 91101}
\altaffiltext{5}{Lunatic Asylum, Division of Geological and Planetary
Sciences, California Institute of Technology, Pasadena, Ca., 91125}
\altaffiltext{7}{Hubble Fellow}
\altaffiltext{8}{Department of Astronomy and Space Physics, 
University of Uppsala, Box 524, SE-75239 Uppsala, Sweden}

\begin{abstract}

We discuss the detailed composition of 28 extremely metal-poor (EMP)
dwarfs, 22 of which are from the Hamburg/ESO Survey, based on Keck 
Ech\`elle spectra.  Our
sample has a median [Fe/H] of $-2.7$ dex, extends to $-3.5$ dex,
and is somewhat less metal-poor than was expected from 
[Fe/H](HK,HES) determined
from low resolution spectra.
Our analysis supports the existence of a sharp decline in the
distribution of halo stars with metallicity below [Fe/H] = $-3.0$ dex.
So far no additional turnoff stars with
$\mbox{[Fe/H]}<-3.5$ have been identified in our follow up efforts. 

For the best observed elements between Mg and Ni,
we find that the abundance ratios appear to have reached a plateau,
i.e. [X/Fe] is approximately constant as a function of [Fe/H], except
for Cr, Mn and Co, which show trends of abundance ratios varying with [Fe/H].
These abundance ratios at low metallicity
correspond approximately to the 
yield expected from Type II SN with a narrow range in mass and
explosion parameters; high mass Type II SN progenitors are required. The
dispersion of [X/Fe] about this plateau level is surprisingly small, and is
still dominated by measurement errors  rather than intrinsic scatter.
These results place strong constraints
on the characteristics of the contributing SN. 

The dispersion in neutron-capture elements, and the abundance trends for
Cr, Mn and Co are consistent with previous studies of evolved EMP stars.

We find halo-like enhancements for the $\alpha$-elements Mg, Ca and Ti, but
solar Si/Fe ratios for these dwarfs.  This contrasts with studies of EMP 
giant stars, which show Si enhancements similar to other $\alpha$-elements.
Sc/Fe is another case where the results from EMP dwarfs and from EMP giants
disagree; our
Sc/Fe ratios are enhanced compared to the solar value by $\sim$0.2 dex. 
Although this conflicts with the solar Sc/Fe values seen in EMP giants, we 
note that $\alpha$-like Sc/Fe ratios have been claimed for dwarfs at higher 
metallicity.

Two dwarfs in the sample are carbon stars, while two
others have significant C enhancements, all with \ciso\ $\sim$7
and with C/N between 10 and 150.
Three of these  C-rich stars have
large enhancements of the heavy neutron capture
elements, including  lead, which implies a strong
$s$-process contribution, presumably from binary mass transfer;
the fourth shows no excess of Sr or Ba.

\end{abstract}


\section{Introduction}

The most metal deficient stars in the Galaxy provide crucial evidence
on the early epochs of star formation, 
the environments in which various elements
were produced, and the production of elements subsequent to the Big 
Bang and
prior to contributions from lower mass stars to the ISM. 
The major existing survey for very metal-poor stars is the HK survey
described in detail by Beers, Preston \& Shectman (1985, 1992). 
The stellar inventory of this survey has been scrutinized with
considerable care over the past decade, 
but, as summarized by \cite{beers99}, only roughly 100
are believed to be extremely metal poor (henceforth EMP), 
with [Fe/H] $\le -3.0$ dex\footnote{The 
standard nomenclature is adopted; the abundance of
element $X$ is given by $\epsilon(X) = N(X)/N(H)$ on a scale where
$N(H) = 10^{12}$ H atoms.  Then
[X/H] = log$_{10}$[N(X)/N(H)] $-$ log$_{10}$[N(X)/N(H)]\subsun, and similarly
for [X/Fe].}.

We are engaged in a large scale project to find additional 
extremely metal poor stars exploiting the database of the
Hamburg/ESO Survey (HES).   The HES is an 
objective prism survey
primarily targeting bright quasars \citep{wis00}.
However,  it
      is also possible to efficiently select a variety of interesting
      \emph{stellar} objects in the HES (Christlieb 2000; Christlieb \etal\ 2001a,b),
among them EMP stars \citep{christlieb03}.
The existence of a new
list of candidates for EMP stars 
with [Fe/H] $< -3$ dex 
selected in an automated and unbiased manner
from the HES, coupled with
the very large collection area and efficient high resolution Ech\`elle
spectrographs of the new generation of large telescopes,
offers the possibility for a large
increase in the number of EMP stars known
and in our understanding of their properties.  The results of our
successful
Keck Pilot Project to  determine the effective yield of the HES for
EMP stars through high dispersion abundance analyses of a sample of
stars selected from the HES are presented in \cite{cohen02} and
\cite{carretta02}, while \cite{lucatello03} discuss in detail the
star of greatest interest found among that group.

Since the completion of the Keck Pilot Project, we have continued our efforts
at isolating a large sample of EMP stars from the HES. 
In this paper we discuss
a sample of 28 candidate EMP dwarfs, including 14
previously unpublished candidate EMP dwarfs selected from the HES.
We study their abundances,
their abundance ratios, the spread of these, 
and the implications thereof for nucleosynthesis
and supernovae in the early Galaxy.  We then compare
our results for EMP dwarfs with those published by the First Stars Project
\citep{cayrel03}
for a large sample of brighter EMP giants from the HK survey, and with the
abundance ratios seen in Galactic globular clusters and in damped
Ly$\alpha$ absorbers.  After a brief preliminary investigation of the binary
fraction, the paper concludes with a discussion
of the implications of our results for the
overall characteristics of the much larger HES sample.

\section{Sample of Stars}

Selection of EMP stars in the HES, reviewed by \cite{christlieb03},
is carried out by automatic spectral
classification, using classical statistical methods.
As described in Christlieb \etal\ (2001a), $B-V$ colors can be estimated
directly from the digital HES spectra with an accuracy of $\sim
0.1$\,mag, so that these samples can be selected not only on the 
basis of
spectroscopic criteria but also with restrictions on $B-V$ color.

The principal spectroscopic criterion used for sample selection for EMP
stars is the same as that used by the HK project, the absence/weakness
      of the 3933\,{\AA} line of Ca~II.
A visual check of the HES spectrum is then made
to eliminate the small fraction of spurious objects (plate defects,
misidentifications, etc.) 

The present sample was selected from the HES database 
to have $0.3<B-V<0.5$ to focus on main sequence turnoff stars.
The pool of candidates in the turnoff region consists of those stars
which show in the HES spectra
a Ca~K line weaker than expected for a star with $\mbox{[Fe/H]}=-3.0$ at given
$B-V$ color. 
For the hotter dwarf stars the Ca K line for [Fe/H] $< -3$ is not detected
in the HES spectra; thus the initial sample contains some stars 
more metal-rich than [Fe/H]=$-3$ dex.

To make the best use of the
limited observing time available on the largest telescopes, these
candidates from the HES database must first be verified through 
moderate-resolution ($\sim$1--2\,{\AA}) follow-up spectroscopy 
at 4-m class telescopes.  At this spectral resolution, [Fe/H](HK), a measure of
the metallicity of the candidate
based on the strength of the Ca II line at 3933\AA\ 
and H$\delta$ (a temperature indicator)
 \citep*[see, e.g.,][]{bee92,beers99b}, can be determined and used to   
select out the genuine EMP stars from the much more numerous
stars of slightly higher metallicity -- of interest in their own right -- but
not relevant for our present study.
It is the overall efficiency of this multi-stage selection 
process for isolating genuine EMP stars which we tested in the Keck Pilot
Project. 
 
We note here that
stars were selected for the present sample from among more than 1000
objects for which moderate resolution spectra were obtained
at the ESO, Palomar or Las Campanas Observatories.
The results from follow-up observations of
more than 2000 metal-poor candidates from the HES will be described
elsewhere (Christlieb \etal\, in preparation).  
The upper panel of Figure~\ref{figure_fehist} shows a histogram of
[Fe/H](HK) for the dwarf sample from the HES observed to date
at at ESO by Christlieb.
Note the (expected) very wide distribution in metallicity
and the significant fraction of stars at relatively high metallicities,
which, however, is considerably reduced with respect to that present in the
HK Survey candidate sample \citep*[see Figure 5 of][]{christlieb03}.

The sample studied here includes 28 candidate EMP dwarfs in the region
of the main sequence turnoff having \teff $> 6000$ K\footnote{This is
slightly bluer than the B-V cutoff in the selection for the HES.
Analyses of the somewhat cooler EMP subgiants will appear in
a subsequent paper.}.   These
stars have such weak lines that even the best 
moderate resolution follow up spectra
cannot discern  much more than their Balmer and 3933~\AA\ (Ca~II) line
strengths; the G band of CH is undetectable in most
of them.  While dwarfs have weak  metal lines, they are unevolved, with
no internal nuclear processing beyond H burning and
no known processes that could bring any products of nucleosynthesis
from the stellar interior to the surface.  We thus avoid
the issue of mixing to the surface of the products of internal
nuclear burning
that might afflict EMP red giants, and definitely 
plague the study of red giants in globular clusters, 
discussed, for example, in \cite{cohen02a} and \cite{cannon03}.

The middle panel of Figure~\ref{figure_fehist} shows a histogram of
[Fe/H](HK) for the sample of candidate EMP dwarfs
chosen from the HES and the HK Survey for which
we have obtained high resolution observations.
Our sample includes 14 previously unpublished stars  from the HES, two
candidate EMP dwarfs from the HK
survey and one high proper motion EMP dwarf from the NLTT catalog
\citep{luyten80} analyzed by \cite{ryan91}.
% Fe/H(Ryan 91) -3.4 dex  6000 4.0 
In addition, we add the data
from the Keck Pilot Project for stars with suitable \teff, 
as well as the peculiar dwarf HE2148$-$1247, discussed in detail in 
\cite{cohen03}, for a total sample of 28 candidate EMP dwarfs.

\section{HIRES Observations \label{hires} }

Once a list of vetted candidates for EMP dwarfs from the HES
database was created, observations were obtained
at high dispersion with HIRES \citep{vogt94} at the Keck I telescope
for a detailed abundance analysis.
A spectral resolution of 45,000
was achieved using a 0.86 arcsec wide slit projecting to 3 pixels in
the HIRES focal plane CCD detector.  For those stars presented here from
the run of Sep. 2002, all of which have $V > 15$ mag,
a spectral resolution of 34,000 was used. 

The spectra cover the region from 3840 to 5320\,{\AA} with essentially
no gaps\footnote{This HIRES configuration is shifted one order bluer than that
used for the Keck Pilot Project.}.  Each exposure 
for the HES stars was broken up into 1200 sec segments.  
The spectra were exposed until a SNR of 100 per spectral
resolution element in the continuum at
4500\,{\AA} was achieved; a few spectra, particularly from 
the May 2002 run, when the weather was very poor, have 
lower SNR.   This SNR calculation utilizes only
Poisson statistics, ignoring issues of cosmic ray removal,
night sky subtraction, flattening, etc.   The observations
were carried out with the slit length aligned to
the parallactic angle.

The list of new stars in the sample, their $V$ mags, and the detailed
parameters 
of their HIRES exposures, including the
exposure times and signal to noise ratios per spectral
resolution element in the continuum, are
given in Table~\ref{table_newstars}. 

This set of HIRES data was reduced
using a combination of Figaro scripts and
the software package MAKEE\footnote{MAKEE was developed
by T.A. Barlow specifically for reduction of Keck HIRES data.  It is
freely available on the world wide web at the
Keck Observatory home page, http://www2.keck.hawaii.edu:3636/.}.
MAKEE automatically applies heliocentric corrections to each of the
individual spectra.
The bias removal, flattening, 
sky subtraction, object extraction and
wavelength solutions with the Th-Ar arc were performed within MAKEE,
after which further processing and analysis was carried out within
Figaro, where the individual spectra were summed.
The continuum fitting to the sum of the individual
spectra (already approximately corrected via the mean signal level
in the flat field spectrum)
uses a 4th-order polynomial to line-free regions of
the spectrum in each order.  The degree of the polynomial was
reduced in orders with H and K of Ca~II or Balmer lines where the
fraction of the order available to define the continuum decreased
significantly.
A scheme of using adjacent orders to help define the polynomial
under such conditions is included in the codes.
The suite of routines for analyzing Ech\`elle
spectra was written by \cite{mccarthy88} within the 
Figaro image processing package
\citep{shortridge93}.

% The gain setting of the HIRES
% CCD detector is 2.4 e$^-$/ADU; to reach the desired SNR of 100 
% per spectral resolution element thus requires 
% 1390 ADU/pixel.

\subsection{Equivalent Widths \label{equiv_widths} }

The search for absorption features present in our HIRES data and the
measurement of their equivalent width (\eqw) was done automatically with
a FORTRAN code, EWDET, developed for a globular cluster project. 
Details of this code and its features are given in \citet{ramirez01}.
Except in regions affected by molecular bands, 
the determination
of the continuum level in these very metal poor stars was easy
as the crowding of lines is minimal.  Hence
the equivalent widths
measured automatically should be quite reliable, and we  
initially use the automatic Gaussian fits for \eqw.

The list of lines identified and measured by EWDET was then correlated,
taking the radial velocity into account, 
to a template list of suitable unblended lines 
with atomic parameters similar to that described in \cite{cohen03}
to specifically identify the various atomic lines. 
The automatic identifications were accepted as valid for
lines with \eqw\ $\ge 15$ m\AA.  They were checked by hand
for all lines with smaller \eqw\ and for all the rare earths.
The resulting \eqw\ for 167 lines in the spectra of the 14 
previously unpublished candidate
EMP dwarfs selected from the HES and in the three additional EMP stars are 
listed in Table~\ref{table_eqw}.

Occasionally, for crucial elements where no line was securely detected
in a star, we
tabulate upper limits to \eqw.  These are indicated as negative entries in 
Table~\ref{table_eqw}; the upper limit to \eqw\ is the absolute value of
the entry.

\section{Atomic Data and Solar Abundances\label{at_data}}

To the maximum extent possible,
the  the atomic data and the analysis procedures
used here are identical to those developed for the
Keck Pilot Project.  The provenance of
the transition probabilities of the lines in 
the template list is described in detail in
\cite{cohen03} and is a slightly modified and updated
version of the template list used in the Keck Pilot
Project \citep{carretta02}.  Many of the $gf$ values are
taken from the NIST Atomic Spectra 
Database Version 2.0 (NIST Standard Reference Database \#78),
see \cite{wiese69}, \cite{mar88}, \cite{fuh88} and \cite{wiese96}.

In many of the program stars, the absorption lines are so weak
that no correction for hyperfine structure is necessary. 
However, when required,
for ions with hyperfine structure, we synthesize the spectrum for each line
including the appropriate HFS and isotopic components.
We use the HFS components from \cite{prochaska00} for the lines we utilize 
here of
Sc~II, V~I, Mn~I, Co~I.  For Ba~II, we adopt the HFS from \cite{mcwilliam98}.
We use the laboratory spectroscopy of \cite{lawler01a}
and \cite{lawler01b} to calculate the HFS patterns
for La~II and for Eu~II.  
We have updated our Nd~II
$gf$ values to those of \cite{denhartog03}.

Based on the
work of \cite{bau96} and \cite{bau97} we adopt a fixed 
non-LTE correction of
$-$0.6 dex for Al~I in these EMP dwarfs.  No other non-LTE corrections
were applied nor initially deemed necessary.

We use damping constants for Mg~I from the detailed analysis 
of the Solar spectrum by \cite{zhao98}.
For lines of Si~I, Al~I, Ca~I, Sr~II and Ba~II,
we use the damping constants of \cite{barklem00} which were
calculated using the
theory of Anstee, Barklem \& O'Mara \citep{barklem98}.
For all other elements,
the damping constants were set to twice
that of the Uns\"{o}ld approximation for van der Waals broadening
following \citet{holweger91}.

The regime in which we are operating is so metal poor that we cannot
attempt to calculate Solar abundances corresponding to our particular choices of
atomic data because the lines seen in the EMP dwarfs are far too strong
in the Sun.  We must rely on the accuracy of the
$gf$ values for each element across the large relevant range of
line strength and wavelength.
We adopt the Solar abundances of \cite{anders89} for most elements.
For Ti and for Sr, we adopt the slightly modified values
given in \cite{grevesse98}.  For the special cases 
of La~II, Nd~II and Eu~II we use the results found
by the respective recent laboratory studies
cited above.  For Mg, we adopt the slightly updated value
suggested by \cite{holweger01}, ignoring the small suggested non-LTE
and granulation corrections, since we do not implement such in our analyses.

For the CNO elements we use the recent results of 
\cite{allende02}, \cite{asplund03} and \cite{asplund04}.  These
values are considerably ($\sim$0.2 dex) lower than those of  \cite{anders89} and
somewhat lower than those of \cite{grevesse98}.  The CNO elements
play only a small role in the present work; we obtain
approximate C abundances from the CH bands, and
rough N abundances from the CN bands.  Changes of a factor of two are
small compared to the variations in C and N to be discussed here.

We adopt log$\epsilon$(Fe) = 7.45 dex for iron  
following the revisions in the Solar photospheric abundances
suggested by \cite{asplund00} and by \cite{holweger01}.
This value is somewhat lower
than that given by \cite{grevesse98} and considerably lower
than that recommended by \cite{anders89}.
Some papers in the literature use the \cite{grevesse98} value
and some older ones use 7.67 dex, the value recommended by
\cite{anders89}. In such cases, their values of [Fe/H]
will be 0.1 to 0.2 dex smaller than ours while their
abundance ratios [X/Fe] the same amount
larger than ours.

\section{Stellar Parameters \label{section_params} }

The procedure used to derive effective temperatures for the EMP 
dwarfs is fully
explained in \S 4-6 of \cite{cohen02}. Very briefly, \teff\ is derived from
broad-band colors, taking the mean estimates deduced from the de-reddened 
$V-K$ and $V-J$ colors, where the infrared colors are from
2MASS \citep{skrutskie97}.
The corrections due to slight differences in filter bandpasses
between the 2MASS and the CIT JHK systems are small
($\le0.02$ mag)
\citep{carpenter01}, and we ignore them.
The $V$ photometry is largely
from three runs with the Swope 1-m telescope at the Las Campanas
Observatory.  Other sources for individual
stars can be found in the notes to Table~\ref{table_newstars}.
We 
corrected the
colors for reddening, adopting the extinction maps of 
\cite{schlegel98}.  Since the HES is restricted to $|b| \ge 30$ deg,
the reddenings are low.

We used the grid of predicted broad-band colors and
bolometric corrections of \cite{houdashelt00}, based on the
MARCS stellar atmosphere code \citep{gustafsson75}
to determine the \teff\ for each star.  This scale, comparing only
$V-K$, is identical to that of the widely used 
scale of \cite{alonso96} for [Fe/H] = $-3$ dex,
and is hotter than that scale by 70 to 150 K at [Fe/H] = $-2$ dex.

There are now sufficient stellar angular diameter measurements
from interferometers to provide a preliminary check on our \teff\
scale.  \cite{mozur03} used the Mark III interferometer to determine
limb-darkenening corrected radii for a sample of very bright nearby stars.
These are combined with
a parallax to yield \teff, then the observed colors of
each interferometrically observed star are corrected for reddening to define
a \teff,~$V-K$ relation.  There is as yet insufficient data
to split their sample into giants and dwarfs,
low metallicity stars, etc.  However, if we  compare our
adopted \teff,~$V-K$ relation from \cite{houdashelt00}, 
assuming a mean metallicity of $-$0.2 dex for the bright field stars with
interferometric angular diameters, 
with that derived by \cite{mozur03}
we find good agreement, to within 50 K, at all values of \teff\
from 4000 to 6500 K.

Ignoring any systematic errors in the color-\teff\ relations,
which are believed to be, as suggested by the above comparison, small,
the uncertainty in \teff\ depends on the accuracy of the photometry,
i.e. on the brightness of the star, with the errors at $J$ and at $K$ 
dominating.  
For the brightest stars considered
here this is an uncertainty of 30~K, while for the fainter HES stars,
an uncertainty of $\pm100$~K results.

We adopt the surface gravity corresponding to our choice
of \teff\ for each star from the 12 Gyr, 
[Fe/H] = $-3.3$ dex isochrone of \cite{yi01}. There is little
sensitivity to the choice of [Fe/H] in this range of \teff\
and [Fe/H]; a change in [Fe/H] of the isochrone of +1.0 dex
produces, for a fixed age, a decrease in log(g) of $\sim$0.1 dex.
Except at the TO itself, for a fixed \teff, there are, however, 
two solutions for log(g), one more luminous than the TO, and one
less luminous.  Adopting the higher luminosity
results in the inclusion of more distant objects
in the magnitude limited HES sample, but there are far fewer subgiants
than main sequence stars in a 12 Gyr isochrone.  We initially
take the higher luminosity (lower surface gravity) case, but sometimes
the abundance analysis itself leads us to subsequently choose
the solution with luminosity below that of the TO.  
The uncertainty in \grav, once the choice of luminosity above or below
that of the TO is made, is small, $\pm0.1$ dex.
The resulting stellar parameters are listed in
Table~\ref{table_params}.  

We emphasize again that this procedure is completely 
consistent with that used in
all our earlier work on globular clusters stars, e.g.
\cite{cohen02,ramirez01,ramirez02} for M71; 
\cite{ramirez03} for M5,
and \cite{cohen04} for Pal~12, and is identical
to that we used in our earlier papers on EMP
field stars from the HES, i.e. \cite{cohen02}, \cite{carretta02} 
and \cite{cohen03}.  Use of such a procedure eliminates many
of the degeneracies between choice of damping constants, $v_t$ and
\teff\ for dwarfs described by \cite{ryan98}.

The wavelength scale of our spectra is set by observations of a Th-Ar
arc at least twice per night.
The radial velocity measurement scheme developed for very
metal poor stars in the HES relies upon a set of accurate
laboratory wavelengths for very strong isolated 
features within the wavelength range of
the HIRES spectra.  The wavelengths were taken from the NIST Atomic Spectra 
Database Version 2.0 (NIST Standard Reference Database \#78).
Using an approximate initial $v_r$,
the list of automatically detected lines, restricted to the
strongest detected lines only in the spectrum of each star,
was then searched for
each of these features.  A $v_r$ for each line was determined
from the central wavelength of the best-fit Gaussian, and the
average of these, with a 2.5$\sigma$ clipping reject cycle, 
defined the $v_r$ for the star.
Appropriate heliocentric corrections are then applied.
Because of  concerns regarding slit
filling in periods of good seeing
and because of
the complex reduction scheme involving
two data reduction packages with different algorithms for representing
the wavelength scale in the 2-D echelle spectra which
we adopted here, these $v_r$  must be assigned
an uncertainty of $\pm$1.5~\kms. 
Very metal poor stars which have been observed repeatedly by
George Preston \citep*[see][]{preston01}
are used as radial velocity standards.  The
observations of these stars are treated identically as the
sample stars; they serve to monitor the accuracy of the analysis.

The maximum $|{v_r}|$ listed in Table~\ref{table_params} is 359~\kms.
However, when the reflex of the Galactic rotation, assumed to be
220~\kms, is removed, HE0458$-$1346 has a galactocentric 
$v_r$ of +256~\kms.  Analysis of the radial velocity distribution of the much
larger sample of HES candidate EMP stars with follow up spectra,
most of which were taken with 4m or larger telescopes and have 
velocity accuracy of better than 10~\kms, separated
into giants and dwarfs using 2MASS colors, should provide a strong
constraint on the velocity ellipsoid of the Galactic halo
and the escape velocity from the Galaxy.

\section{Abundance Analysis \label{section_anal}}

We rely heavily in the present work on the procedures and atomic data
for abundance analyses of very metal poor stars
described in our earlier papers reporting the results
of the Keck Pilot Program on Extremely Metal-Poor 
Stars from the HES \citep{cohen02, carretta02,lucatello03}.

Given the derived stellar parameters from Table~\ref{table_params}, we 
determined the abundances using the equivalent widths obtained as 
described above.
The abundance analysis is carried out using a current version of the LTE
spectral synthesis program MOOG \citep{sneden73}.
We employ the grid of stellar atmospheres from \cite{kurucz93} 
without convective overshoot, when available.  Plane parallel model
atmospheres
are an excellent approximation for dwarfs.  We compute the
abundances of the species observed in each star using 
the four stellar atmosphere
models with the closest \teff\ and log($g$) to each star's parameters.
The abundances were interpolated using results from the closest stellar model
atmospheres to the appropriate \teff\ and log($g$) for each star given
in Table~\ref{table_params}.

The microturbulent velocity ($v_t$) of a star can be determined 
spectroscopically by requiring the abundance to be independent of the 
strength of the lines, see, e.g. \cite{magain84}.  
The uncertainty
in our derived $v_t$ is estimated to be +0.4,$-$0.2 \kms\
based on repeated trials with the same line list for several stars
varying $v_t$. However, since the lines
in these EMP candidate dwarfs are in general very weak, the exact
choice of $v_t$ is not crucial. 
We apply this technique here to the large sample of detected
Fe~I lines in each star; the results are listed with the
stellar parameters in Table~\ref{table_params}.  

At this point it became clear that the \teff\ assigned to the
brightest stars were much higher than their excitation temperatures $T(exc)$
determined from the abundances derived from their
Fe~I lines, which cover a range in $\chi$ of $\sim$3.5 eV.
The 14 new dwarfs from the HES have a mean $\Delta(T)$,
where $\Delta(T) =$ \teff\ $- T(exc)$, of $+95 \pm175$K, while the
two stars from the HK Survey presented here each showed $\Delta(T) \sim +500$K.
This is not expected for an accurate abundance analysis with valid
stellar parameters.
We obtained the visual photometry for the HES stars in the present sample
ourselves.
Many stars were observed on multiple nights, so we are sure
the $V$ mags of Table~\ref{table_newstars} are reasonably accurate.
A change
in $V-K$ of 0.15 mag corresponds roughly to a change in derived \teff\ of
250~K, and it appears
that the published optical photometry referenced in this table
is not sufficiently accurate
for present purposes for the three brightest stars.  We have therefore
set their \teff\ to be $T(exc) + 95$ K,
then redetermined their surface gravities.

The abundance analysis was carried out twice.  The first iteration used the
equivalent widths measured automatically as described above.  
The \eqw\ for lines in the template list
which were not picked up automatically were set to 5 m\AA.
The [Fe/H] for the model atmosphere was set to the metallicity
inferred from the moderate resolution follow up spectra,
[Fe/H](HK).
The results of the first trial were used to guide a
search by hand for
additional lines which should have been picked up automatically
but for various reasons were not,
specifically
those which gave abundances substantially lower than those of the
detected lines of the relevant species.  (The usual 
reasons for failure of the automatic \eqw\ routine to
pick up a feature are marginal
detections or, for stronger lines, assymetry
in the line profile.) The equivalent widths for the
set of additional lines which could be measured by hand
were added, and all remaining lines with \eqw\ still at the
default value for non-detections were then deleted from the line list for
the star.  At this time also we adjusted the [Fe/H] of the
stellar atmosphere model used to reflect the value of [Ca/H]
determined from the first trial. This was done in an effort to partially
take into account
the large $\alpha$-element enhancements seen in some EMP stars
and their importance in setting the opacities.  A similar
scheme for rescaling
of Solar composition isochrones to obtain $\alpha$-enhanced
isochrones, originally suggested by \cite{chieffi91}, has been widely used.

The results for the abundances of
these species in the 14 previously unpublished 
candidate EMP dwarf stars from the HES,
the two additional candidate EMP stars from the HK Survey in our sample 
and the EMP dwarf selected from the NLTT \citep{luyten80} 
proper motion survey 
are given in 
Table~\ref{table_abunda} to Table~\ref{table_abundd}.
We tabulate log$\epsilon(X)$, i.e. 
all abundances in these tables are given with respect to H = 12.0 dex.
Our adopted Solar abundances, described in \S\ref{at_data},
are given as the last set in Table~\ref{table_abundd};
note that we adopt 
log$\epsilon$(Fe) = 7.45 dex for the Sun. 
Upper limits are provided
in some cases when no lines of a key element could be detected;
they are indicated
by ``U'' in the tables.

Table~\ref{table_feabund} gives a comparison of the [Fe/H] determined
from our high resolution analysis compared to [Fe/H](HK) which is
discussed in detail in \S\ref{section_hescomp}.
The origin of the moderate 
resolution spectrum for each star is also indicated in this table.

Table~\ref{table_sens} gives the changes in the deduced abundances
for small changes in \teff, \grav, $v_t$ and
in the [Fe/H] of the model atmosphere used.  These again are changes
in log$\epsilon(X)$.  One is usually interested in abundance ratios;
changes in [X/Fe] can be derived by subtracting the
relevant entries.  The last column gives 
expected random uncertainties for [X/Fe] as determined from
the present data for a single star, combining in quadrature 
the uncertainties in [X/Fe] resulting from
the errors in stellar parameters established in
\S\ref{section_params}, i.e. an uncertainty of 
$\pm$100~K in \teff, of $\pm$0.1 dex in \grav, of $\pm$0.5 dex
in the metallicity assumed in the model atmosphere used for the
analysis, of $\pm$0.2 \kms\ for $v_t$, and a contribution
representing the errors in the measured equivalent widths.
This last term is set at 20\% (0.08 dex) for a single
detected line, and is scaled based on the number of
detected lines.  For Fe~I and Fe~II, the table lists the
total uncertainty in log$\epsilon$(Fe).  In some cases (i.e. Ni and Sr)
the dominant term in the error budget arises from the
uncertainty in $v_t$.  Systematic uncertainties, such as might arise
from errors in the scale of the transition probabilities for an element,
are not included in the entries in Table~\ref{table_sens}.

\subsection{Ionization Equilibrium and non-LTE}

Since we have not used the high resolution spectra themselves to determine
\teff\ or \grav, the ionization equilibrium is a stringent test of our analysis
and procedures, including in particular the assumption of LTE\footnote{This
statement ignores the issue of the choice of luminosity and hence of
\grav, above or below the TO.}.
The ionization equilibrium for Fe~I versus Fe~II is extremely good.
Excluding the one dwarf which is so metal poor that no Fe~II lines
could be detected, the average difference for the remaining 27
candidate EMP dwarfs in our sample
between [Fe/H] as inferred from Fe~II lines
and from Fe~I lines is $-0.02\pm0.10$ dex.  
A plot of the Fe ionization equilibrium as a function of \teff, with different
symbols used for the stars above and below the main sequence turnoff, is shown
in Figure~\ref{figure_feioneq}. 

For Ti, where both Ti~I and Ti~II are sometimes detected, 
the errors are larger as the absorption lines
from neutral Ti in such hot stars
are all very weak in the optical; even in the best spectra only a few 
Ti~I lines can be detected.
Excluding those stars for which 
no Ti~I lines could be detected, the ionization equilibrium
from Ti is almost as good;
the average difference from 15 stars between [Ti/H] as inferred 
from Ti~II lines and from Ti~I lines
$-0.15\pm0.15$ dex.

The Fe abundances  derived from the
        neutral and ionized lines shift out of equilibrium by $\sim$0.2 
        dex for a 250~K change in \teff\ in this temperature regime (see
Table~\ref{table_sens}).  Our uncertainty in \teff\ of $\pm$100~K can thus give
        rise to most of the dispersion observed in the Fe abundances
        between the neutral and ionized lines observed among the sample
        stars.

Among stars with almost Solar metallicity,
\cite{yong04} found 
in their extensive study of Hyades dwarfs with \teff\ between 4000 
and 6200 K
that inconsistencies in simple classical
LTE analyses appear to develop only at \teff\ $<$ 5000 K, a regime we
do not reach here. The very careful analysis
by \cite{allende02b} of Procyon also
shows no sign of such problems.
The most careful analyses of metal poor
globular cluster and field stars (but still of higher metallicity 
than those considered here)
in the range $4000 <$ \teff\ $< 6200$ K, such as that
of  \cite{cohen01} and \cite{ramirez01}
for a large sample of stars
over a wide range in luminosity in M71,
show that departures from LTE in the formation of Fe lines are
relatively small for these stars. 

In the still lower metallicity range considered here, the Keck Pilot Project 
\citep{carretta02} found, as we do again, that non-LTE does not 
appear to significantly
alter the results of a classical abundance analysis such as presented here.
The theoretical situation
is somewhat unclear, as the results of recent theoretical analyses 
\citep{gratton99,thevenin99} disagree on the amplitude to be expected.
\citet{gratton99} found that non-LTE corrections for Fe lines are very small in
dwarfs of any \teff, and only small corrections ($<$ 0.1 dex) are expected for
stars on the red giant branch.
\citet{thevenin99} found that non-LTE corrections become more important as [Fe/H]
decreases, being about 0.2 dex for stars with \fe $\sim -$1.25 dex, and that
lines from singly ionized species are not significantly affected by non-LTE.  
Recently, \cite{geh01a} and \cite{geh01b} have carefully calculated the kinetic
equilibrium of Fe, and present in \cite{korn02}
a critique of earlier calculations.  They suggest 
that non-LTE corrections intermediate between
the above sets of values are appropriate 
for Fe~I.

\subsection{Comparison with Previous High Dispersion Analyses}

The only star presented here that has been analyzed previously is
the brightest of the three very bright comparison stars,
LP 0831-07, studied by \cite{ryan91} and by 
\cite{thevenin99}.  The agreement  between the 
stellar parameters and metallicity derived here for this star
(see Table~\ref{table_params} and \ref{table_abunda}) 
and those of \cite{ryan91} is poor.  Their derived [Fe/H] is $-3.4$ dex, 
0.5 dex lower than our deduced value log$\epsilon$(Fe) = 4.59 dex, 
corresponding to [Fe/H] = $-2.86$ dex\footnote{This value is derived
from the Fe~I lines. The more uncertain value from the small number
of Fe~II lines yields a slightly higher [Fe/H].}. Half of this difference
is due to their adoption of the \cite{anders89} Fe abundance for the Sun.
The other half is due to their adoption of
a \teff\ which is 270 K cooler than our value.  (Note that we obtained
$T(exc) = 6175$ K for this star, and adopt \teff = 6270~K.)   Since
 \cite{ryan91} could not detect Fe~II or Ti~I lines, they were totally
dependent on the calibration  they adopted of their photometry with
\teff.  If we had adopted \teff = 6000~K, we would have obtained 
log$\epsilon$(Fe) ${\sim}4.25$ dex, which reproduces their Fe abundance.

Abundance ratios are less sensitive to the choice of stellar parameters
(see Table~\ref{table_sens}),
and hence there is reasonable
agreement, given their stated uncertainties, between the two analyses
for the values of [X/Fe] in LP 0831-07,
except for Ca/Fe,
% their ratio is 0.48 for Ca/Fe
where their ratio is $\sim$0.4 dex higher than ours.
This too is largely a result of the difference in
adopted stellar parameters since the \eqw\ for the 
Ca~I lines in common agree well.

\cite{thevenin99} analyzed just Fe~I in this star 
using the  \eqw\ and stellar parameters from \cite{ryan91}.
They obtained 
[Fe/H](LTE) = $-3.15$ dex, in accord with their adoption of a Solar
Fe abundance of log$\epsilon$(Fe) = 7.46 dex, quite different from that 
adopted by \cite{ryan91}.

Previous analyses of the brighter comparison stars included in 
the Keck Pilot Project are discussed in \cite{carretta02}.

\section{Comments on Individual Elements \label{sec_individual}}

\subsection{Iron \label{section_abundfe} }

\begin{figure}
\epsscale{0.8}
% Comment out the following line to embed the PS figure into the manuscript
% \plotone{/scr2/jlc/hamburg_survey/hires_summary/programs/dwarf_fehist.ps}
\plotone{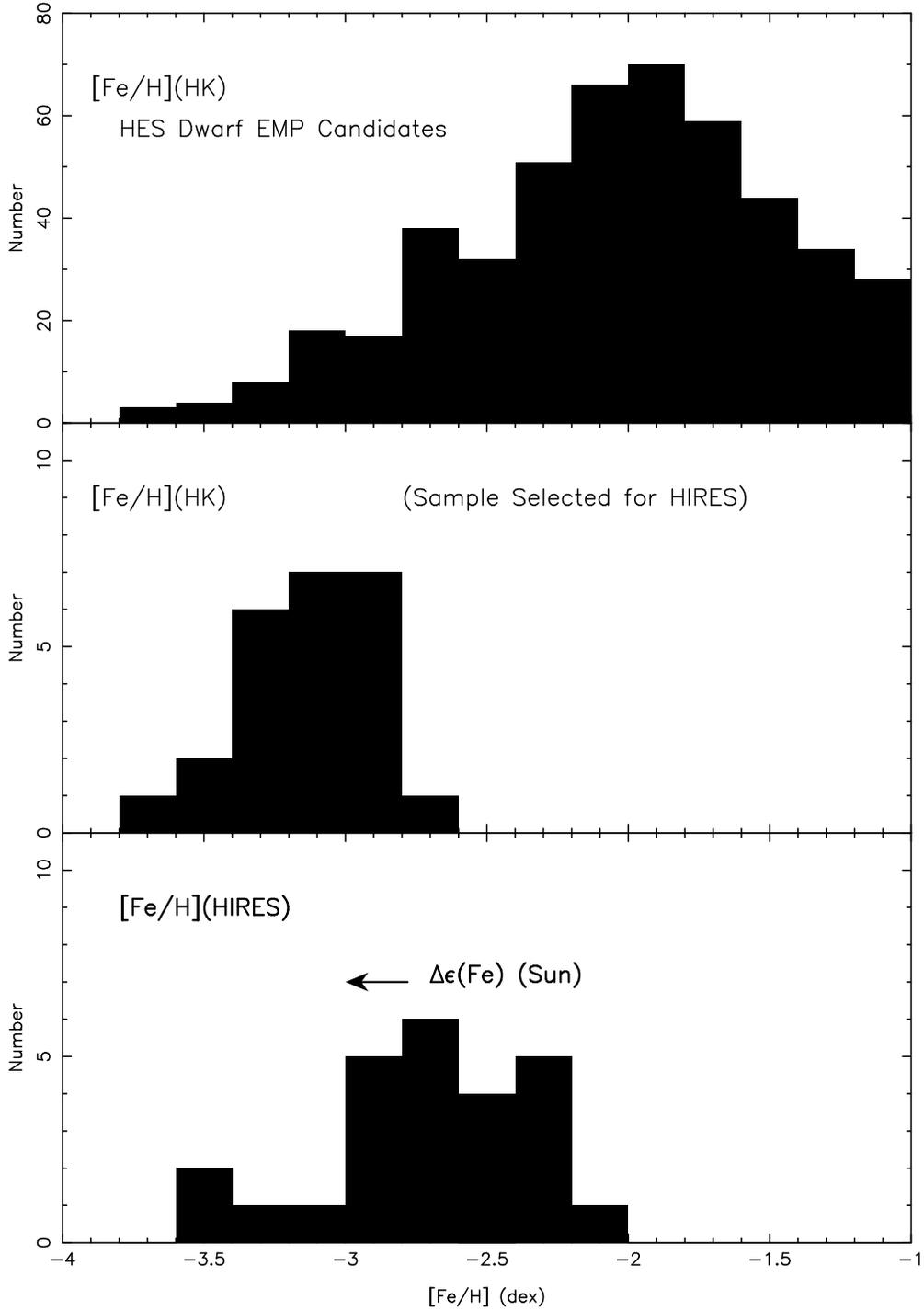}
\caption[]{A histogram in metallicity as indicated
by [Fe/H](HK) for the entire set of dwarf
EMP stars isolated from the HES is shown in the upper panel.  The middle
panel shows the sample chosen for high dispersion observations at Keck.
The lower panel shows the [Fe/H] determined from the detailed abundance
analysis for the set of stars displayed in the middle panel.
The arrow in the lower panel indicates the probable magnitude of the
shift of the metallicity scales due to different choices of the Solar Fe
abundance.
\label{figure_fehist}}
\end{figure}

\begin{figure}
\epsscale{1.0}
% Comment out the following line to embed the PS figure into the manuscript
% \plotone{/scr2/jlc/hamburg_survey/hires_summary/programs/dwarf_feioneq.ps}
\plotone{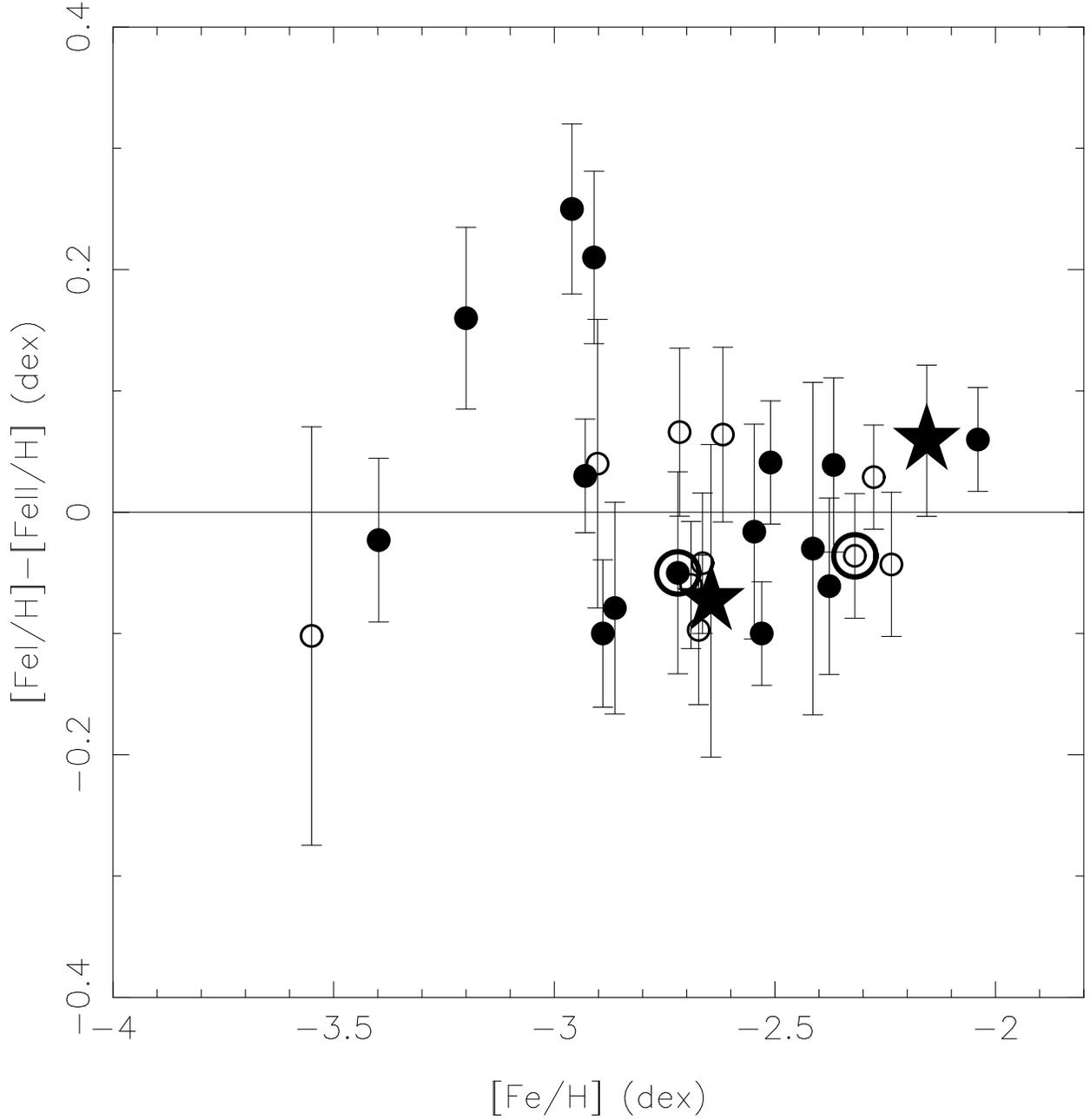}
\caption[]{The ionization equilibrium of Fe is shown as a function of 
Fe abundance for 
27 of the candidate EMP dwarfs in our sample with 1$\sigma$ error bars.  
The open circles denote the subgiants,
while the filled circles indicate the stars below the main sequence turn off.
The symbol key for the C-enhanced stars is that used 
throughout this paper; C stars are shown as large stars,
while the C-rich stars  are circled.  
\label{figure_feioneq}}
\end{figure}

The Fe abundances of our sample of EMP candidate dwarfs from the HES
reveal the behavior typical of halo stars from the HK and earlier
surveys \citep{bee92}, specifically a metallicity distribution showing a
steep decrease in the number of stars below
[Fe/H] ${\sim} -3$ dex as compared to metallicities only slightly
higher than this value.  A histogram of the [Fe/H] values for these 28 
stars as inferred
from our detailed abundance analyses is shown in the bottom panel of
Figure~\ref{figure_fehist}.
The arrow in the bottom panel of this figure denotes the appropriate shift
in deduced [Fe/H] for the abundance scale of the
high dispersion analyses to match the
choice made for the Solar iron abundance in the low dispersion study,
see \S\ref{at_data}.

The uncertainty expected in the determination of [Fe/H] 
for each star is given in the last column of
Table~\ref{table_sens}, and is 0.13 dex.  To demonstrate that this
rather low value
is a realistic error estimate we look at the dispersion in [Fe/H]
achieved by Cohen and her collaborators in their analyses of HIRES
spectra of large samples of globular cluster stars.  These
spectra are of similar precision but somewhat lower dispersion
than those used for the EMP dwarfs from the HES.
Using  an analysis procedure close to that used here,
\cite{ramirez02} found
a 1$\sigma$ rms dispersion in [Fe/H] of 0.12 dex, comparable
to the slightly larger predicted error of 0.14 dex,
for a sample of
25 stars in M5 covering a range in brightness similar to that of
the stars discused
here.   Similarly low $\sigma$ determinations are already available for 
globular clusters both more metal rich and
more metal poor than M5.  Thus we believe that the distribution in
the lower panel of
Figure~\ref{figure_fehist} reflects the true [Fe/H] distribution of the
sample of EMP dwarfs to within this very small error.  This distribution
from the high resolution spectra
is shifted towards higher metallicity and broadened compared to
the middle panel of Figure~\ref{figure_fehist}, an issue which is
discussed at length in \S\ref{section_hescomp}.

Our sample of 28 candidate EMP dwarfs has a median [Fe/H] of $-2.7$ dex,
corresponding roughly to [Fe/H](HK) = $-2.9$ dex, with two stars
reaching slightly below $-3.5$ dex.
All are below $-2$ dex,
and more than 75\% of the sample from the HES has [Fe/H] below 
$-2.5$ dex.

\subsection{Carbon and Nitrogen}

There are five stars in the sample with detectable features at the
G band of CH.  Two of these (HE0007$-$1832 and HE0143$-$0441)
also show weak C$_2$ bands and
hence are carbon stars.  Two others (HE0024$-$2523
and HE2148$-$1247) have [C/Fe] $> 1.0$ dex,
but C$_2$ bands are not detected,
and we denote them as C-enhanced stars. The two C-enhanced stars
are extremely peculiar and
have been discussed in great detail elsewhere, \mystar\ by \cite{cohen03} and
\sarastar\ by \cite{lucatello03}.  Weak CH is seen in G139$-$8 corresponding
to a Solar ratio of C/Fe, and not of interest here.

\begin{figure}
\epsscale{0.8}
% Comment out the following line to embed the PS figure into the manuscript
% \plotone{/scr2/jlc/hamburg_survey/hires_sep2001/moog_files/he0007_chsyn.ps}
\plotone{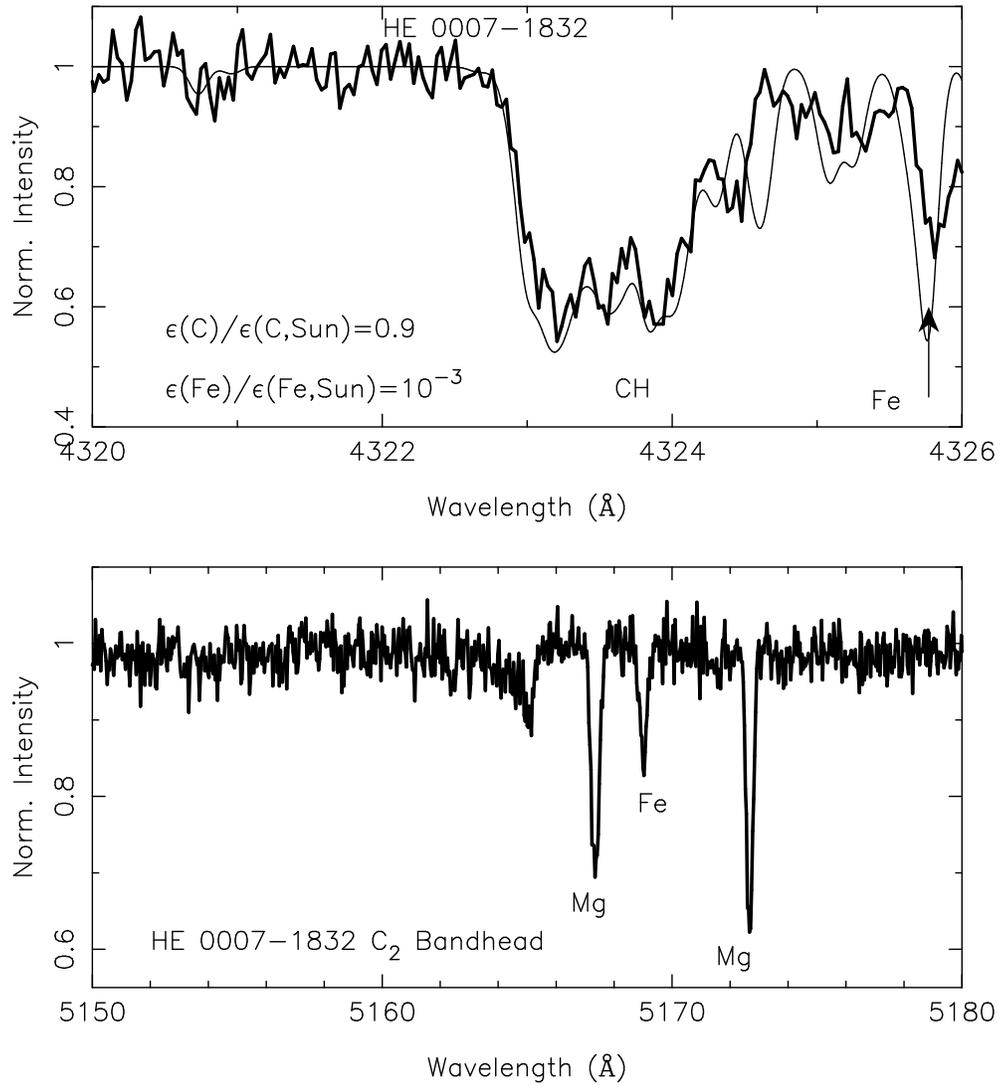}
\caption[]{A synthesis for the region of the CH band near 4320~\AA\ (thin line) is shown
superposed on the spectrum of the dwarf C star 
HE~0007$-$1832 (thick line).  The region of the
bandhead of C$_2$ near 5160~\AA\ is shown in the lower panel.
\label{figure_chsyn}}
\end{figure}

The C abundance for the two carbon stars
stars was determined from syntheses of the G band of CH.
The molecular line data for CH, including the $gf$ values and the
isotope shifts,
were taken from \cite{jorgensen94} and \cite{jorgensen96}.  
The synthesis was carried out first with an initial guess at \ciso, then
with the value determined from the spectra of each star.
However, the main bandhead of the G band at 4305 \AA\ was not used
due to concerns about continuum placement given the strength of the band
and the relatively short length of the Ech\`elle orders.  Furthermore
the 0-0 vibrational band is formed higher in the atmosphere, and thus
more subject to any errors in the temperature distribution at high layers.
The region from 4318 to 4329 \AA\ was used instead.
Figure~\ref{figure_chsyn} shows the spectrum of
HE~0007$-$1832, a carbon star, in this region with a synthesis
superposed, and in the region of the C$_2$ bandhead near 5160~\AA.

The O abundance is required to calculate the molecular equilibrium, but this is
not known for three of these four stars. Based
on the characteristics of other heavily C enhanced metal poor stars
\citep*[see, e.g.][]{lucatello03},
we adopt the larger of [O/Fe] = +0.5 or [C/O] = $-$0.5 dex for the calculation. 

The resulting C abundances for the four stars
range from log$\epsilon$(C) = 7.8 to 8.3 dex, corresponding to C/Fe enhancements 
between a factor of 40 and 400 compared to the Solar ratio. 
C abundances smaller than log$\epsilon$(C) $\sim$7.0 dex would produce features
at the G band that are not easily detectable 
in the spectra of dwarfs in this \teff\ range.

% \footnote{These were computed in the
% scale of \cite{grevesse98}, as it is with those abundances that
% our set of atomic data reproduce well the Solar CH and CN bands.
% Constants were then subtracted to move into the new Solar
% abundance values for C and N discussed in \S\ref{at_data}.}

% he2148-1247 log($\epsilon$(C)) = $8.08\pm0.2$ dex,just 0.5 dex below the solar value.
% he2148-1247  c12/c13 about 10, A ratio smaller than 5 can be ruled out
% he0024-2523  log($\epsilon$(C)) = 8.47 \pm 0.2 dex, C12/C13
%  from Lucatello, C12/C13 about 6
% old scale with Solar C about 8.56 dex.

The ratio of ~\ciso\ was determined by synthesizing selected regions
between 4210 and 4225 \AA. 
The resulting values of ~\ciso\ range from 6 to 9 for all four stars
(identical to within the errors),
again indicating substantial nuclear processing.

The N abundances were determined from syntheses in the region of the
3885~\AA\ CN band.  The resulting C/N ratios range from 10 to 150 while
the Solar ratio is four.

\subsection{Magnesium to Zinc \label{section_mg_to_zn}}

\begin{figure}
\epsscale{0.8}
% Comment out the following line to embed the PS figure into the manuscript
% \plotone{/scr2/jlc/hamburg_survey/hires_summary/programs/dwarfs_mgalsi.ps}
\plotone{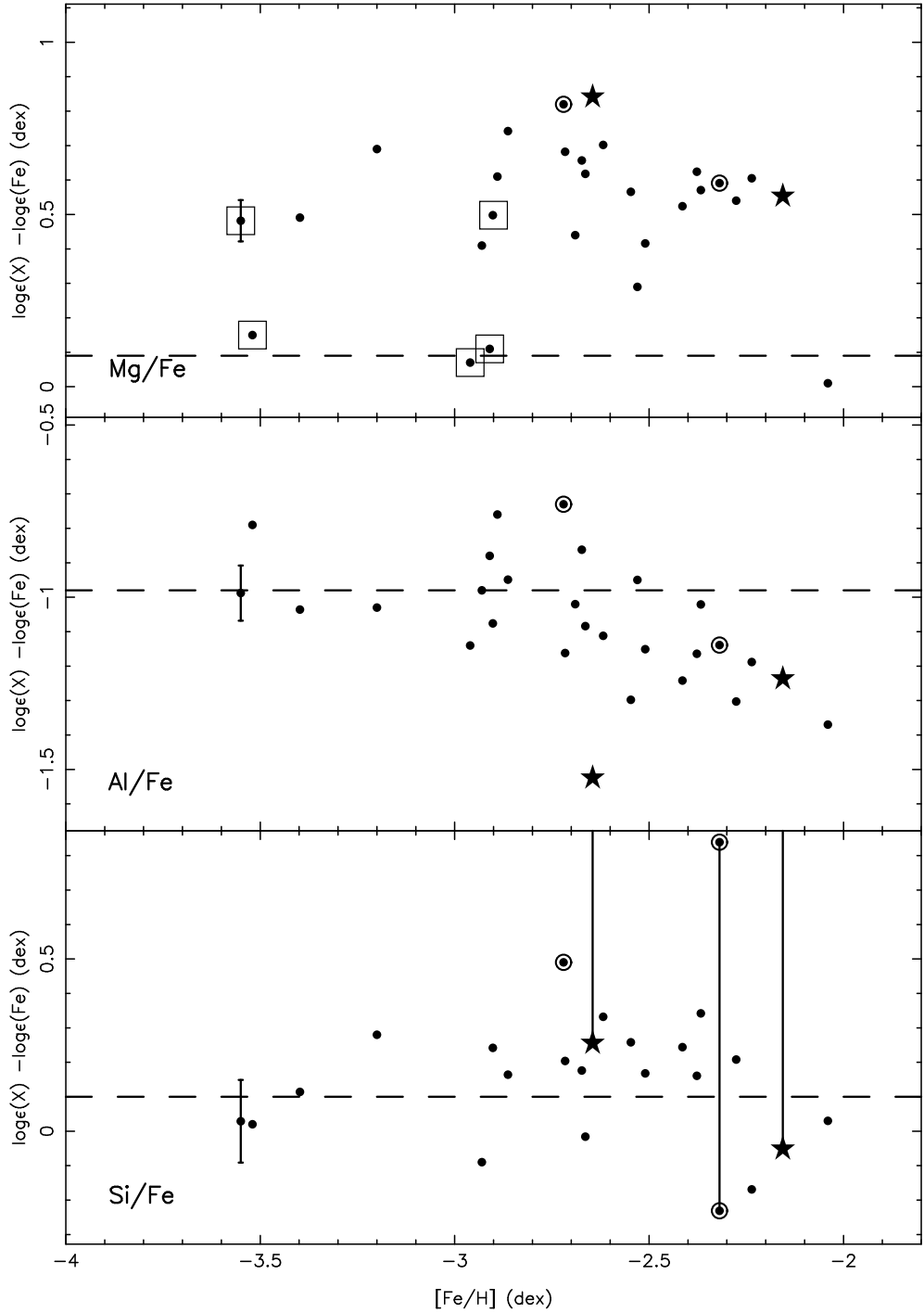}
\caption[]{The abundance ratios of Mg (top panel), Al (middle panel) and
Si (bottom panel) are shown with respect to Fe for the sample of 28
candidate EMP dwarfs.
The symbol key for the C-enhanced stars is that of Figure~\ref{figure_feioneq}.
The dashed horizontal line is the Solar ratio. 
A typical 1$\sigma$ uncertainty for each abundance ratio is shown
for the most metal poor star in each panel.
The vertical 
range is fixed at 1.2 dex for each panel. The stars where 
the only lines of Mg~I detected are those of the triplet at 5170~\AA\
are enclosed in rectangles in the upper panel. 
The [Si/Fe] values for the C-enhanced stars from the standard analysis 
of \eqw\ are connected to those obtained via spectral synthesis by vertical lines.
The vertical range is fixed at 1.2 dex for each panel.
\label{figure_mgalsi}}
\end{figure}

\begin{figure}
\epsscale{0.8}
% Comment out the following line to embed the PS figure into the manuscript
% \plotone{/scr2/jlc/hamburg_survey/hires_summary/programs/dwarf_cascti.ps}
\plotone{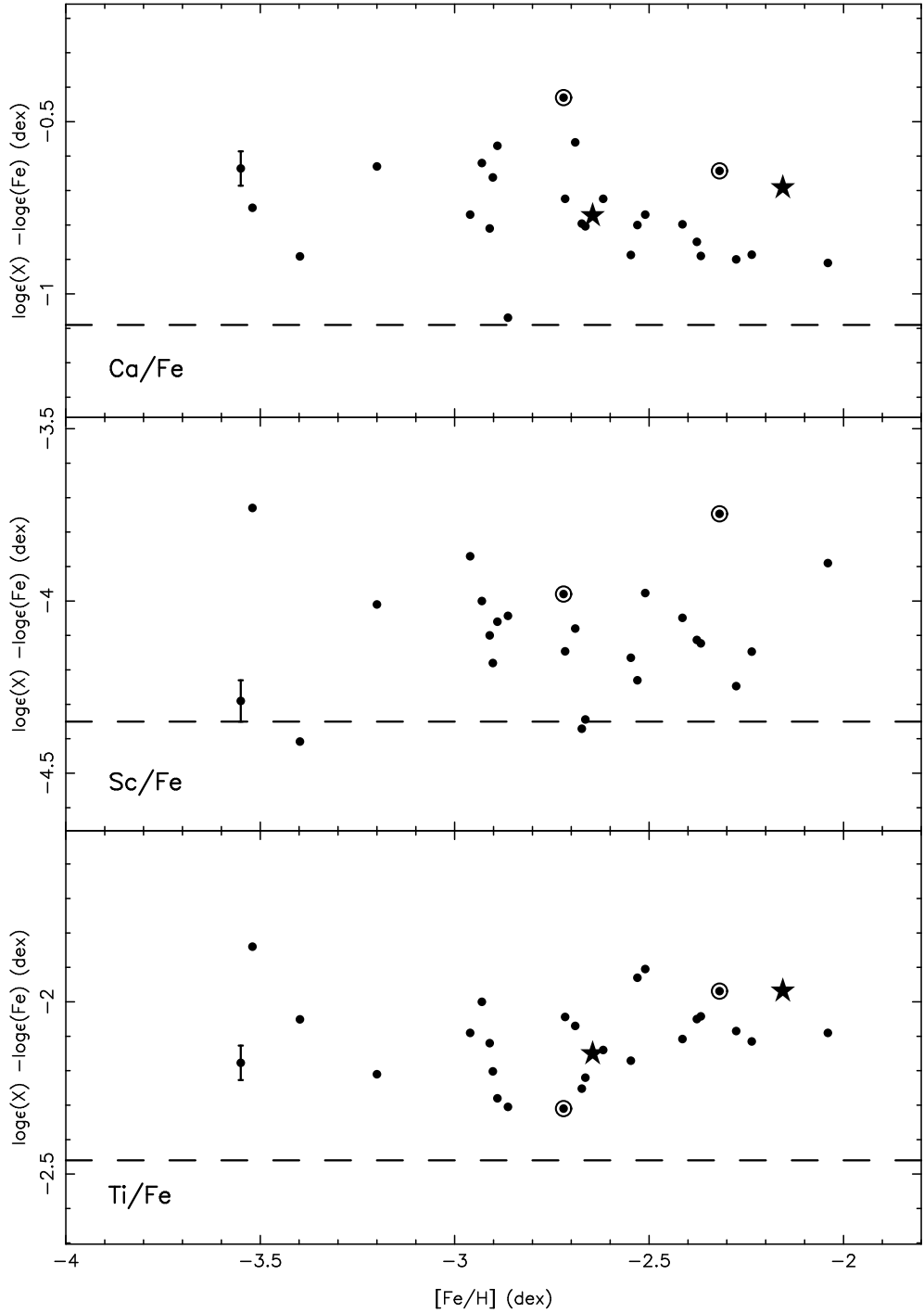}
\caption[]{The abundance ratios of Ca (top panel), Sc (middle panel, using
Sc~II and Fe~II) and
Ti (bottom panel, using Ti~II and Fe~II) are shown with respect 
to Fe for the sample of 28
candidate EMP dwarfs.
The symbol key is that of Figure~\ref{figure_mgalsi}.
The dashed horizontal line is the Solar ratio.
A typical 1$\sigma$ uncertainty for each abundance ratio is shown
for the most metal poor star in each panel.
The vertical range is fixed at 1.2 dex for each panel.
\label{figure_cascti}}
\end{figure}

\begin{figure}
\epsscale{0.8}
% Comment out the following line to embed the PS figure into the manuscript
% \plotone{/scr2/jlc/hamburg_survey/hires_summary/programs/dwarf_crmnco.ps}
\plotone{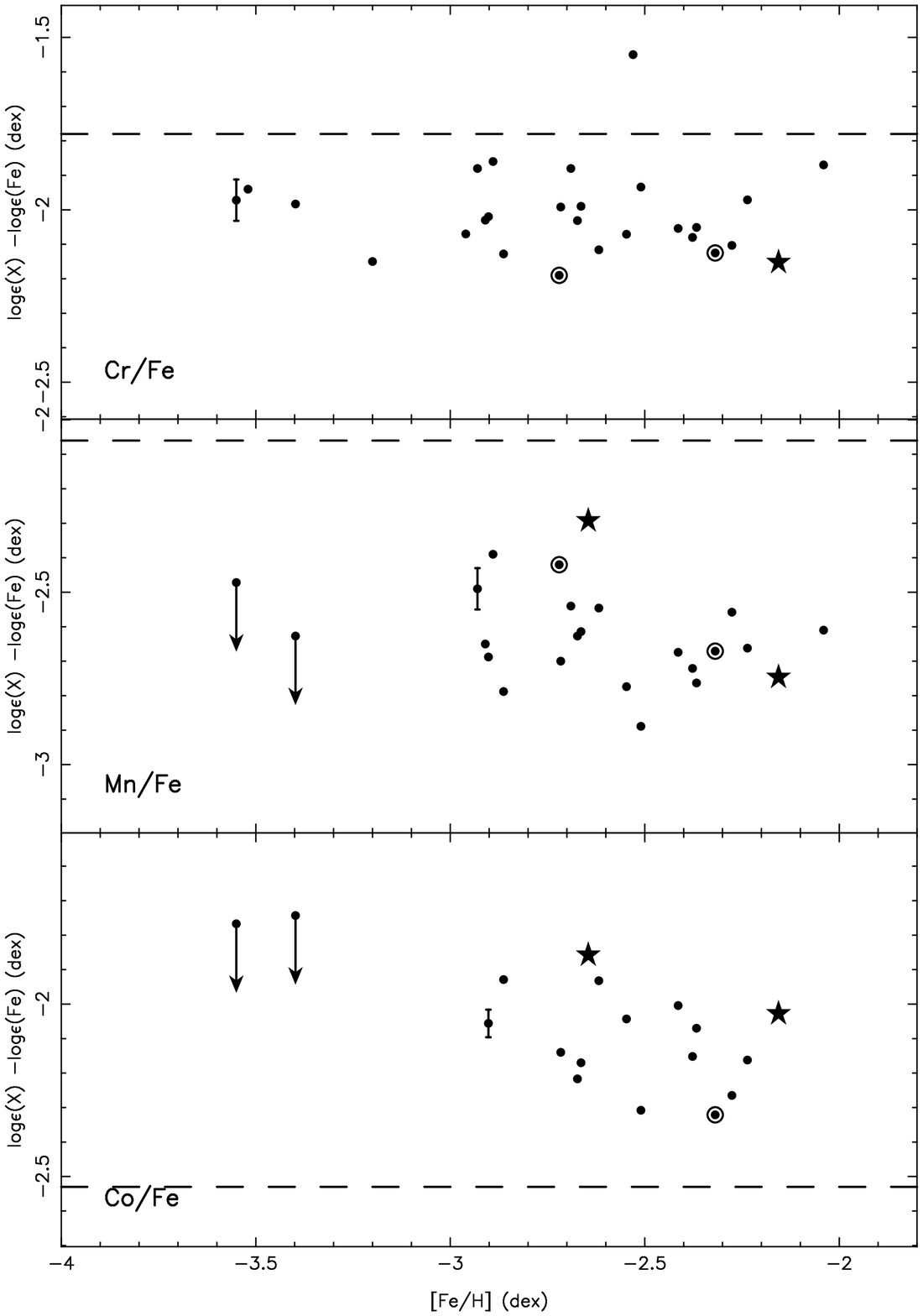}
\caption[]{The abundance ratios of Cr (top panel), Mn (middle panel)
and Co (bottom panel) are shown with respect to Fe for the sample of 28
candidate EMP dwarfs.
The symbol key is that of Figure~\ref{figure_mgalsi}.  
The dashed horizontal line is the Solar ratio.
A typical 1$\sigma$ uncertainty for each abundance ratio is shown
for the most metal poor star in each panel.
The vertical range is fixed at 1.2 dex for each panel.
\label{figure_crmnco}}
\end{figure}

\begin{figure}
\epsscale{1.0}
% Comment out the following line to embed the PS figure into the manuscript
% \plotone{/scr2/jlc/hamburg_survey/hires_summary/programs/dwarf_nizn.ps}
\plotone{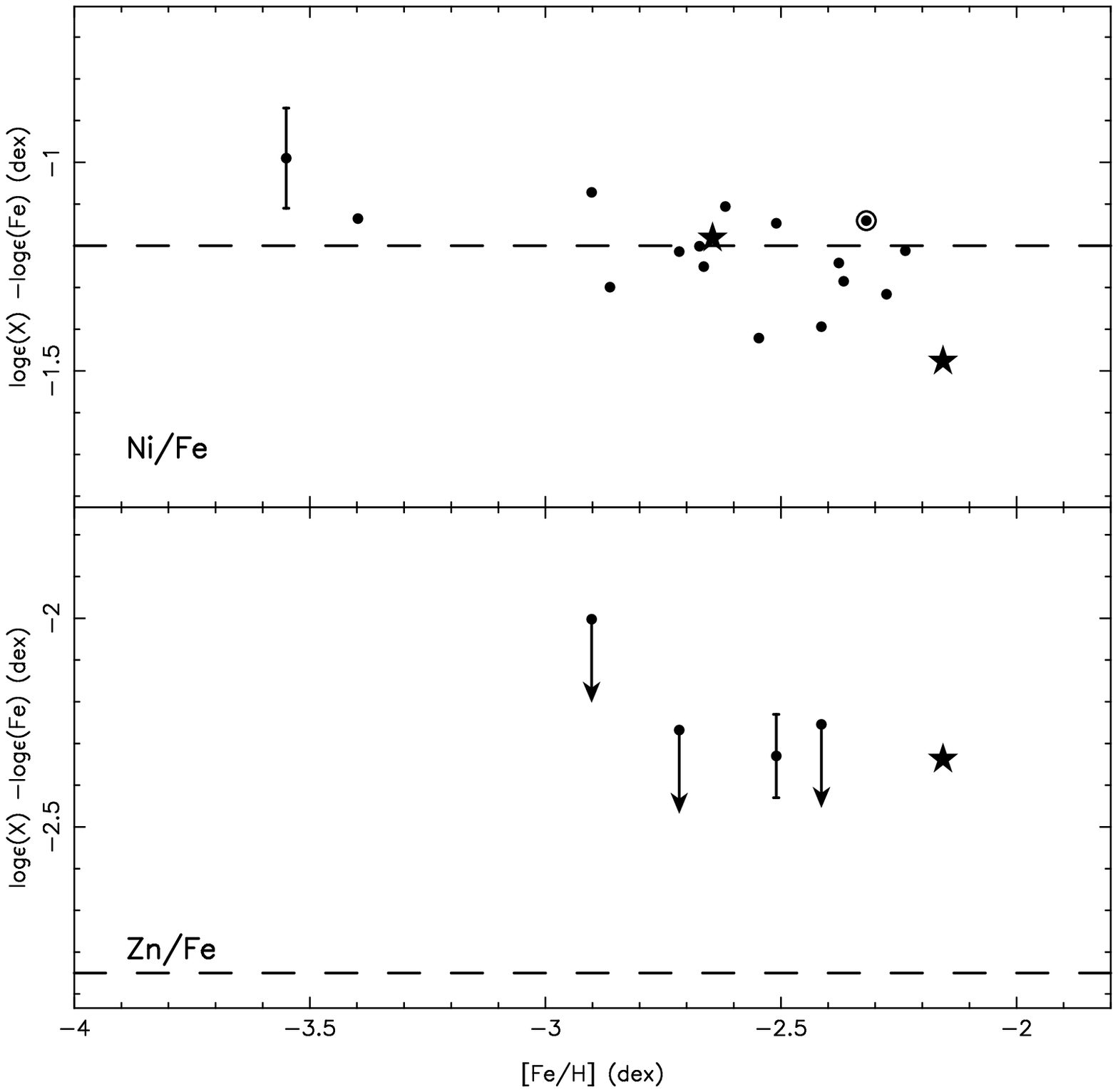}
\caption[]{The abundance ratios of Ni (top panel)
and Zn (bottom panel) are shown with respect to Fe for the sample of 28
candidate EMP dwarfs.
The symbol key is that of Figure~\ref{figure_mgalsi}. 
The dashed horizontal line is the Solar ratio.
A typical 1$\sigma$ uncertainty for each abundance ratio is shown
for the most metal poor star in each panel.
The vertical range is fixed at 1.2 dex for each panel.
\label{figure_nizn}}
\end{figure}

The abundance ratios for the 28 candidate EMP dwarfs in our
sample of Mg, Al and Si with respect to Fe
are shown in Figure~\ref{figure_mgalsi}.  The vertical axis in each of
Figures~\ref{figure_mgalsi} through ~\ref{figure_nizn}
% , Figure~\ref{figure_cascti},
% Figure~\ref{figure_crmnco} and Figure~\ref{figure_nizn}
is $AB \equiv$ log$\epsilon$(X) $-$ log$\epsilon$(Fe), so that [X/Fe] is 
$AB$(star) $ - AB$(Sun).
The vertical scale in each of these figures
is set to $\pm$0.6 from the mean value 
of the sample, excluding upper limits. 
Most of the dwarfs form a tight group at [Mg/Fe] = +0.5 dex
(which we will call the main group). There are four stars 
with [Mg/Fe]~$\sim 0$ which seem to form a separate small class (the
``small group''). The difference of $\sim 0.5$ dex in [Mg/Fe] between this group of 
four stars and the main group is too large to be due to observational error.
\cite{mcwilliam95} also saw evidence of a similar range in [Mg/Fe].

Because of possible evolutionary shifts in the light metals with Fe, we 
have calculated the slope of the best fit least squares line to the [Fe/H] 
versus [Mg/Fe] relation for the sample with the C-rich\footnote{We use
this term to include the two C-enhanced stars and the 
two carbon stars in our sample.}
stars excluded. The result is  given in Table~\ref{table_slopes}
and is indistinguishable 
from zero. The dispersion of [Mg/Fe] calculated about the best linear fit
is also given
Table~\ref{table_slopes}; it differs little from that calculated
assuming a constant [Mg/Fe].

We calculate the dispersion of [Mg/Fe] for all the stars, 
then excluding the C-rich stars, and finally after also excluding the
``small group''.  Since the slopes are small and uncertain, these are calculated
about the mean  ratio for [Mg/Fe] as is the case for all elements considered here.
The results are listed in Table~\ref{table_disp}.
The values from this table
should be compared to those expected from random uncertainties
in the stellar parameters and equivalent widths, denoted $\sigma$(pred)
and given in the last
column of Table~\ref{table_sens}.  If 
$\sigma$(obs)/$\sigma$(pred) $\le 1.5$, we assert that
observational error is dominating the observed scatter in the abundance ratios
for a particular species [X/Fe].
For Mg, even after culling out the ``small group'' mentioned above,
this ratio is 2.2, suggesting that there may still
be a hint of a genuine spread in [Mg/Fe],
a subject to which we return in \S\ref{section_disp_real}. 

For [Al/Fe] we note that there appears to be a trend
for an increase in this ratio as [Fe/H] decreases,
corresponding to the negative slope given in Table~\ref{table_slopes}.
We are using a constant correction for non-LTE 
for Al~I, which should be acceptable over the small range in
stellar parameters considered here.  The ratio of the observed
to expected dispersion is 2.0, but there are only 1 or 2 Al~I lines
and hence the expected dispersion may be underestimated.

The C-rich stars initially appeared to have high Si/Fe.  However, 
we only use
a single Si~I line at 3905~\AA, which is subject to
blending by CH lines, so spectral syntheses were used to determine
the abundance of Si in the C-rich stars.  These yield
abundances of Si substantially lower than those obtained
with the standard analysis and are indicated by ``S''
in Tables~\ref{table_abunda} to 
\ref{table_abundd}\footnote{Since HE2148$-$1247
was analyzed in \cite{cohen03}, it is not included in
Table~4a to 4d of the present paper.  However, the synthesis indicates
that the Si abundance given in the published analysis needs to be
reduced by a factor of $\sim$5.}.
An arrow indicates the change
in the [Si/Fe] values from the standard analysis to 
those obtained from spectral syntheses for the C-enhanced stars
in the lower panel of Figure~\ref{figure_mgalsi}.
As a check, syntheses of the 3905~\AA\ line
were also carried out for three EMP dwarfs with
no sign of C enhancement.  The mean difference in the derived
log$\epsilon$(Si) from the syntheses and from the 
standard analysis of equivalent widths
was only $-0.03$ dex, confirming that the standard approach is adequate
for most of the stars.
With these improved Si abundances, a small dispersion in the 
Si abundance for the entire sample follows;
if the C-rich stars are excluded, the dispersion
in [Si/Fe] falls slightly further to the value
expected from the known sources of random error given in the last column of
Table~\ref{table_sens}.

The abundance ratios for Ca, Sc and Ti with respect to Fe
are shown in Figure~\ref{figure_cascti}.
The dispersion
in [Ca/Fe] is roughly twice that expected from known 
random uncertainties when the C-rich stars are excluded.
We find a 
a slight negative slope to the relationship of [Ca/Fe]
versus [Fe/H], significant at the 2.6$\sigma$
level.
We calculate [Sc/Fe] using the Fe~II abundances to match the
use of ionized Sc lines.  

For Ti, we only use the Ti~II lines, as the Ti~I lines are 
weak and unreliable.  Hence we
again use the Fe~II abundances to 
better compensate against small errors in \teff\ or \grav.
The C-rich stars lie within
the main distribution.  The dispersion is roughly twice the value
expected from the known sources of random error given in the last column of
Table~\ref{table_sens}.

\begin{figure}
\epsscale{1.0}
% Comment out the following line to embed the PS figure into the manuscript
% \plotone{/scr2/jlc/hamburg_survey/hires_summary/programs/dwarf_simnfull.ps}
\plotone{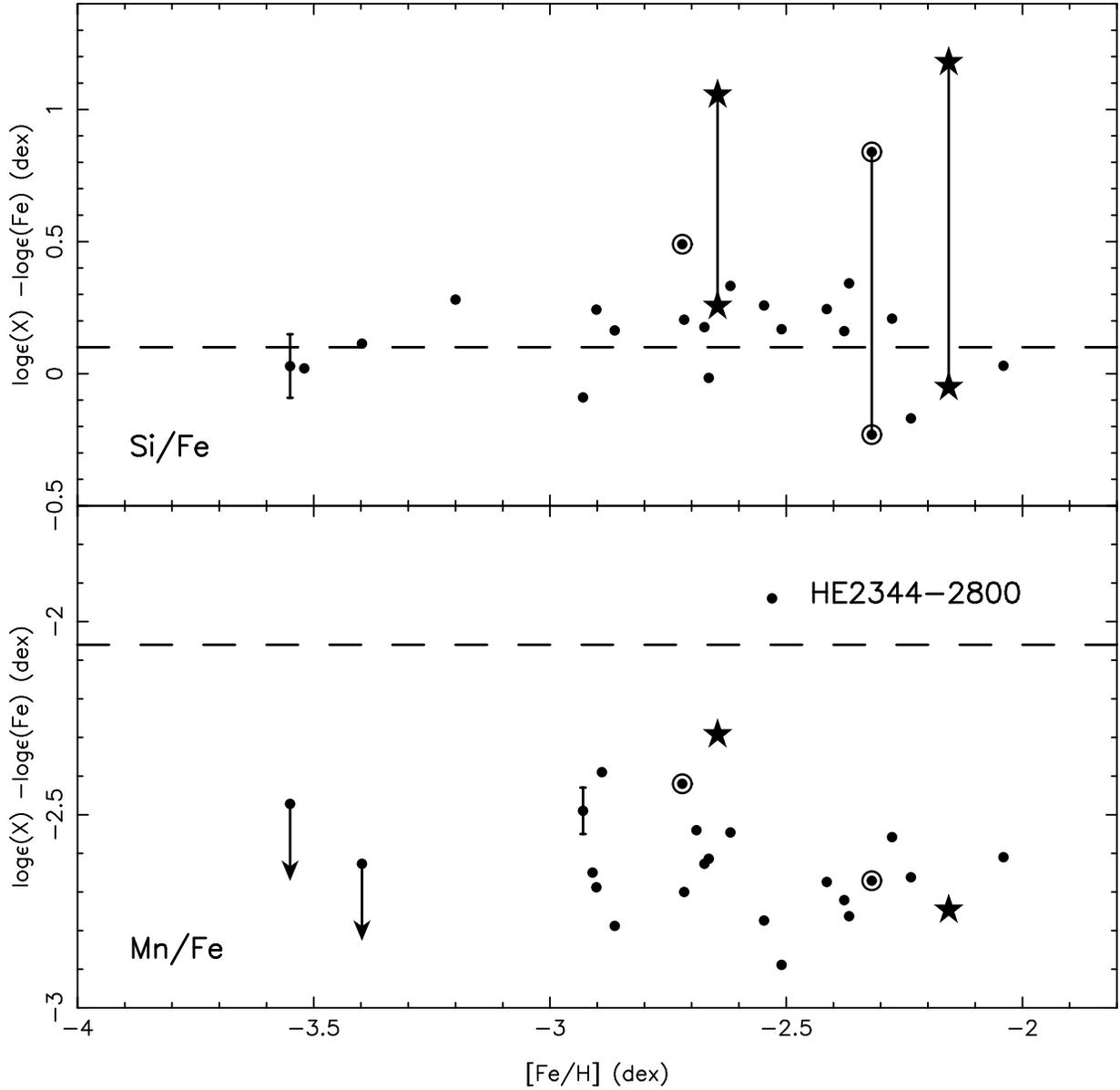}
\caption[]{The abundance ratios of Si (top panel)
and Mn (bottom panel) are shown with respect to Fe for the sample of 28
candidate EMP dwarfs.
The symbol key is that of Figure~\ref{figure_mgalsi}.  
The dashed horizontal line is the Solar ratio.  
A typical 1$\sigma$ uncertainty for each abundance ratio is shown
for the most metal poor star in each panel.
The vertical range is larger here than in the previous figures, 
so that the full range of the data can be displayed. 
The [Si/Fe] values for the C-enhanced stars from the standard analysis 
of \eqw\ are connected to those obtained via spectral synthesis 
by vertical lines.
\label{figure_simnfull}}
\end{figure}

The abundance ratios for Cr, Mn and Co with respect to Fe
are shown in Figure~\ref{figure_crmnco}.
For Cr, the C-enhanced stars lie at the low end of the
distribution, but not far off.  There is one star,
HE2344$-$2800, which has an abnormally high Cr abundance.
This star, part of the Keck Pilot Project, also has an
extremely high Mn abundance, which easily stands out in its
spectrum as illustrated in Figure~11 of \cite{carretta02}.
The wavelength range of the Keck Pilot Project does not include the
strong Co~I lines around 3850~\AA\ which are those most often
seen among these EMP stars.
The range of [Mn/Fe] over the entire sample is too large
to be encompassed within the vertical range of the middle panel in
Figure~\ref{figure_crmnco}; a complete display
of the data can be found in  Figure~\ref{figure_simnfull}.
The dispersion in [Cr/Fe]
becomes 1.5 times that expected from the known sources of
random errors once the C-rich stars and
HE2344$-$2800 are excluded.

Once HE2344$-$2800 and the C-rich stars are excluded,
the dispersion in [Mn/Fe], ignoring the dwarfs with
only upper limits for the Mn abundance, is twice that 
from the known sources of
random errors.
The expected dispersion of [Co/Fe] is unrealistically  small, only 0.04 dex;
not surprisingly, the observed dispersion is larger.

The abundance ratios for the 28 candidate EMP dwarfs in our
sample of Ni and Zn with respect to Fe
are shown in Figure~\ref{figure_nizn}.
Ni shows a dispersion at the value expected from the known sources of
random errors,
with the mean of [Ni/Fe] being
indistinguishable from the Solar ratio.  [Ni/Fe] versus [Fe/H]
has a slope of
$-0.20$ dex/dex, different from 0.0 by 3$\sigma$, assuming
the uncertainties per star have been accurately estimated.  Only one
Ni~I line was detected in almost all of these stars, and it
is at 3858~\AA, where the spectra are somewhat noisier.  This line is
fairly strong, so the $v_t$ contribution to the uncertainty
in [Ni/Fe] pushes the expected
$\sigma$ to a value somewhat higher than that for most other elements. 

There are only two
detections of Zn, one of which is in a carbon star.
The lines of Zn in the optical spectral region  
of the spectra of these dwarfs are  very weak and difficult
to detect.  

In summary, Table~\ref{table_disp} contains 
10 entries for dispersions in abundance
ratios [X/H] for elements in the range discussed here.
Eight of these are smaller than 0.15 dex when the C-rich stars
and a small number of outliers
are excluded.  The largest is only 0.16 dex.
These are remarkably small.
They should be compared with the uncertainty in the
determination of [X/H] for a single star, given for each species
in the last column of Table~\ref{table_sens}.  It is clear
that for at least some of the elements
the dispersion is still dominated by observational errors
rather than intrinsic scatter.
The linear least squares fits to the
relations [X/Fe] versus [Fe/H] are flat (have zero slope)
to within 1.5$\sigma$ for the elements with more than a few detected lines,
equivalent to reaching a plateau in these abundance ratios.
For some of the elements with only a few detected lines,
the slopes differ from zero by 2.5 to 4$\sigma$, but the
uncertainties in abundance ratios may have been underestimated in
these specific cases.  See Tables~\ref{table_slopes} and
\ref{table_disp} for details.

\subsection{A Cautionary Tale: The Origin of the Dispersions in Abundance Ratios 
\label{section_disp_real} }

Many of the dispersions about the relationship
between abundance ratio [X/Fe] and overall metallicity [Fe/H]
for the elements presented in Table~\ref{table_disp},
discussed  in the previous section, are about
twice the expected values; Si, Cr and Ni have
have dispersions  closer to the expected values.  The key question is whether
these abnormally large dispersions represent real star-to-star dispersions
in abundance ratios or whether they arise from some effect
neglected thus far in the analysis.  The dispersions given in 
Table~\ref{table_slopes},
calculated about the linear fit to [X/Fe] versus [Fe/H], 
are only slightly smaller
than those taken about the mean of [X/Fe] and given in Table~\ref{table_disp}.
Thus ignoring the trends in [X/Fe] cannot
contribute substantially to the missing factor of two.

To understand whether this 
factor of $\sim$2 excess in $\sigma$
represents small real variations in [X/Fe] for some of the species
or whether the values of the dispersions expected from the known sources of
random errors given in the last column
of Table~\ref{table_sens}, which are quite small, 
have been underestimated, we look at the residuals in several elements.
We define $R(X) = $[X/Fe] $- <$[X/Fe]$>$.  If the factor of two apparent
excess in dispersion is a result of genuine star-to-star variations
in abundance ratios, then we would expect $R(X)$ to be correlated
with $R(Y)$, where $X$ and $Y$ are elements close together in the periodic
table with similar nucleosynthetic histories.  If these are instead the
result of an underestimate of our predicted errors, then $R(X)$ and $R(Y)$
should not be correlated.

Figure~\ref{figure_cati_disp} displays $R$(Ti) (from Ti~II lines) 
versus $R$(Ca).  The distribution is roughly circular and
is centered on the origin when the C-rich stars are ignored.  This
suggests that the expected random errors for these two elements have
been slightly underestimated, producing a dispersion in both [Ca/Fe]
and in [Ti/Fe] which is roughly twice the predicted value.
A similar plot is shown for $R$(Cr) versus $R$(Ti)
(Figure~\ref{figure_ticr_disp}).  Again the
2D-distribution of the residuals is roughly circular 
with the exception of the single
outlier HE~2344$-$2800, which was found by the Keck Pilot Project
to have extremely strong Mn lines, and we have seen here  that this extends
to Cr as well.

Figure~\ref{figure_mgca_disp}, displaying $R$(Mg) versus $R$(Ca),
on the other hand, shows a strong
elongation along the X axis, suggesting real variations in 
[Mg/Fe].   There is a roughly circular distribution, centered at +0.1,+0.1,
not at the origin, with a tail of stars with very low [Mg/Fe] 
(the previously noted ``small group'') and with enhanced
Mg/Fe in some, but not all, of the C-rich stars.  Thus at first glance
it appears that the
range in [Mg/Fe] discussed earlier in \S\ref{section_mg_to_zn} is real
and is not primarily the result of an underestimate of our predicted errors.

However, we have only considered up to now the random errors that occur in
$R$(X). At this point we have to consider the systematic errors that
might occur for the particular case of Mg.  There are five lines
of Mg~I in the template line list, two of the three lines in the Mg~I 
triplet (the third is too blended to use), which have $\chi$=2.7 eV,
and three much weaker lines arising from $\chi$=4.3 eV.  There is a clear
opportunity for a systematic effect in that the triplet lines, being strong,
are detected in all of the sample EMP dwarfs, while the weaker 
higher excitation lines are not detected in the most metal-poor stars in
the sample nor in the stars with spectra noisier than typical.

We consider whether errors in \teff\ could be important given the
difference in $\chi$ of the Mg lines we use.  
The difference in abundance
for our nominal 100~K \teff\ uncertainty for the triplet versus the
higher lines is 0.07 dex in the sense that if
\teff\ is overestimated by 100~K, the triplet lines will give an abundance
which is higher than that of the other three lines by +0.07 dex.
However, this will manifest itself as a random error, and cannot
produce the ``small group'' seen in Figure~\ref{figure_mgca_disp}.

To check whether any such systematic bias has occured in
our sample, we look at the group of 7 stars for
which all five Mg~I lines were detected, for which the SNR of the HIRES
spectra is 100 or more, and which are not C-rich.  We only consider the
previously unpublished stars presented here to ensure full
access to all analysis data.
We check the line-by-line deviations from the mean Mg~I abundance for each
of those stars and average the results, which are given in Table~\ref{table_mg}.
The entries in this table clearly show that the Mg triplet lines yield
lower Mg abundances than do the higher excitation blue Mg~I lines.  If only
the triplet lines are detected, we expect the deduced Mg abundance 
for a star to be, on average, 0.12 dex lower than if all five Mg lines
were detected.
The data for the full sample of stars confirm the reality of this difference. 
The mean of [Mg/Fe] is 0.38 dex for
the sample ignoring the C-enhanced stars (24 stars) 
(see  Table~\ref{table_disp}), while
for the 5 EMP dwarfs where only the triplet was detected it is 0.24 dex.
This is very close to the predicted difference of 0.12 dex.
The entries in the second row of the table gives the values when
the stars with only the Mg triplet lines detected are omitted.

We next seek to identify the mechanism(s) that could produce these 
line-to-line differences in the deduced Mg abundance.
There are at least two possible sources of 
systematic errors in the atomic data for Mg~I,
the transition probabilities and non-LTE corrections.
We did not include the latter for Mg.  \cite{zhao00}
and \cite{gehren04} have calculated such in considerable detail, and
find typical non-LTE corrections of +0.1 dex for [Mg/Fe] in 
very metal poor stars. Over the small
\teff\ range of the HES dwarf sample, these corrections will, for a given
Mg line, be approximately constant. \cite{zhao99} give non-LTE
corrections for 12 lines of Mg~I, including three of the five we use, for
the range of stellar parameters and metallicity of the stars in our sample
of EMP dwarfs.  They find that the non-LTE corrections are $\sim$0.1 dex
smaller for the Mg triplet than for the weaker higher excitation blue lines,
which is the reverse of that required to reduce our observed
dispersions in [Mg/Fe], but such calculations are extremely difficult.
In any case, the assumption of LTE could in this manner introduce a systematic bias
which could lead to an
increase the observed [Mg/Fe] scatter above its intrinsic value. 

Similarly, if the $gf$ values for the 5 Mg~I lines we use are not correct,
and are scaled differently for the triplet lines than for the higher
excitation blue lines, a systematic bias will also occur. 
We use the Mg~I $gf$ values derived for the Keck Pilot Project whose
provenance is described
in the Appendix to \cite{carretta02}; see, in particular, Table~9
of that paper.  The $gf$ values for Mg~I are, as described there 
\citep*[see also][]{ryan96}, highly
uncertain with significant variations in their values as given by
the various references cited in \cite{carretta02}.  \cite{gratton03}
are now using $gf$ values that are significantly different from those we
adopted in 2002 for the higher excitation blue lines.

Without a careful study of bright stars whose stellar parameters
are similiar to those studied here with an extremely high precision
high resolution 
spectrum with full spectral coverage of the optical regime
it is not possible to determine the exact origin of this systematic
error which we know is present in the data.  If it arises
from the transition probabilities, we cannot determine from the present data
which of the adopted $gf$ values are correct and give the nominal Solar 
abundance.  For purposes of the present discussion of
the dispersion of the abundance ratio [Mg/Fe] within the present sample,
this issue is irrelevant.

Our present work has demonstrated that the abundance ratio dispersions
in EMP dwarfs is very small, $\le$0.15 dex.  To proceed further 
in the future we
must operate at such a high level of precision that effects previously
ignored may become 
significant contributors to the total error budget
in [X/Fe].
In particular, systematic errors in the atomic data for any 
particular line of an element
with only a few detected lines (such as Mg~I) can
combine with a range in line strength such that not all the
lines of the species are detected in all the stars to introduce systematic
biases into the derived abundances.  This can in principle
increase the apparent dispersion of [Mg/Fe] to a level which is
larger than its intrinsic value.  

The stronger Mg triplet lines are observed in all the stars in our sample.
They have the advantage that they are in a region relatively free of
molecular bands and hence should give reliable abundances even in the
C-rich stars.
So we calculate the  [Mg/Fe] ratios for each star using only the
5172 and 5183~\AA\ lines of Mg~I, thus eliminating many of the concerns
about atomic data expressed above.
Figure~\ref{figure_mgtriplet} presents the resulting [Mg/Fe] ratios 
as a function of [Fe/H]. 
We obtain a 
result [Mg/Fe] distribution very similar
to that obtained when all the available Mg~I lines are used,
shown in Figure~\ref{figure_mgalsi} and \ref{figure_mgca_disp}.
The ``small group'' of stars with Mg/Fe approximately Solar is
still present, and we have thus far
found no obvious explanation that would produce
this except real star-to-star variations. 
However, in Figure~\ref{figure_mgtriplet} we differentiate between
the stars whose analyses are presented here and those presented
in the Keck Pilot Project \citep{carretta02}.
This figure strongly suggests that there may be some systematic difference
between the two analyses affecting the derived
[Mg/Fe] abundance ratios.  Since the Mg triplet lines are always
fairly strong, a possible culprit is the determination
of $v_t$.  We thus conclude that we have no credible evidence
at this point for a real range in [Mg/Fe] ratios at
very low [Fe/H].

This cautionary tale suggests that careful attention must be
paid in future efforts to detect 
dispersions in abundance ratios at even lower levels
to the atomic data.   Searches for line-by-line effects, particularly
for species which have only a small number of detected lines, will
have to be undertaken.  Furthermore,
it reminds us of the difficulties of combining for this purpose
multiple published analyses where the set of lines
used for each species, the atomic data and the details of the analysis 
may all differ.

\subsection{The Heavy Elements \label{section_heavy}}

\begin{figure}
\epsscale{1.0}
% Comment out the following line to embed the PS figure into the manuscript
% \plotone{/scr2/jlc/hamburg_survey/hires_summary/programs/dwarf_srba.ps}
\plotone{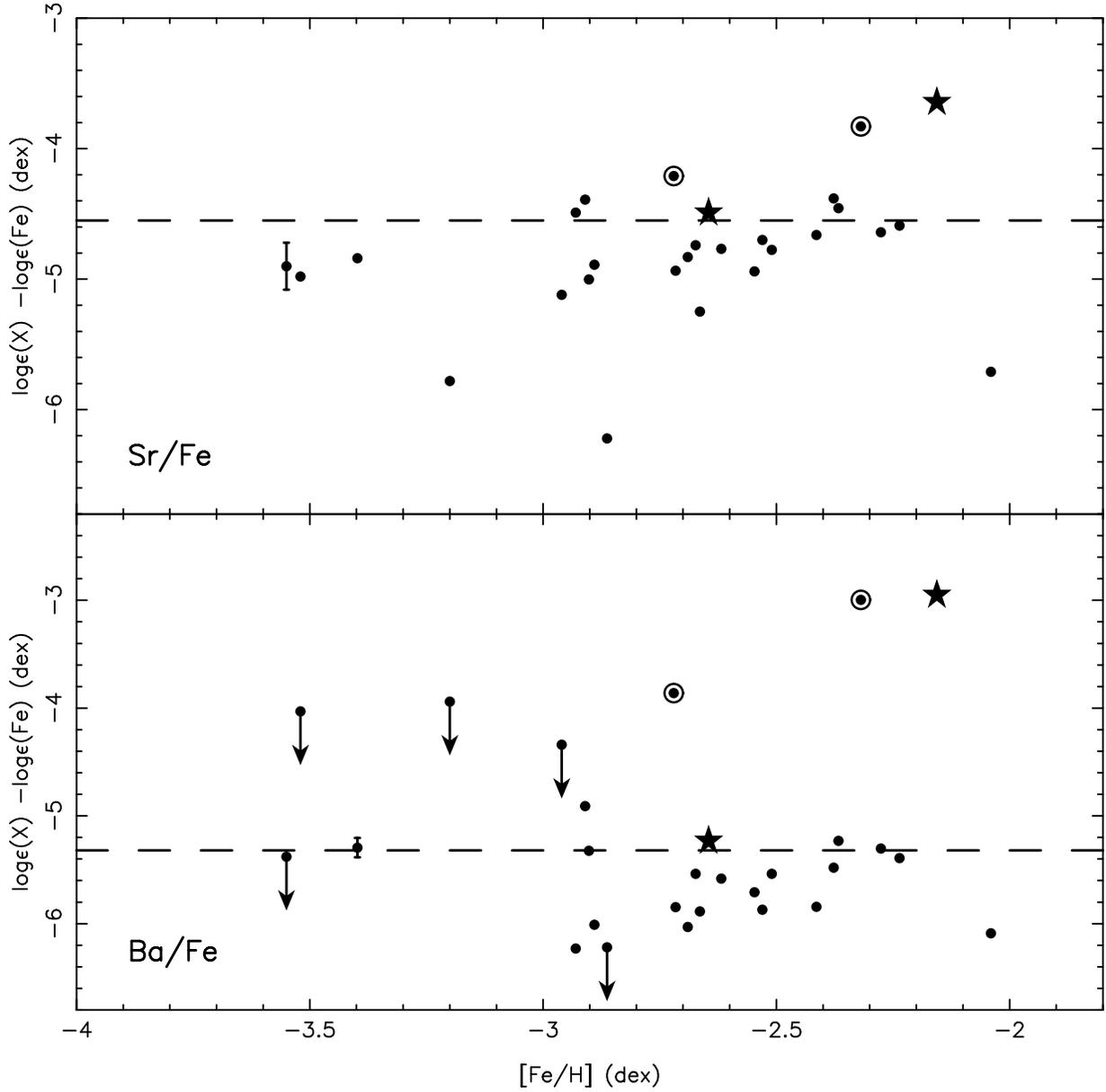}
\caption[]{The abundance ratios of Ba (top panel)
and Sr (bottom panel) are shown with respect to Fe for the sample of 28
candidate EMP dwarfs.
The symbol key is that of Figure~\ref{figure_mgalsi}.  
The dashed horizontal line is the Solar ratio.
A typical 1$\sigma$ uncertainty for each abundance ratio is shown
for the most metal poor star in each panel.
The vertical range for each panel of this figure is much larger than is used  
in Figure~\ref{figure_mgalsi} to Figure~\ref{figure_nizn}, and for the lower
panel, is not centered on the mean of the distribution.
\label{figure_basr}}
\end{figure}

The only heavy elements detected in the majority of the EMP
candidate dwarf stars in our sample are Sr and Ba.
The abundance ratios for the 28 candidate EMP dwarfs in our
sample of these two elements with respect to Fe
are shown in Figure~\ref{figure_basr}.  For both of these
species we again use Fe~II instead of Fe~I.  Sr/Fe shows
a large range.  
(Note that the vertical scale in 
Figure~\ref{figure_basr} is different from that of the previous
set of four figures.)  Three of the four C-rich stars 
have anomalously high Sr abundances, while the carbon star
HE0007$-$1832 does not.  
Then there is a large group of dwarfs at 0.3 dex below the Solar
Sr/Fe ratio, with three stars stars extending down to 
$\sim$1/25 of the Solar abundance.  

Ba shows the same pattern; the same three of the four C-enhanced EMP stars 
show Ba/Fe enhanced by a factor of more than 10, while the
fourth such star, HE0007$-$1832, has Ba/Fe and Sr/Fe at about
the Solar ratio.  \cite{aoki02a} also found several C-enhanced
stars with normal neutron capture element abundances.
Most of the rest of our sample has [Ba/Fe] about 0.3 dex below Solar.
Two of the same three stars which have weak Sr also
have have very low Ba abundances, another factor of three lower,
while the third (and several other stars) have
no detected Ba lines at all\footnote{Upper limits to $\epsilon$(Ba)
have been calculated for the three stars with no detected Ba~II lines from the
Keck Pilot Project.}).  However, HE0130$-$2303 also has
a very low Ba abundance but not such an extremely low Sr
abundance.  Very large spreads in [Ba/Fe] have been reported for low
metallicity stars by many groups, including \cite{mcwilliam95},
\cite{burris00}, \cite{fulbright00} and \cite{johnson02}. Recent surveys, including
all just mentioned, are compiled in
Figure~4 and 5 of \cite{travaglio04}. Because of the significant number of
non-detections of Ba, we cannot compare the dispersion of Ba/Fe in our
sample to that of the published surveys.
Our elimination of
the C-rich stars has significantly reduced  the spread of Sr/Fe and
perhaps Ba/Fe by excluding most of the $s$-process enhancement.

\begin{figure}
\epsscale{1.0}
% Comment out the following line to embed the PS figure into the manuscript
% \plotone{/scr2/jlc/hamburg_survey/hires_summary/programs/dwarf_eupb.ps}
\plotone{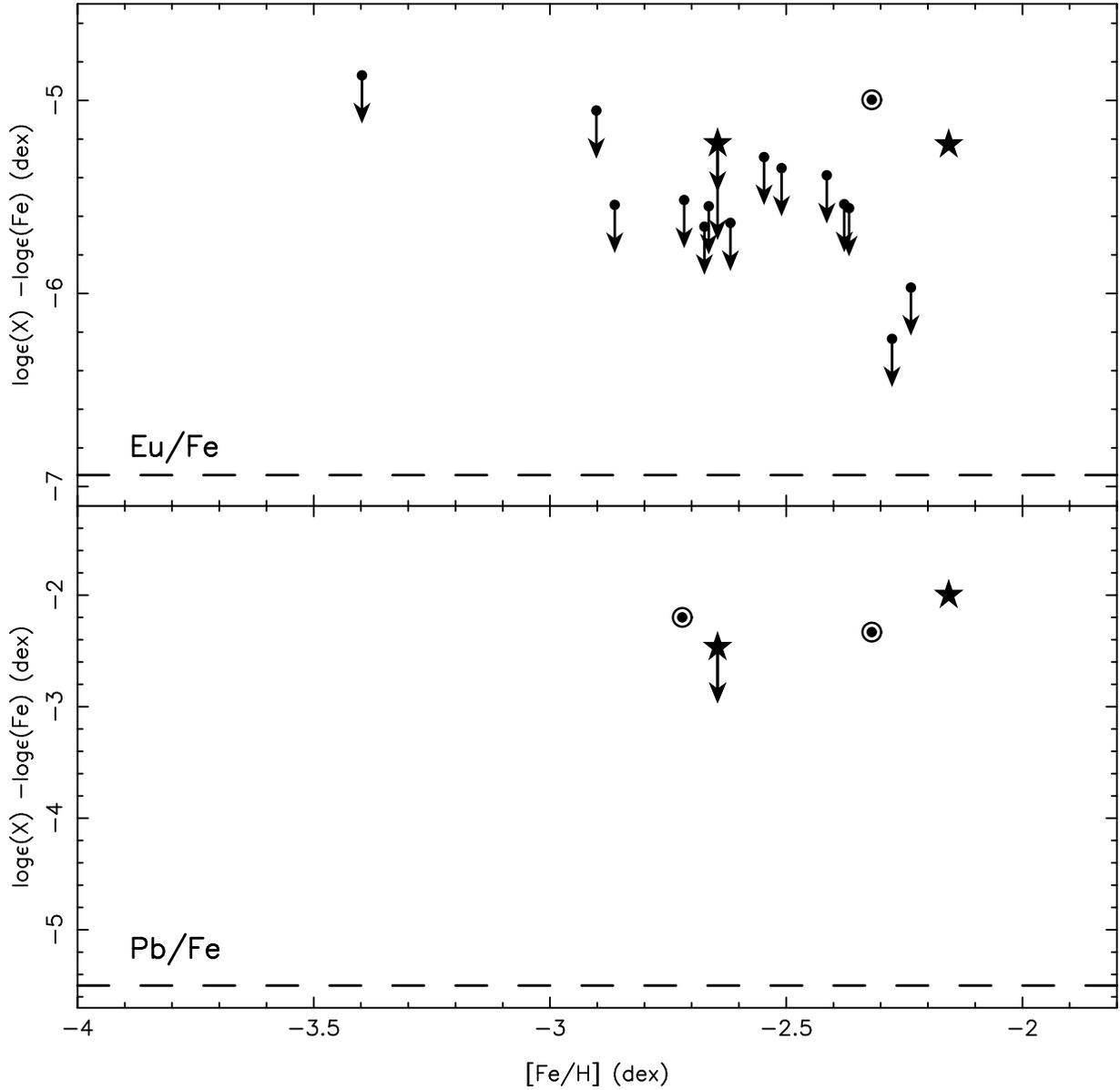}
\caption[]{The abundance ratios of Eu (top panel)
and lead (bottom panel) are shown with respect to Fe for the sample of 28
candidate EMP dwarfs.  Fe~II is used for Eu, while Fe~I is used for Pb.
The symbol key is that of Figure~\ref{figure_mgalsi}.  
The dashed horizontal line is the Solar ratio.
A typical 1$\sigma$ uncertainty for each abundance ratio is shown
for the most metal poor star in each panel.
The vertical range for each panel of this figure is much larger than it is 
in Figure~\ref{figure_mgalsi} to Figure~\ref{figure_nizn}, and 
is not centered on the mean of the distribution.
\label{figure_eupb}}
\end{figure}

There are only three detections of
Y~II lines.  The two high values are in two of the C-rich
stars, the single detection in a normal star is much lower.
Eu is detected  only in two of the C-rich stars, the
carbon star HE0143$-$0441 (with $N$(Ba)/$N$(Eu) $\sim~180$)
and HE2148$-$1247.    The upper limits
for the remaining stars, shown in Figure~\ref{figure_eupb},
are too high to be of interest\footnote{Sums of selected spectra in the region
of the 4129~\AA\ line of Eu~II yielded an upper limit of \eqw $\le 2$ m\AA.
However, since we are on the linear part of the curve of growth, we still do not
reach an interesting regime of Eu abundance.}.
Pb, whose abundance ratios are shown in Figure~\ref{figure_eupb},
is detected in three of the four C-rich stars,
which are similar to the two $s$-process rich EMP subgiants
discussed by \cite{aoki01}.
The EMP carbon star HE0007$-$1832, however, shows no excess of
neutron capture elements.

\begin{figure}
\epsscale{1.0}
% Comment out the following line to embed the PS figure into the manuscript
% \plotone{/scr2/jlc/hamburg_survey/hires_summary/programs/dwarf_eulaba.ps}
\plotone{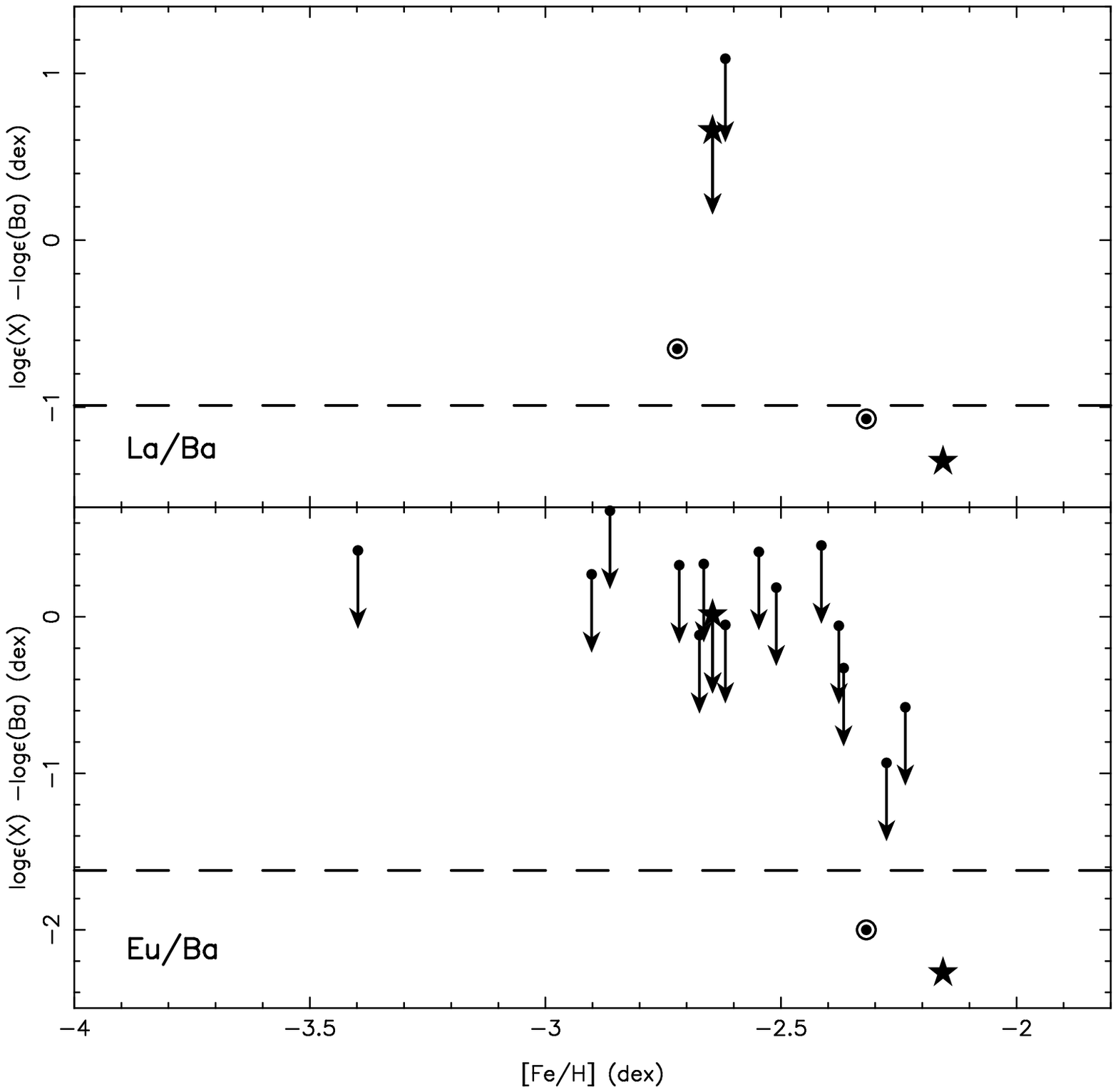}
\caption[]{The abundance ratios of La (top panel)
and Eu (bottom panel) are shown with respect to Ba for the sample of 28
candidate EMP dwarfs, all from the singly ionized species.
The symbol key is that of Figure~\ref{figure_mgalsi}. 
The dashed horizontal line in each panel is the Solar ratio.  
The vertical range for each panel of this figure is much larger than it is 
in Figure~\ref{figure_mgalsi} to Figure~\ref{figure_nizn}, and 
is not centered on the mean of the distribution.
\label{figure_eula}}
\end{figure}

\begin{figure}
\epsscale{1.0}
% Comment out the following line to embed the PS figure into the manuscript
% \plotone{/scr2/jlc/hamburg_survey/hires_summary/programs/dwarf_srpb.ps}
\plotone{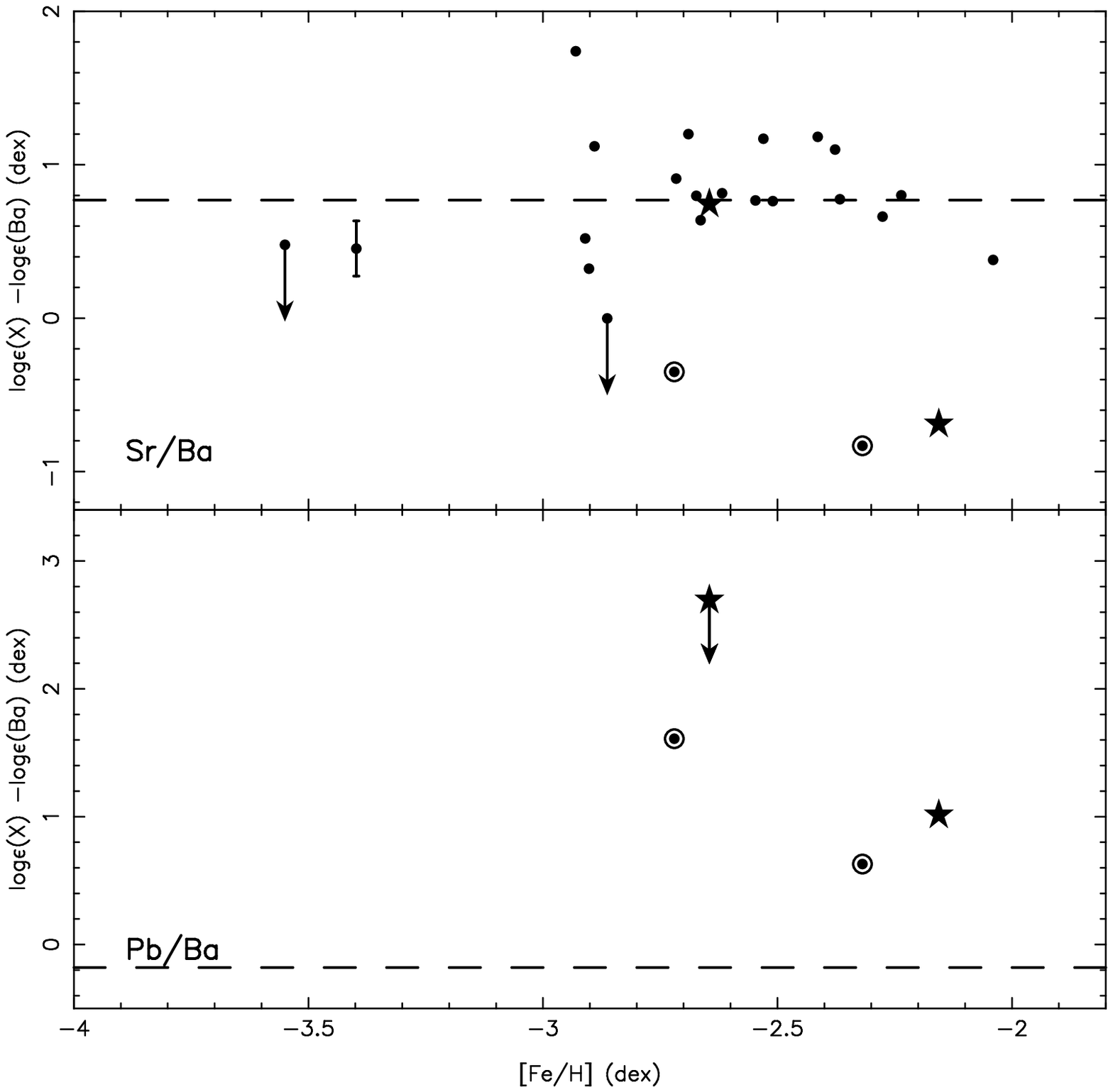}
\caption[]{The abundance ratios of Sr (top panel)
and Pb (bottom panel) are shown with respect to Ba for the sample of 28
candidate EMP dwarfs, all from the singly ionized species.
The symbol key is that of Figure~\ref{figure_mgalsi}.  
The dashed horizontal line in each panel is the Solar ratio.  
The vertical range for each panel of this figure is much larger than it is 
in Figure~\ref{figure_mgalsi} to Figure~\ref{figure_nizn}, and 
is not centered on the mean of the distribution.
\label{figure_srpb}}
\end{figure}

\begin{figure}
\epsscale{1.0}
% Comment out the following line to embed the PS figure into the manuscript
% \plotone{/scr2/jlc/hamburg_survey/hires_summary/programs/ca_ti_disp.ps}
\plotone{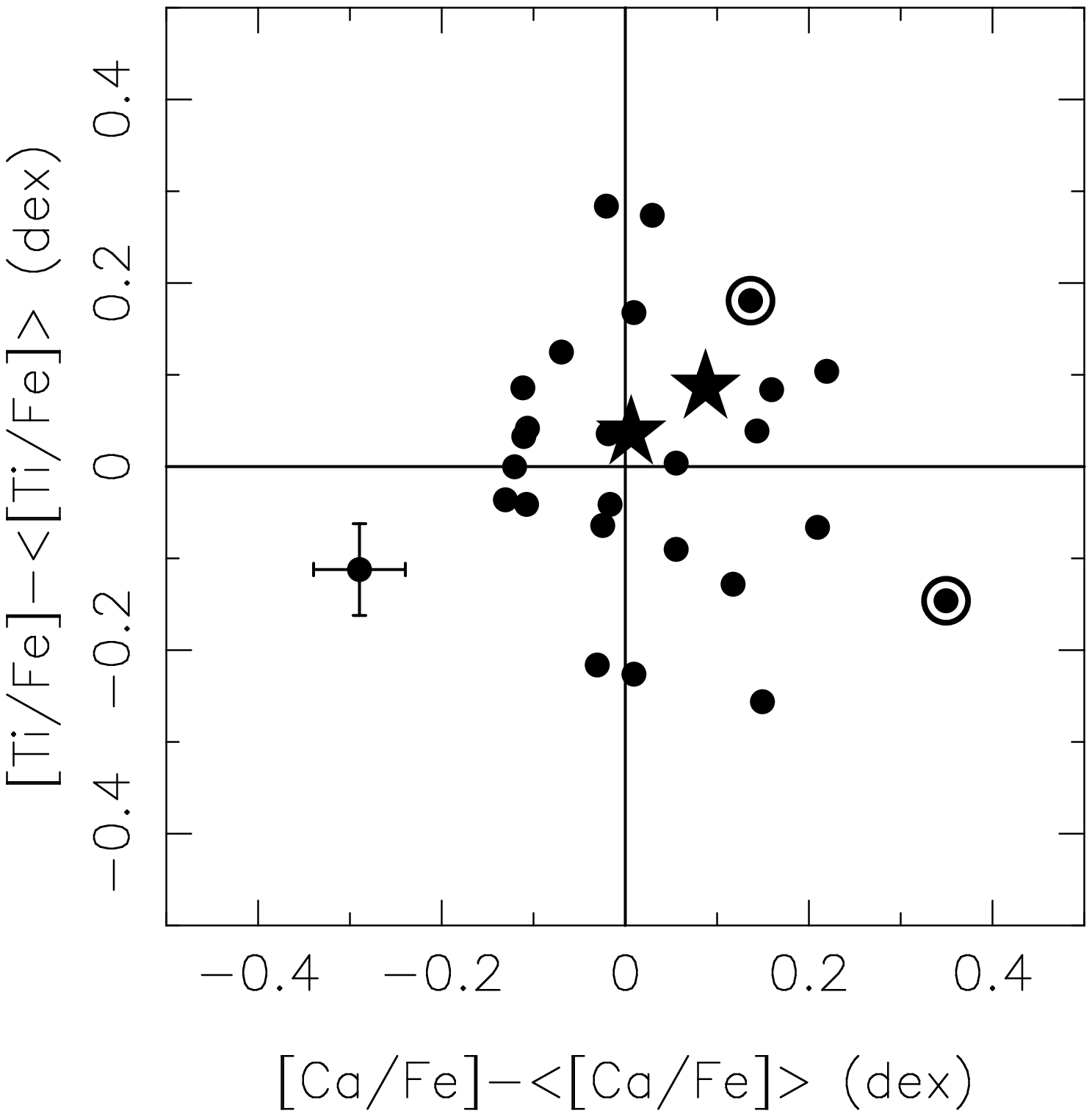}
\caption[]{$R$(Ti) is shown as a function of $R$(Ca) for the sample
of 28 EMP candidate dwarfs where $R(X) \equiv $[X/H] $- <$[X/H]$>$.
The symbol key is that of Figure~\ref{figure_mgalsi}.
\label{figure_cati_disp}}
\end{figure}

\begin{figure}
\epsscale{1.0}
% Comment out the following line to embed the PS figure into the manuscript
% \plotone{/scr2/jlc/hamburg_survey/hires_summary/programs/ti_cr_disp.ps}
\plotone{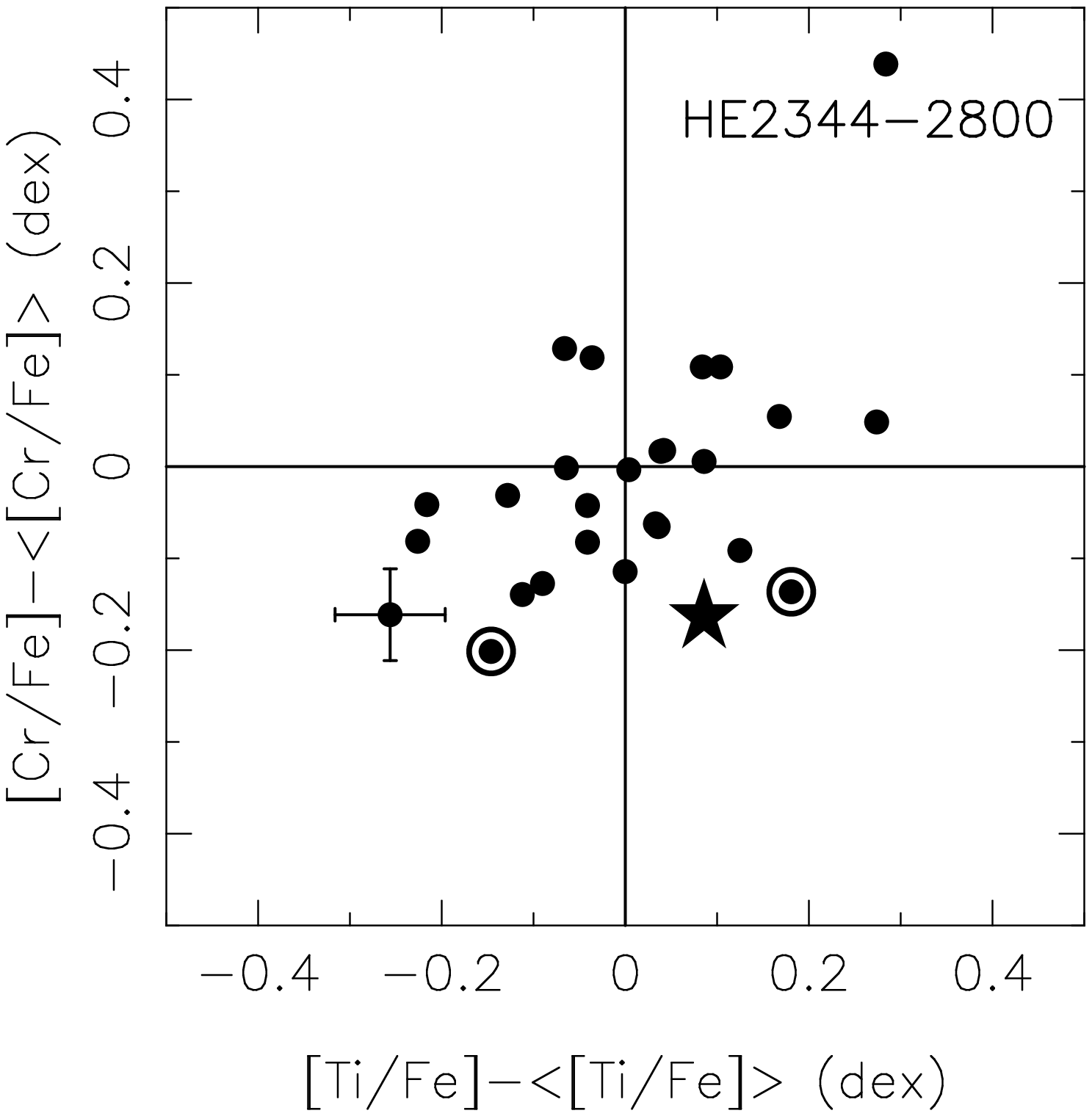}
\caption[]{$R$(Cr) is shown as a function of $R$(Ti) for the sample
of 28 EMP candidate dwarfs where $R(X) \equiv $[X/H] $- <$[X/H]$>$.
The symbol key is that of Figure~\ref{figure_mgalsi}.
\label{figure_ticr_disp}}
\end{figure}

\begin{figure}
\epsscale{1.0}
% Comment out the following line to embed the PS figure into the manuscript
% \plotone{/scr2/jlc/hamburg_survey/hires_summary/programs/mg_ca_disp.ps}
\plotone{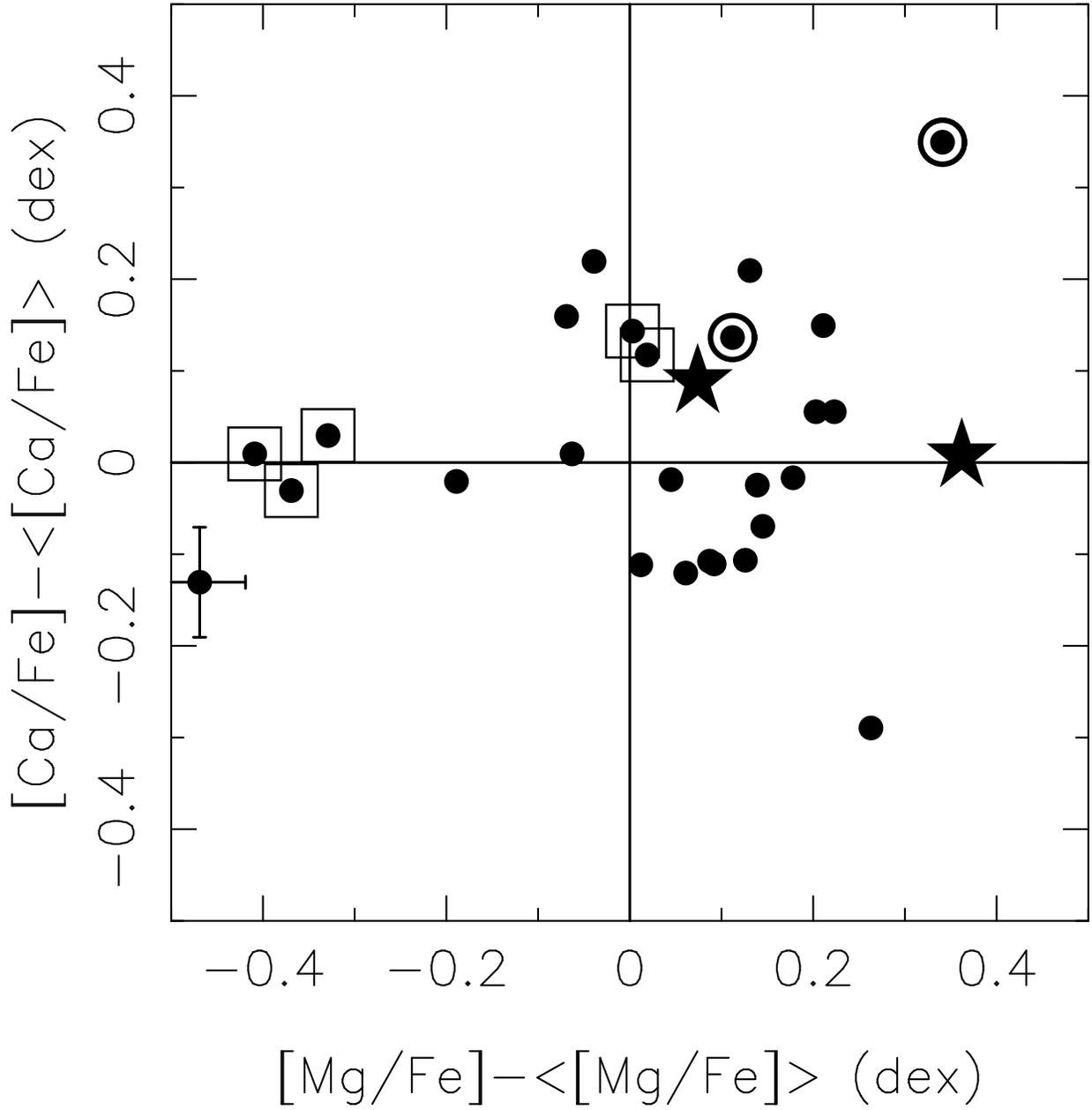}
\caption[]{$R$(Ca) is shown as a function of $R$(Mg) for the sample
of 28 EMP candidate dwarfs where $R(X) \equiv $[X/H] $- <$[X/H]$>$.
The symbol key is that of Figure~\ref{figure_mgalsi}.  Note the
elongation of the distribution for the stars in our sample along the X axis.
The points indicating those program stars where the weaker Mg lines 
were not detected and only those of the Mg triplet were detected
are enclosed within squares.
\label{figure_mgca_disp}}
\end{figure}

\begin{figure}
\epsscale{1.0}
% Comment out the following line to embed the PS figure into the manuscript
% \plotone{/scr2/jlc/hamburg_survey/hires_summary/programs/dwarf_mgtriplet.ps}
\plotone{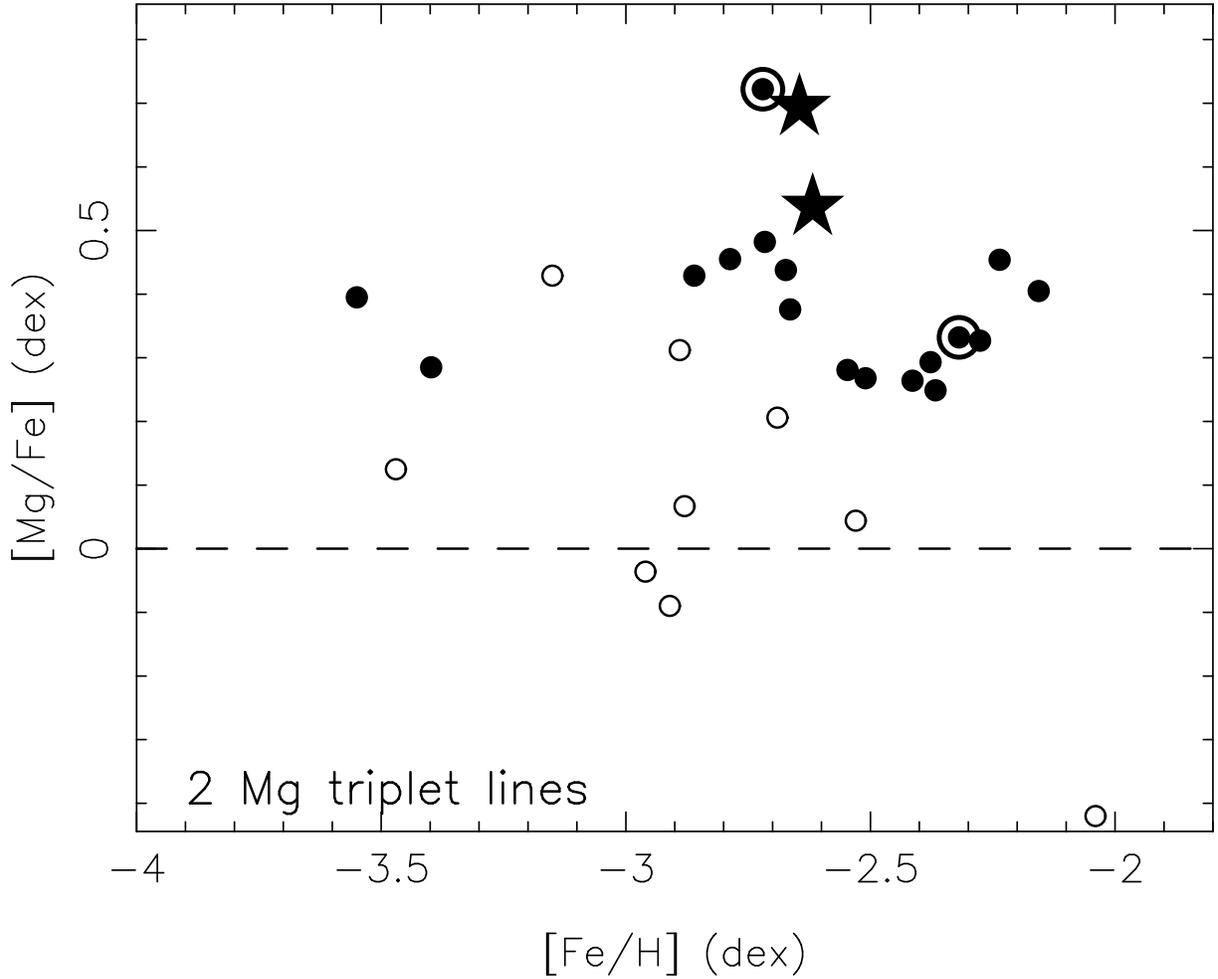}
\caption[]{The abundance ratio [Mg/Fe] when the Mg
abundance is determined 
using only the two lines of the Mg triplet is shown with respect to Fe
for the sample of 28 candidate EMP dwarfs.  
The C-rich stars are denoted as in Figure~\ref{figure_mgalsi}, the filled
circles denote the remaining dwarfs presented here,
while the dwarfs in the Keck Pilot Project are denoted by
open circles.
The dashed horizontal line is the Solar ratio.
\label{figure_mgtriplet}}
\end{figure}

Ba/Sr shows the large enhancement of Ba relative to Sr
among the C-enhanced stars.  Most of the remaining stars 
are grouped with [Ba/Sr] $\sim -0.1$ dex, almost the Solar ratio,
except for HE0130$-$2303,
which is $\sim$0.6 dex lower.
La is only detected in three of the C-rich stars, in which the
ratio La/Ba has
a mean essentially identical to that of the Sun.
Figure~\ref{figure_eula} and \ref{figure_srpb} show the
ratios of Sr, La, Eu and Pb with respect to Ba.

\subsection{Comparison with \cite{cayrel03} \label{section_cayrel} }

In a recent paper describing the First Stars Very Large Program at ESO,
\cite{cayrel03} present their results
from an abundance analysis of 29 EMP giants  selected
from the HK Survey of which 20 have [Fe/H] $\le -3$ dex from
high dispersion UVES spectra,
as well as 6 bright comparison giants. This sample was biased
against  C-enhanced giants, which were largely excluded.

We first note that
this sample is considerably brighter than the stars studied here.
The median $V$ of the 14 previously unpublished 
candidate EMP dwarfs from the HES 
presented here is 15.6, while the median of the EMP giants
from the HK Survey in Cayrel et al's sample is 13.3, a factor
of 8 brighter.  In addition the giants are much cooler,
hence have much stronger lines for a fixed metallicity.  So overall,
the present work on the HES dwarfs
is considerably more demanding of observing time than
is the  ``First Stars'' effort on giants.
Furthermore, at least half of the \cite{cayrel03} sample from the HK Survey
consists of 
EMP giants with previously published high dispersion abundance 
analyses, while the sample from the HES discussed
here was selected only recently by our program. 

The agreement between the two independent analyses
of different luminosity ranges of EMP stars
is extremely gratifying. It is interesting to note that
there is some hint in the data of \cite{cayrel03} for anomalous
stars with Mg/Fe approximately Solar similar to the small group
of outliers we have mentioned above.  We do not know, however, which specific
Mg~I lines are included in their sample and whether all of these lines
are detected in all of their stars, so that  the
problems discussed in \S\ref{section_disp_real} were avoided. 
Table~\ref{table_cayrel}
presents a comparison of the mean abundance ratios of our sample
(with the C-rich stars excluded)
as compared to the best fit relations for [X/Fe] from \cite{cayrel03}
evaluated at [Fe/H] = $-3.0$ dex\footnote{\cite{cayrel03} adopt
log$\epsilon$(Fe)\subsun = 7.50 dex, which is 0.05 dex higher than we do.  No
correction for this difference has been made.}.
The only element with a disagreement
exceeding  0.2 dex is Si.  Sc shows a difference of 0.16 dex,
which may be real as well.  The comparison for these two elements
is discussed in detail below.
The small differences for all other species in common
are most likely due to different choices of transition probabilities,
Solar abundances, and other atomic data. 

Silicon presents a puzzle -- the giants from \cite{cayrel03} 
give abundance ratios Si/Fe consistently larger than the dwarfs studied here.
\cite{ryan96} studied a sample of very metal poor giants and dwarfs.
For the 8 dwarfs in their sample, the average [Si/Fe] is
+0.07 dex, close to our value, and well below the
mean for the giants in their sample. \cite{ryan96} interpreted this
as indicating intrinsic scatter in [Si/Fe] but what we see is
a small dispersion in that quantity coupled with an apparently
real difference between the giants and the dwarfs.  So we must
consider what problems in the analysis might lead to this
difference.  We use only the 3905~\AA\ Si~I
line, which overlaps a CH feature. 
For the range in \teff\ of the EMP dwarfs considered here,
this line is usable as CH is very weak, but
\cite{cayrel03} reject this line in their much cooler giants
as they find it to be too blended with CH even in normal C abundance stars.
They rely instead on a single line of Si~I at 4102.9~\AA, which 
we cannot use  as it is too close to H$\delta$.  Even
in the giants, with their much narrower Balmer lines, they were forced 
to use spectral synthesis techniques to derive a Si abundance.
This difference between the dwarfs and the giants 
has the wrong sign to be due to a decrease in
[Si/Fe] as a function of increasing apogalacticon,
suggested to exist for halo stars by \cite{fulbright02}, 
since our dwarf sample
surely is in the mean closer to the Sun\footnote{It is the
galactocentric radius which is relevant here.} than the giant sample of \cite{cayrel03}.

There appear to be some problems with the transition probabilities
for these very strong Si~I lines, which are not used at all in
Solar abundance studies of Si nor in most stellar analyses; these studies
generally use the much weaker Si~I lines in the 5000 to 8000~\AA\ region.
Our $gf$ value for the Si~I line at 3905~\AA\ is 
that of NIST, which is taken from the laboratory work of
\cite{garz73}.   This is close to the
value of $-1.04$ dex given by \cite{obrian91}, whose work is focused
on the UV and hence do not reach redder than 4110~\AA.
They report only
an upper limit to the $gf$ value for the
4103~\AA\ line of Si~I for which \cite{cayrel03} adopt a log$gf$ value 
of $-3.14$ dex  (Spite 2004, private communication), again taken 
from the 1973 study.  However,
NIST gives $-2.92$ dex as the log$gf$ value for this line.
So the ratio of the transition line strengths
for the 3905 versus the 4103~\AA\ lines of Si~I, the crucial
lines of interest here, differs
by a factor of 1.7 between the values adopted by NIST and those
adopted by the two groups
such that our derived Si abundance is 0.22 dex
lower than that of \cite{cayrel03} for the same line strength.
Furthermore, the errors quoted by
the early laboratory study of \cite{garz73} are rather large,
$\pm$0.08 dex for the 4102~\AA\ line for example, and a modern
higher precision study of the oscillator strengths for 
the blue and red lines of Si~I would
be highly desirable.

If we include
the difference in the adopted value of log$\epsilon$(Fe)\subsun\ between our work
and the First Stars Project, our Si abundance should be 0.27 dex
smaller than theirs.  That
is part, but not all, of the difference between our mean Si abundance
derived from the HES dwarfs and that of the First Stars Project
for the HK giants.
It  is impossible from the present data to ascertain
which Si~I $gf$ value gives the standard Solar abundance
and hence should be adopted as correct. 
Then there is the issue of non-LTE. 
\cite{wedemeyer01} has demonstrated that 
non-LTE for Si in the Sun is negligably small. 
Our preliminary result for [Si/Fe] for a large sample of 
giants from the HES
is about 0.4 dex larger than that obtained here for the dwarfs.  
Most of this
difference probably arises from
contamination of the Si~I line at 3905~\AA\ by blending
CH features in the giants; spectral syntheses have not yet been carried out
for the HES giants.

We find enhanced [Sc/Fe] ratios for our sample, with a mean of +0.24 dex
($\sigma$=0.16).  This  is significantly higher than the [Sc/Fe]
ratio found by \cite{cayrel03} for EMP giants,
but is consistent with observations of dwarfs at higher metallicity, 
which indicate that the trend of [Sc/Fe] with [Fe/H] is similar to the
so-called $\alpha$ elements

\cite{nissen00} found an alpha-like trend of increasing [Sc/Fe] with
decreasing [Fe/H] for 119 F and G main sequence stars,
in the range $-$1<[Fe/H]<0.1 dex.  However, \cite{prochaska00}
showed that the use of incorrect HFS parameters by Nissen \etal\
caused an exaggeration of the [Sc/Fe] slope.  
In a study of Galactic disk F and G dwarf stars \cite{reddy03}
found a slight slope in [Sc/Fe] versus [Fe/H],
indicative of an alpha-like behavior for Sc.  In general their [$\alpha$/Fe]
trends with [Fe/H] have much shallower slopes than found by previous
studies; thus, although Sc has a very gentle slope with metallicity it is comparable to
Ca and Ti, but with greater scatter.
Recently, \cite{allende04} have studied the composition of
118 stars within 15pc of the sun.  They found an alpha-like slope
for Sc, with [Sc/Fe]$\sim$+0.4 dex near [Fe/H]=$-$1; much steeper
than found by \cite{reddy03}.  Curiously they found a steeper
slope for [Sc/Fe] than for [Ca/Fe].  These results for local dwarfs suggest agreement
with the halo dwarf star results of \cite{zhao90}, who found [Sc/Fe]=$+$0.27$\pm$0.10
for a sample of 20 dwarfs with $-$2.6$\leq$[Fe/H]$\leq$$-$1.4.  It should be
noted that Zhao \& Magain did not perform proper HFS, but instead 
employed an approximate
method to compute HFS abundance corrections. 

For metal-poor red giant stars in the Galactic halo the [Sc/Fe] abundances
provide no evidence for overall alpha-like enhancements.  
For example, studies of field giants by \cite{luck85}, \cite{gratton88},
\cite{gratton91}, and \cite{mcwilliam95}
indicate that [Sc/Fe] is close to the solar value in the range 
$-$3.5$\leq$[Fe/H]$\leq$0.
Solar [Sc/Fe] values are also seen in globular cluster red giants 
(see Table~\ref{table_globcomp} and the references cited therein, and note
that the sample studied for NGC~6397 is composed of dwarfs).
Recently, \cite{johnson02} measured [Sc/Fe] for 23 halo stars with [Fe/H]$<-$1.7, 
using $\sim$10 Sc~II lines, and found a mean [Sc/Fe]=$+$0.08 ($\sigma$=0.07 dex).  

Thus, while abundances for dwarf stars, including this work, suggest that Sc behaves 
similar to the $\alpha$-elements, the compositions of red giant stars generally do not
confirm this conclusion.   It seems more likely that the difference is due to analysis
problems rather than to an evolutionary process that has 
depleted Sc in red giants.  It is ironic that the HFS desaturation effects 
for Sc~II lines are relatively small and unlikely to be the causal agent. 

It may be safest to favor the Sc abundance  trend derived for dwarfs over the red giant
results, because of the similarity of the dwarf and solar model atmospheres; thus,
we favor the idea that Sc behaves as an $\alpha$-element.

A comparison of the dispersions in abundance ratio for the EMP sample
of dwarfs from the HES discussed here (see Table~\ref{table_disp})
and for the First Stars Project
sample of EMP giants \citep{cayrel03} is given in the last two columns of 
Table~\ref{table_cayrel}.  The dispersions are of comparable size
for most of the elements in common, with \cite{cayrel03} achieving
somewhat smaller values of $\sigma$ for the abundance ratios of Sc
and Cr with respect to Fe.  The low values achieved are a testimony
to the high precision of both of these efforts.

\subsection{Comparison With Galactic Globular Clusters \label{section_globs}}

The metal-poor Galactic globular clusters, with few exceptions, 
are believed to be extremely old halo objects. Thus their abundance ratios
should also be representative of the halo of the Galaxy
in its early stages of formation.  Although individual field halo stars
may be brighter, the stellar parameters for
globular cluster stars are easier to determine, since strong constraints
are imposed by them being located within
a cluster of uniform distance, age and 
(at least approximately) reddening.   Furthermore, we do not
need to worry about the issue of halo versus thick disk stars;
we can choose globular clusters which are believed to be halo objects.

We therefore compare the
abundance ratios derived here for our sample of 27 candidate EMP dwarfs 
from the HES with a median [Fe/H] of $-2.7$ dex with the results
from recent detailed analyses using Keck/HIRES or VLT/UVES spectra of large
samples of stars in M71 ([Fe/H] = $-0.7$ dex)
\citep{ramirez01}, M5 ([Fe/H] = $-1.3$ dex) \citep{ramirez03} 
M3 ([Fe/H] = $-1.45 $ dex) \citep{sneden04,cohen04a},
47~Tuc ([Fe/H] = $-0.67$ dex) \citep{carretta04},
and NGC~6397 ([Fe/H] = $-2.02$ dex)\citep{thevenin01}.  
In each of these cases, the internal errors in the mean
abundances for any element are small.  The actual
uncertainties however may be dominated by systematic effects as we will see
shortly.

Table~\ref{table_globcomp} presents this comparison
for our candidate EMP dwarfs and for the five Galactic
globular clusters for elements from
Mg to Ni. The abundance ratios given in the table for the
present sample are those excluding the C-rich stars.
This table shows that [X/Fe] is approximately constant from
[Fe/H] of $-0.7$ to $\sim-3$ dex for most
of the elements in common, i.e. 
the elements Mg, Ca,
Ti, Cr, Mn and Ni.  [Al/Fe] appears to have 
a large range, but much of this arises
from the use or neglect of the substanial non-LTE corrections
suggested for Al~I by \cite{bau96} and by \cite{bau97}.
\cite{sneden04} and his collaborators
appear not to use any non-LTE correction for Al~I,
while JGC and her collaborators  include such in their analyses.
Once this difference is removed, the
data in Table~\ref{table_globcomp}
are consistent with constant [Al/Fe] over this range of [Fe/H].
We give values for [Ba/Fe] in Table~\ref{table_globcomp} (ignoring
the upper limits), but
note that this quantity shows 
large star-to-star variations in the EMP dwarf sample.  

The abundance ratio [Si/Fe] is lower in the dwarfs from the HES than
it is in the globular cluster giants.  This is similar
to the difference between them and the field EMP giants
of \cite{cayrel03}  discussed in
\S\ref{section_cayrel}.

Co and Mn are the only elements where a definite change is seen, in the sense
that [Co/Fe] is higher in lower metallicity 
systems while [Mn/Fe] is lower, a trend already noticed 
by \cite{mcwilliam97}.  
Scandium, which has hyperfine structure, as do Co and Mn, 
shows a possible trend towards 
higher values in lower
metallicity systems, but the trend is not large.  

To summarize the result of this comparison, the halo globular
clusters, beginning at [Fe/H] = $-0.7$ dex and ranging downwards
in metallicity, show an abundance distribution [X/Fe] which 
is essentially identical to that of the field EMP dwarfs and giants
for most of the elements between Mg and Ni.  Co, Mn, and perhaps Sc,
do show genuine differences.  Si shows differences as well, but
we do not know yet if they are real; see the discussion in
\S~\ref{section_cayrel}.

\subsection{Comparison with DLA Abundances} 

Damped Ly$\alpha$ absorption systems seen in the spectra of QSOs
are presumably the result of absorption of light from a background
QSO by the outer parts of a foreground
galaxy.  Study of such systems yields abundances integrated
along the line of sight through the intervening object, which may
be at any redshift up to that of the QSO.  The 
set of elements that can be observed in such systems 
consists of those that have suitably
placed resonance lines within the wavelength regime that
can be observed at high dispersion.  These are not necessarily
the species that can easily be observed in optical spectra of local stars.
As reviewed by \cite{prochaska04}, the uncertainties in such
an analysis are the ionization corrections to convert from the
abundance inferred for an
observed species to that of the element and the correction
for depletion of various species from the gas onto dust grains.
Iron, the benchmark for stellar chemical analyses, 
is often considered to be depleted onto the dust grains in DLA systems.

The chemistry of high redshift DLA systems should represent at least
crudely the state of the ISM gas at an early stage in the formation of
the halo of the galaxy.  Although the metallicities of DLA
systems do not reach as low as those of individual Galactic halo
stars, we would hope that the abundance ratios
of elements determined from DLAs would be consistent with those deduced from
spectra of Galactic EMP stars.
There are now at least three DLA systems with $1.7 < z < 2.6$
where absorption from 15 or more elements has been detected; see
\cite{prochaska03,dess04}.  The abundances thus derived are
quite uncertain, but the abundance ratios of
[Mg/Fe] (+0.24 dex), [Si/Fe] (+0.08 dex),
[Cr/Fe] (+0.01 dex), [Mn/Fe] ($-0.16$ dex) and [Ni/Fe] (+0.01 dex)
derived from averaging the results of these two investigations (when
the results are given as values and not as upper or lower limits) are
in reasonable agreement with the values deduced from EMP stars given
in Table~\ref{table_disp}.

Due to the same technical factors as apply for EMP stars,
i.e. the new
generation of large telescopes and efficient spectrographs, 
a leap forward in our understanding of the chemistry of DLA systems
as well as in the search for the nature of the material giving rise to 
the DLA systems has occurred recently.  Further rapid progress in this area, 
including improved comparisons between early nucleosynthesis as observed 
in the very distant DLAs and in the local Universe 
in the halo of our Galaxy through EMP stars, is certain.

\subsection{The Bigger Picture}

Our data cover the range from $-3.6 <$[Fe/H] $< -2.0$ dex with
a median [Fe/H] of $-2.7$ dex.  Over this limited range, calculations
of dispersions in abundance ratios are straightforward, but,
as shown in Table~\ref{table_slopes},
it is very hard to determine accurately the trends
in abundance ratios with metallicity  given the
uncertainties in the data.  We therefore construct a composite
of the behavior of selected abundance ratios over the full range
of abundance
extending up to Solar metallicity by collecting the results of 
some of the many other 
published surveys of abundances in halo stars\footnote{The 
stars with [Fe/H] $> -1.0$ dex in the samples of \cite{fulbright00} 
and of \cite{gratton01} may not be in the halo.}. 
In doing so, we must be cognizant of the issues discussed 
in \S~\ref{section_disp_real}.  To avoid potential systematic
differences between surveys, we ideally would focus on those
elements where the changes are  expected to be largest,
where there are many useful lines, minimal non-LTE, etc.  
This is difficult to achieve
in practice.  For example, Mn and Co appear to be among the best cases for
real changes in abundance ratio with metallicity, but
both have strong hyperfine structure.  Most analyses
of very metal poor stars use the Mn triplet at 4030~\AA, which
is much stronger than other optical Mn lines. \cite{cayrel03}
claim that a correction of 0.4 dex is necessary to bring the
abundance deduced from these triplet lines into accord with that
deduced from the other weaker Mn lines.

In combining the samples, the only adjustment that was made was
to take into account the difference in the Solar Fe
abundance adopted in each publication.  The
surveys included are \cite{cayrel03}, \cite{mcwilliam95},
\cite{fulbright00}, \cite{norris01}, \cite{johnson02}, \cite{ryan96},
\cite{gratton01}
and \cite{nissen00}. 
The extremely low metallicity
star found by \cite{christlieb03} is not included.
We add in the abundance ratios for the five 
Galactic globular 
clusters from  Table~\ref{table_globcomp}.
No attempt was made
to eliminate duplicate analyses of the same star except
among the most metal poor ones, where the very recent work 
of \cite{cayrel03} was used in preference to that of any
other group when stars were included in multiple samples.
Most of these surveys at low metallicity
are of giants; ours is the only one in this metallicity range focused
on dwarfs. 

Figures~\ref{figure_surveyni} to \ref{figure_surveyco}
present the abundance ratios of Ni, Mg, Ca, Cr, Mn and Co
with respect to Fe for this large sample of halo stars over a very wide
range in metallicity using our work (ignoring the C-rich stars), 
that of \cite{cayrel03}
and the other surveys referenced above.  The first of these figures
shows the relationship between [Ni/Fe] and metallicity
(Figure~\ref{figure_surveyni}).  [Ni/Fe] is constant over the full
range of [Fe/H] with a
mean value within 0.1 dex of the Solar value and a scatter consistent
with the observational errors.  Since Ni and Fe are very closely
related in terms of the nucleosynthetic origin, their constant
ratio is gratifying but expected.  The relationship between
[Cr/Fe] and metallicity (Figure~\ref{figure_surveycr}) is reasonably
tight and shows a gradual increase of this abundance ratio
towards higher metallicity, which explains the difference
between the [Cr/Fe] seen among globular clusters  
and those of the even more metal poor dwarfs from the HES
(see Table~\ref{table_globcomp}). The scatter is consistent with the
measurement errors at all values of [Fe/H].

A similar plot for [Mn/Fe] (Figure~\ref{figure_surveymn})
shows a similar trend, but also illustrates the tremendous
difficulty of combining surveys without delving deeply into the
adopted atomic data of each.  There clearly are systematic differences
in abundance scale between the various surveys displayed.
The same comment regarding the difference between the EMP dwarfs
and the globular clusters made above for [Cr/Fe] apply here as well.

Plots of the $\alpha$-elements [Ca/Fe] and [Mg/Fe] show an
elevated flat
plateau extending over most of the metallicity range.  
Assuming that the stars included with [Fe/H] $> -1$ dex actually
are halo stars, these
ratios drop to the  Solar value at [Fe/H] $> -1$ dex.  The [Mg/Fe] plot shows
a rather large dispersion at all metallicities, which suggests problems
in the atomic data reminiscent of those discussed earlier and increases
our reluctance to consider the dispersion and apparent existence of the
``small group'' with Solar [Mg/Fe] we detect in the HES dwarfs
as arising from star-to-star variations in [Mg/Fe].

Figure~\ref{figure_surveyco} shows the behavior of [Co/Fe]
as a function of metallicity.  It is above Solar at the lowest metallicites,
but appears to fall to near the Solar value somewhat earlier than do the
curves of the $\alpha$-elements.

Real trends in abundance ratios do exist, but they are relatively small
and large samples of stars over a wide range in [Fe/H] must be assembled
to identify them.  Considerable care is required in merging different
surveys.

\section{Binarity \label{section_binary}}

Field stars of Solar metallicity are well known to often be members
of binary systems.  \cite{latham02} and \cite{carney03} find no difference 
in the binary frequency they establish for
moderately metal poor halo populations compared to that of 
nearby more metal-rich populations; 
the formal frequency of spectroscopic
binaries and multiple systems per target is 17.0$\pm$1.0\%. 
There are several methods to search for such using our data.
Examination of the spectra by eye established that 
HE0218$-$2738 from the Keck Pilot Project is a double lined
spectroscopic binary, and that the star HE0024$-$2523
has broadened lines, both reported in \cite{cohen02}.
\cite{lucatello03} determined the orbital parameters for
the latter star.  A more quantitative test is to examine the
cross-correlation of the spectrum over spectral regions
selected to be
without strong lines to determine the width of the spectral lines.
This width is presumably from stellar rotation, but
the rotation of main sequence turn off stars is expected to be
undetectably small under normal circumstances.  For the C-enhanced star
HE0024$-$2523, \cite{lucatello03}
demonstrate that the high rotation is due to
tidal synchronization in a close binary.

\begin{figure}
\epsscale{1.0}
% Comment out the following line to embed the PS figure into the manuscript
% \plotone{/scr2/jlc/hamburg_survey/hires_sep2001/crosscor/fwhm.ps}
\plotone{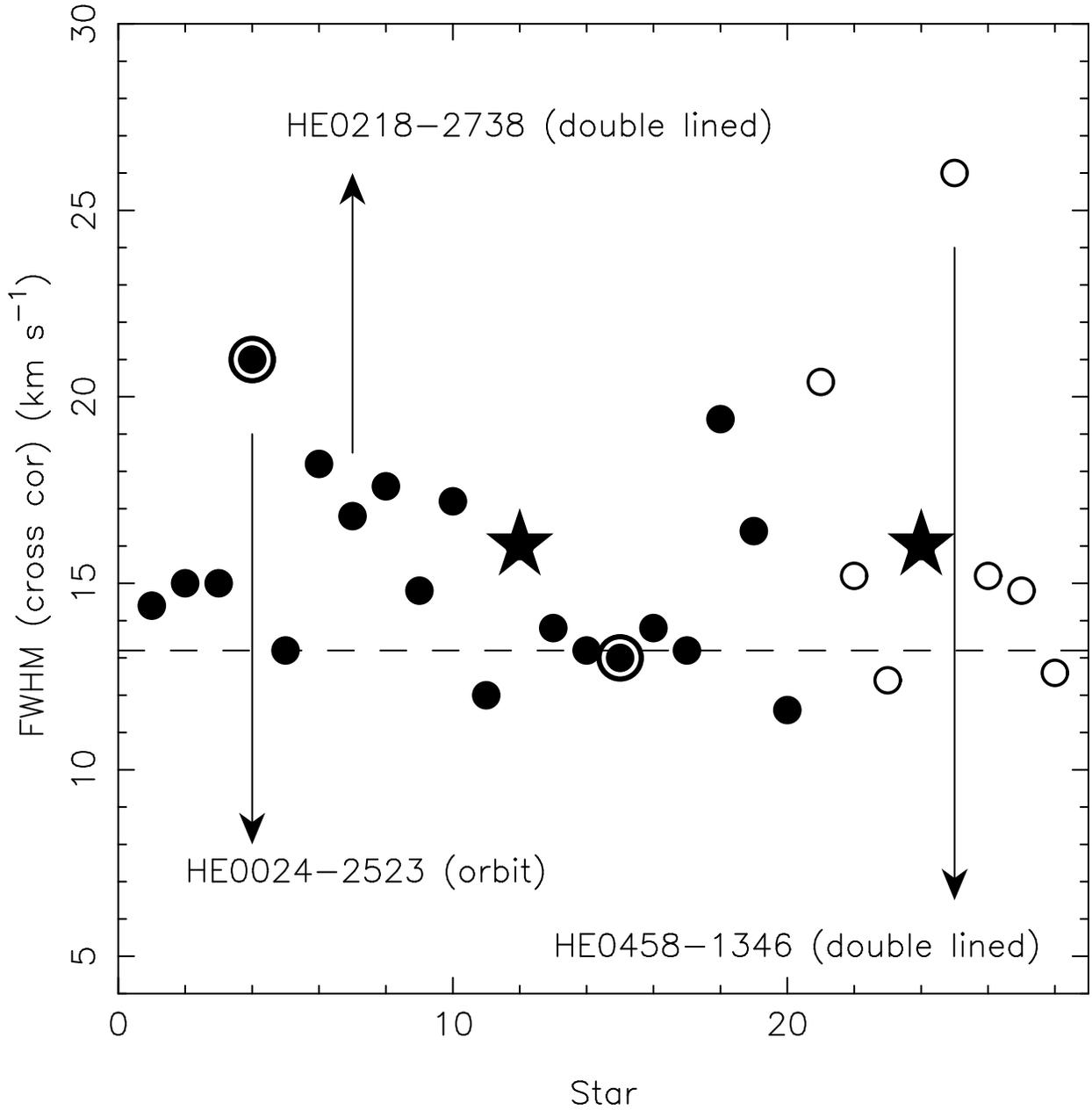}
\caption[]{The FWHM in \kms\ of the peak of the cross-correlation function for
regions of the HIRES spectrum of each sample star selected to be free
of strong spectral features.  The spectrum of LP 0831$-$07 was used
as a template.
The symbol key is that of Figure~\ref{figure_feioneq}.
The spectral resolution for most
of the spectra corresponds to a FWHM of 13.2~\kms~(twice the resolution),
indicated by the dashed line.  For the spectra of May 2002 only,
which are indicated by open circles,
a slightly wider slit was used.  The three confirmed binaries are marked.
\label{figure_fwhm}}
\end{figure}

Figure~\ref{figure_fwhm} shows the mean
FWHM of the peak of the cross-correlation
function, which we denote as $W$, for each star  over selected 
spectral regions devoid of very strong lines.
The bright  dwarf LP0831$-$07 was used as a template.  The stars are
ordered by observing run, so any secular trend in the FWHM
would be apparent.
Recall that the nominal resolving power for the
HIRES configuration we use is 
$R=\lambda/\Delta\lambda =$45,000, which corresponds to 
$W$ for a single spectrum of 13.3~\kms, since the
cross-correlation has an expected FWHM twice that of a single
star (with unresolved lines).
Open circles denote the measurements
for the stars from the May 2002 run taken with a slightly
wider slit.  Most of the stars scatter around the expected
value; the few cases slightly below are either measurement error
or represent spectra taken during periods of
unusually good seeing where the stellar image did not totally
fill the slit.  But there are some cases with noticably
larger $W$, one of which is HE0024$-$2523.  

This process led to the discovery of
HE0458$-$1346, the star with the largest
$W$ in the sample, as a double lined spectroscopic binary, with the
secondary component contributing a small fraction of the total light
and with a velocity separation at the epoch of observation smaller than that
of HE0218$-$2738, which had at the epoch of observation
a wider velocity separation, but the second star contributes a very
small fraction of the light at optical wavelengths.
There are several other stars with unusually large values of
$W$, but nothing suspicous was seen when their spectra were checked.

We are monitoring some of 
these dwarfs, particularly those with C-enhancement, 
for radial velocity variations. Establishing the
the fraction of binary stars for our sample of EMP
dwarfs will require substantial additional observational efforts.

\section{Comments on Nucleosynthesis}

This study
of EMP stars together with the very recent work of \cite{cayrel03}
offer significant improvements upon
previous studies in two critically
  important areas.  One, we detect a plateau in the 
  abundance ratios relative to [Fe/H] of very metal-poor dwarf stars
   for the $\alpha$-elements included in our work between Mg and Ni.
In contrast Cr, Mn and Co show definite trends with
metallicity.     Two, 
  in contrast to previous results showing a large 
  scatter in the abundances, we find a small dispersion 
  in [X/Fe] for these elements.  These dispersions are still for many species limited
by internal or systematic errors rather than by star-to-star variations
in abundance ratios.  
We discuss the abundance ratios first.  
The values we infer for our
sample of EMP dwarfs from the HES
are similar to those found in the previous
studies of very metal poor stars by 
\cite{mcwilliam95}, \cite{carretta02} and the many
additional surveys referred to in \S\ref{section_heavy}.
We adopt the simplest possible hypothesis initially.  Such metal
poor stars must be very old, and hence the contribution of Type Ia SN
to their chemical inventory must be essentially zero.  We therefore 
assume that only Type II SN have contributed to the stellar inventory.

\begin{figure}
\epsscale{1.0}
% Comment out the following line to embed the PS figure into the manuscript
% \plotone{/scr2/jlc/hamburg_survey/hires_summary/programs/dwarf_nucleo1.ps}
\plotone{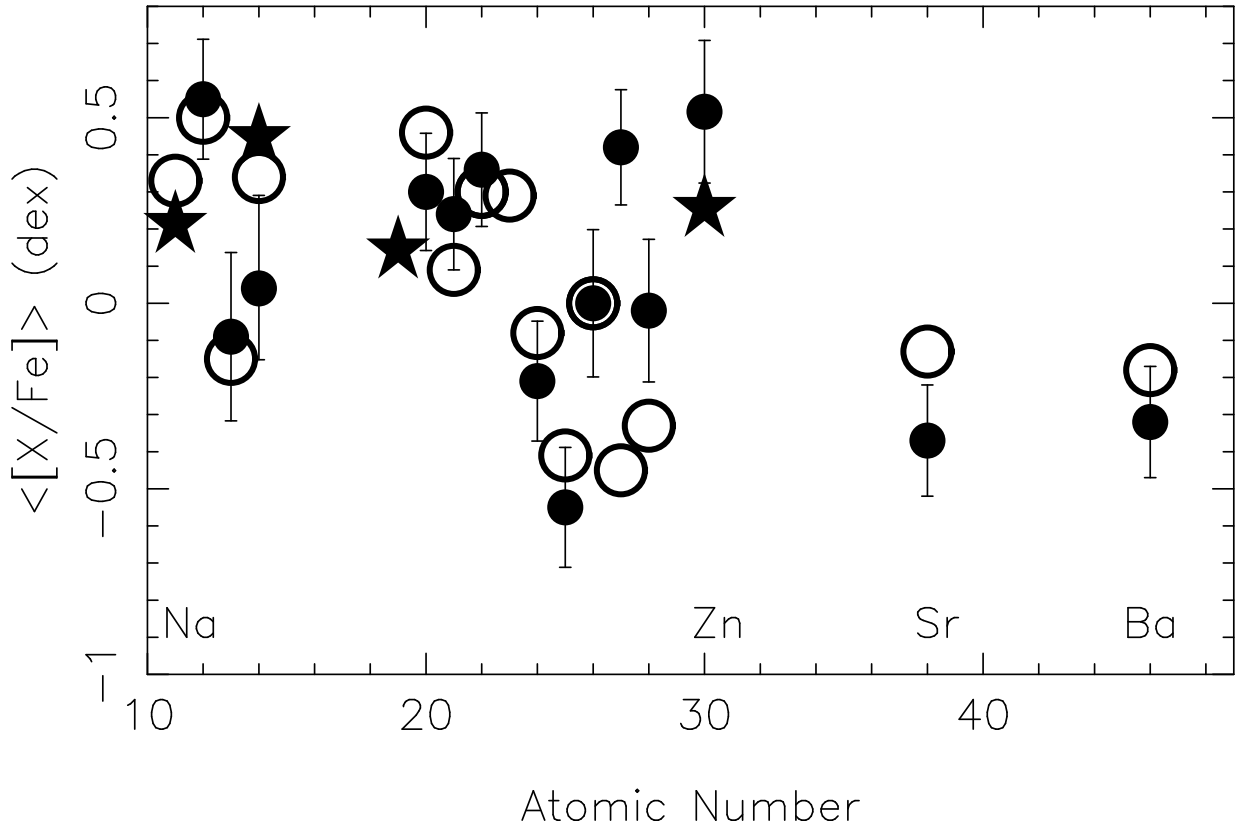}
\caption[]{The mean abundance ratios are plotted as a function of atomic
number, with Ba plotted at atomic number 46.  The filled circles are from 
our analyses and the stars are from that of \cite{cayrel03}.  The open 
circles represent the abundance ratios predicted
the phenomenological model for the yield of Type II SN by \cite{qian02}.
\label{figure_nucleo_gjw}}
\end{figure}

We first compare our deduced abundances with 
the yields for type II SN taken from 
the semi-empirical phenomenological calculations of \cite{qian02}, which are
in fact based on earlier analyses of EMP stars.
Figure~\ref{figure_nucleo_gjw} plots the mean abundance ratios 
[X/Fe] for our sample
of EMP dwarfs as a function of atomic number, with barium plotted at
atomic number 46.  The large stars denote values
from \cite{cayrel03} for elements where we have no data (Na and K) or
cases where our
values disagree with theirs (Si) or our value is particularly uncertain (Zn).
In comparing the mean abundances with model predictions, the systematic
errors as well as the random ones in the abundance ratios must be included.
Hence we combine in quadrature a systematic error of 0.15 dex, allowing an
upward change of 0.22 dex as well for Si only (see \S\ref{section_cayrel}),
with the error in the mean [X/Fe] from random errors.  An additional error
term of $\pm$0.15 dex was added for [Al/Fe], due
to the large non-LTE corrections applied to Al.  These errors are indicated
in Figure~\ref{figure_nucleo_gjw} and also in the next figure.
The predictions of \cite{qian02} are indicated
in Figure~\ref{figure_nucleo_gjw}  by open circles.
These fit the data quite well in general, although Co has a large  
discrepancy ($\sim0.8$ dex, more than 4$\sigma$).

\begin{figure}
\epsscale{1.0}
% Comment out the following line to embed the PS figure into the manuscript
% \plotone{/scr2/jlc/hamburg_survey/hires_summary/programs/dwarf_nucleo2.ps}
\plotone{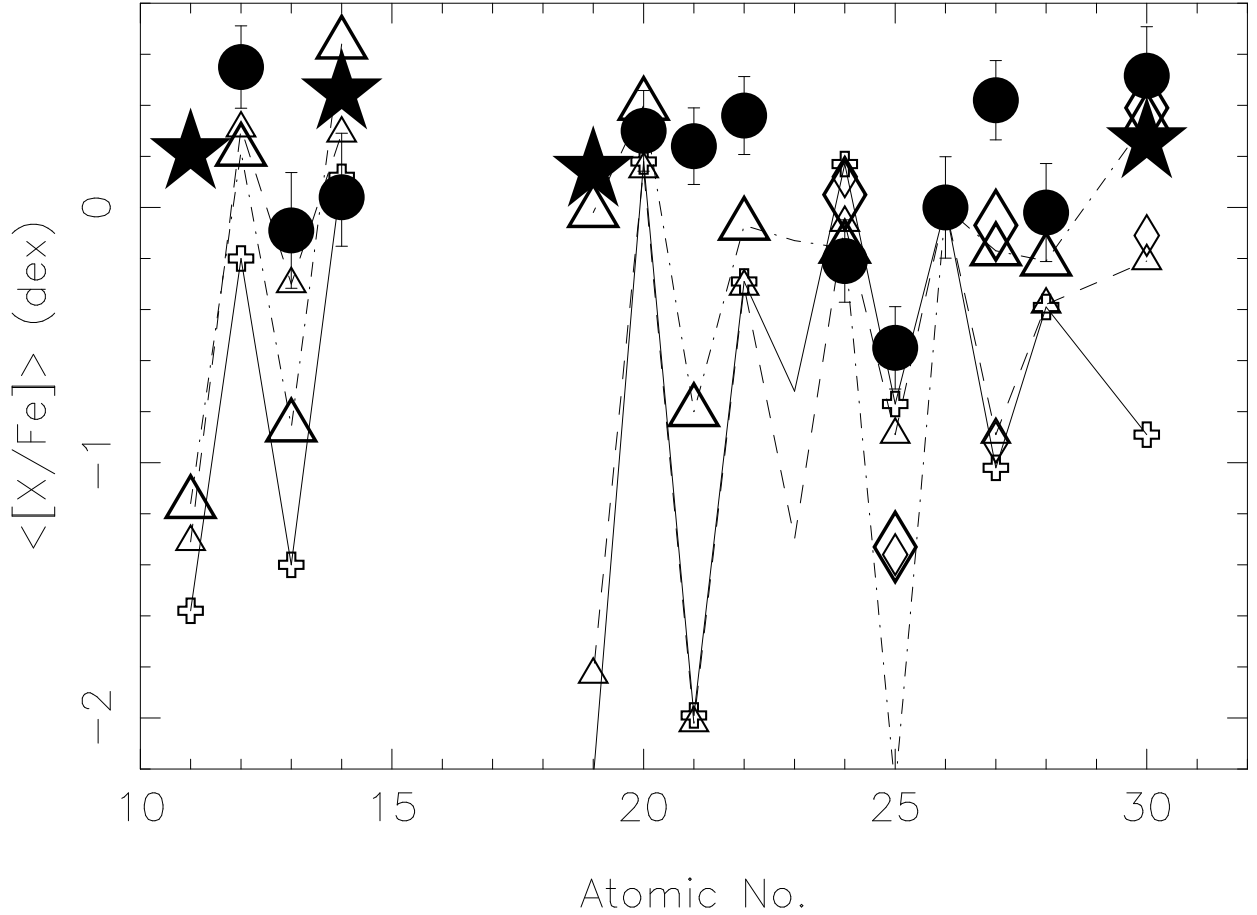}
\caption[]{The mean abundance ratios from Na to Zn are plotted as a 
function of atomic
number.  The filled circles are from our analyses and
the stars are from that of \cite{cayrel03}.  The crosses denote predictions
for selected Population III high mass SN from \cite{umeda02} and 
\cite{umeda04}.  The small open crosses are from a 15 M\subsun\
model with explosion energy 10$^{51}$ ergs and are connected
by line segments. The small open triangles
are from a 30 M\subsun\ model with explosion energy 10$^{51}$ ergs,
while the large triangles are from a model with
explosion energy $5 \times 10^{52}$ ergs.  The predictions
for several Fe-peak elements from 50 M\subsun\ models with
energies of 10$^{52}$ and 10$^{53}$ ergs are shown as small and large
diamonds.
\label{figure_nucleo_nomoto}}
\end{figure}

\begin{figure}
\epsscale{1.0}
% Comment out the following line to embed the PS figure into the manuscript
% \plotone{/scr2/jlc/hamburg_survey/hires_summary/programs/surveys_nife.ps}
\plotone{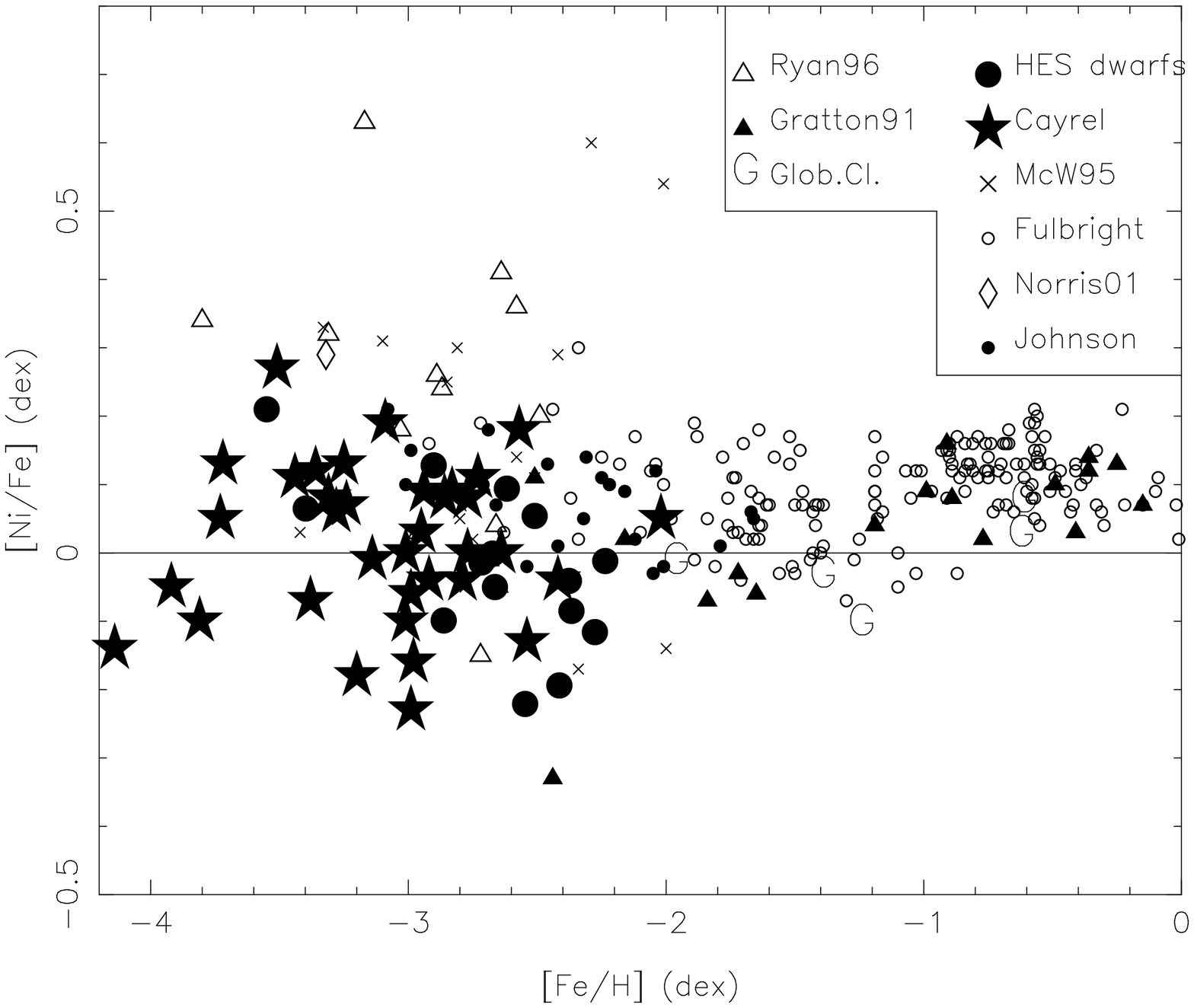}
\caption[]{[Ni/Fe] is shown as a function of metallicity for
a large sample of halo stars compiled from the literature.  
The symbol
key is shown on the figure and the full references
are given in the text. 
see \ref{section_globs}.
\label{figure_surveyni}}
\end{figure}

\begin{figure}
\epsscale{1.0}
% Comment out the following line to embed the PS figure into the manuscript
% \plotone{/scr2/jlc/hamburg_survey/hires_summary/programs/surveys_crfe.ps}
\plotone{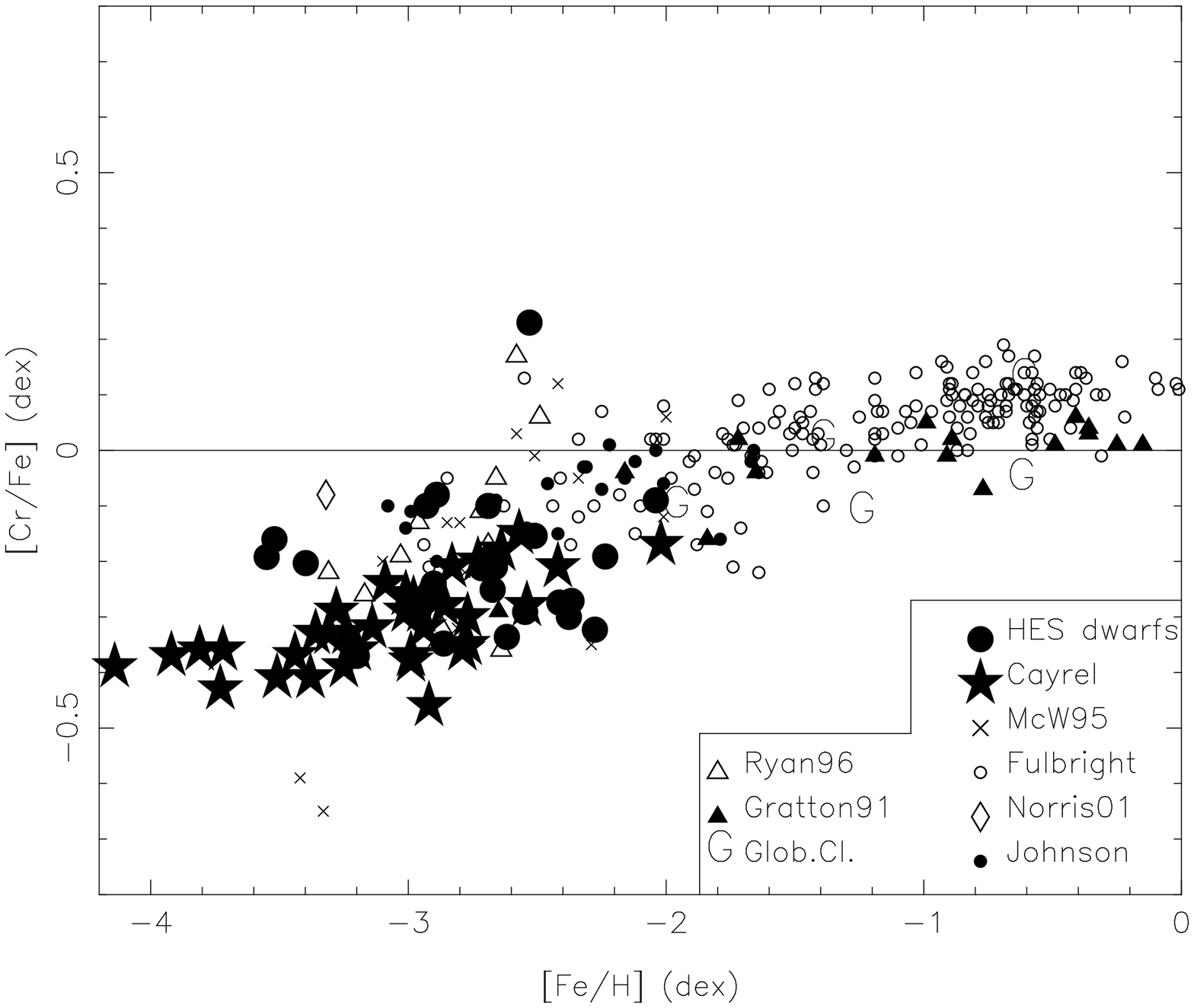}
\caption[]{[Cr/Fe] is shown as a function of metallicity for
a large sample of halo stars compiled from the literature.  
The symbol
key is shown on the figure and the full references
are given in the text. 
\label{figure_surveycr}}
\end{figure}

\begin{figure}
\epsscale{1.0}
% Comment out the following line to embed the PS figure into the manuscript
% \plotone{/scr2/jlc/hamburg_survey/hires_summary/programs/surveys_mgfe.ps}
\plotone{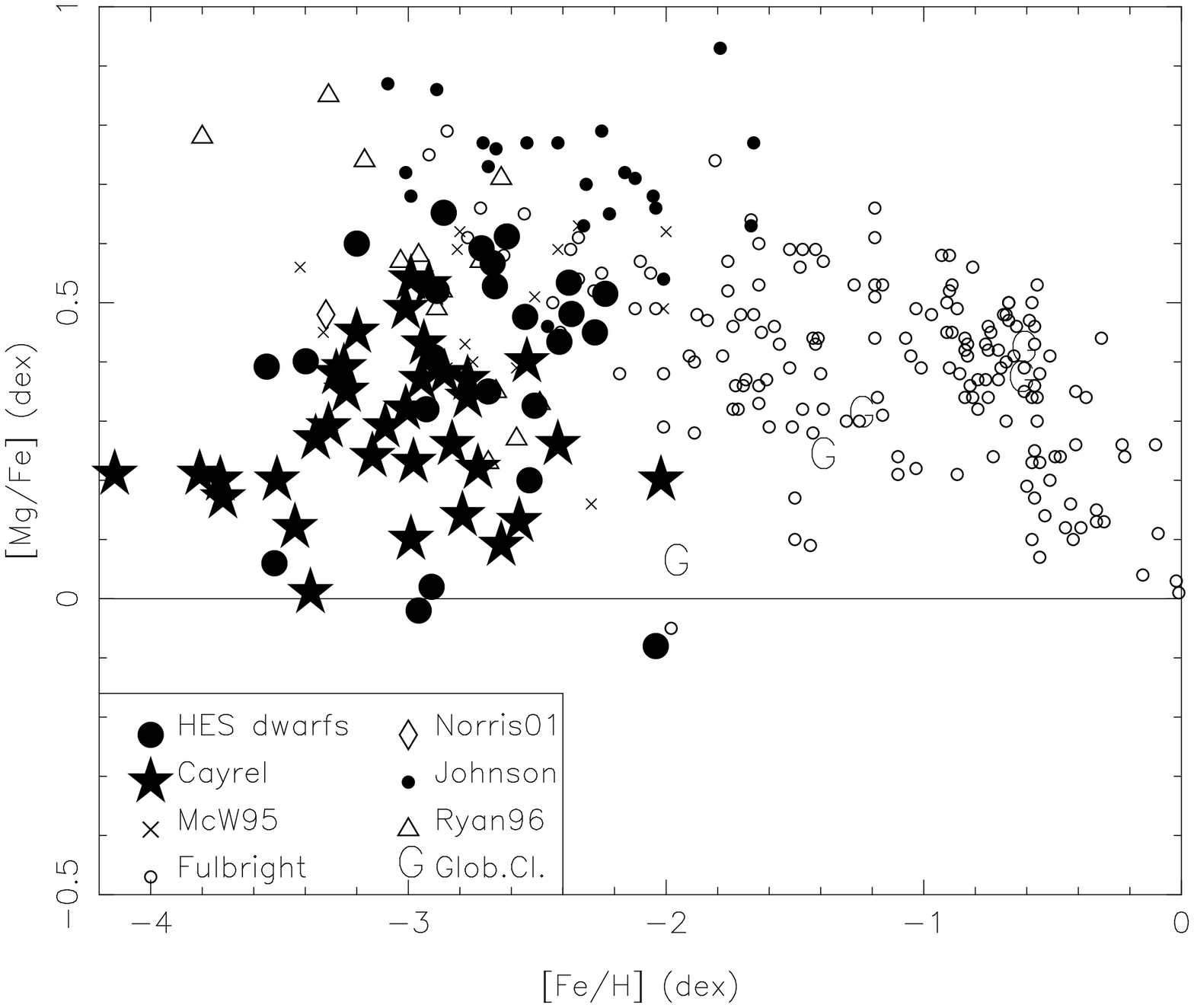}
\caption[]{[Mg/Fe] is shown as a function of metallicity for
a large sample of halo stars compiled from the literature.   The symbol
key is shown on the figure and the full references
are given in the text. 
\label{figure_surveymg}}
\end{figure}

\begin{figure}
\epsscale{1.0}
% Comment out the following line to embed the PS figure into the manuscript
% \plotone{/scr2/jlc/hamburg_survey/hires_summary/programs/surveys_cafe.ps}
\plotone{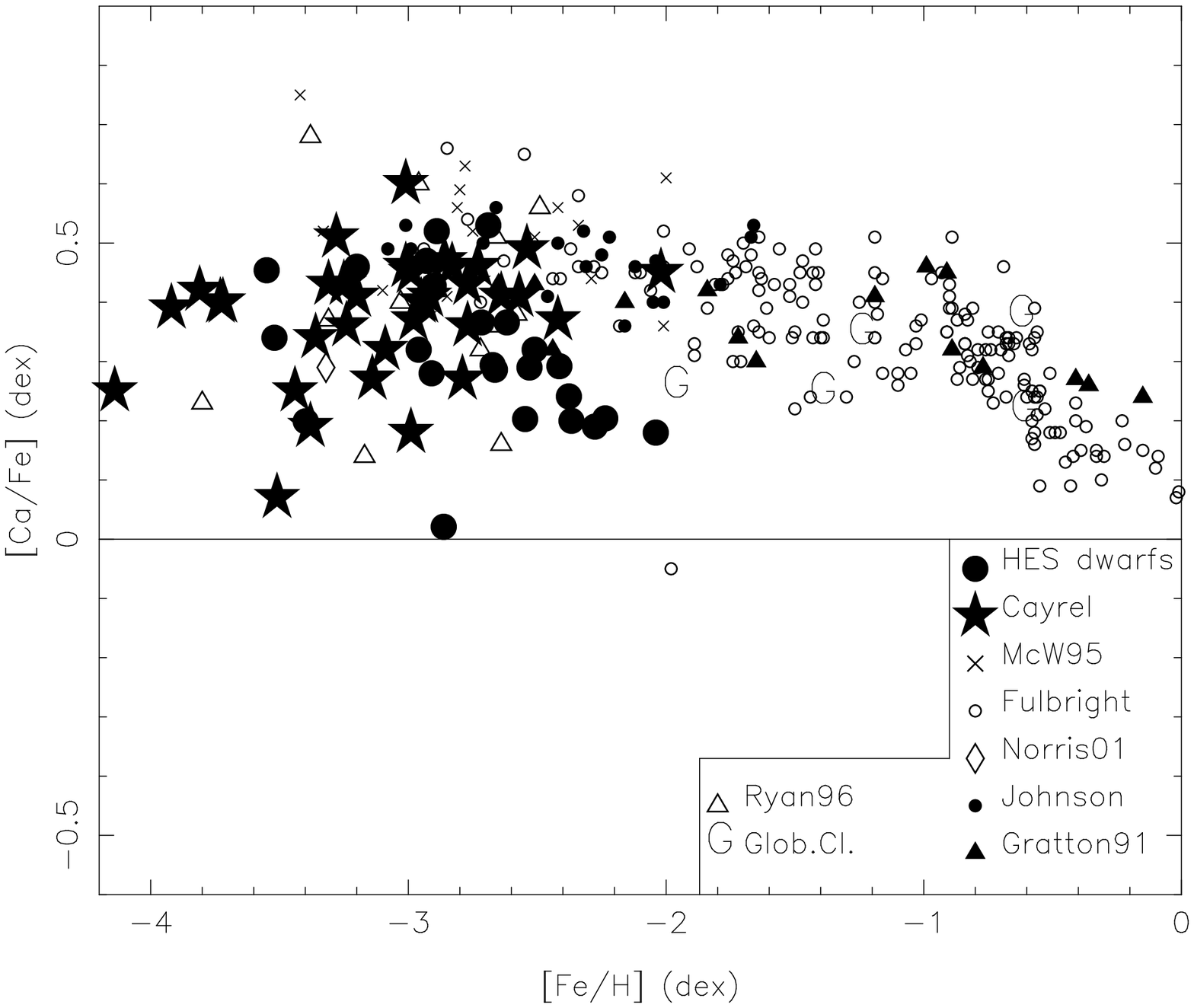}
\caption[]{[Ca/Fe] is shown as a function of metallicity for
a large sample of halo stars compiled from the literature. The symbol
key is shown on the figure and the full references
are given in the text.  
\label{figure_surveyca}}
\end{figure}

\begin{figure}
\epsscale{1.0}
% Comment out the following line to embed the PS figure into the manuscript
% \plotone{/scr2/jlc/hamburg_survey/hires_summary/programs/surveys_mnfe.ps}
\plotone{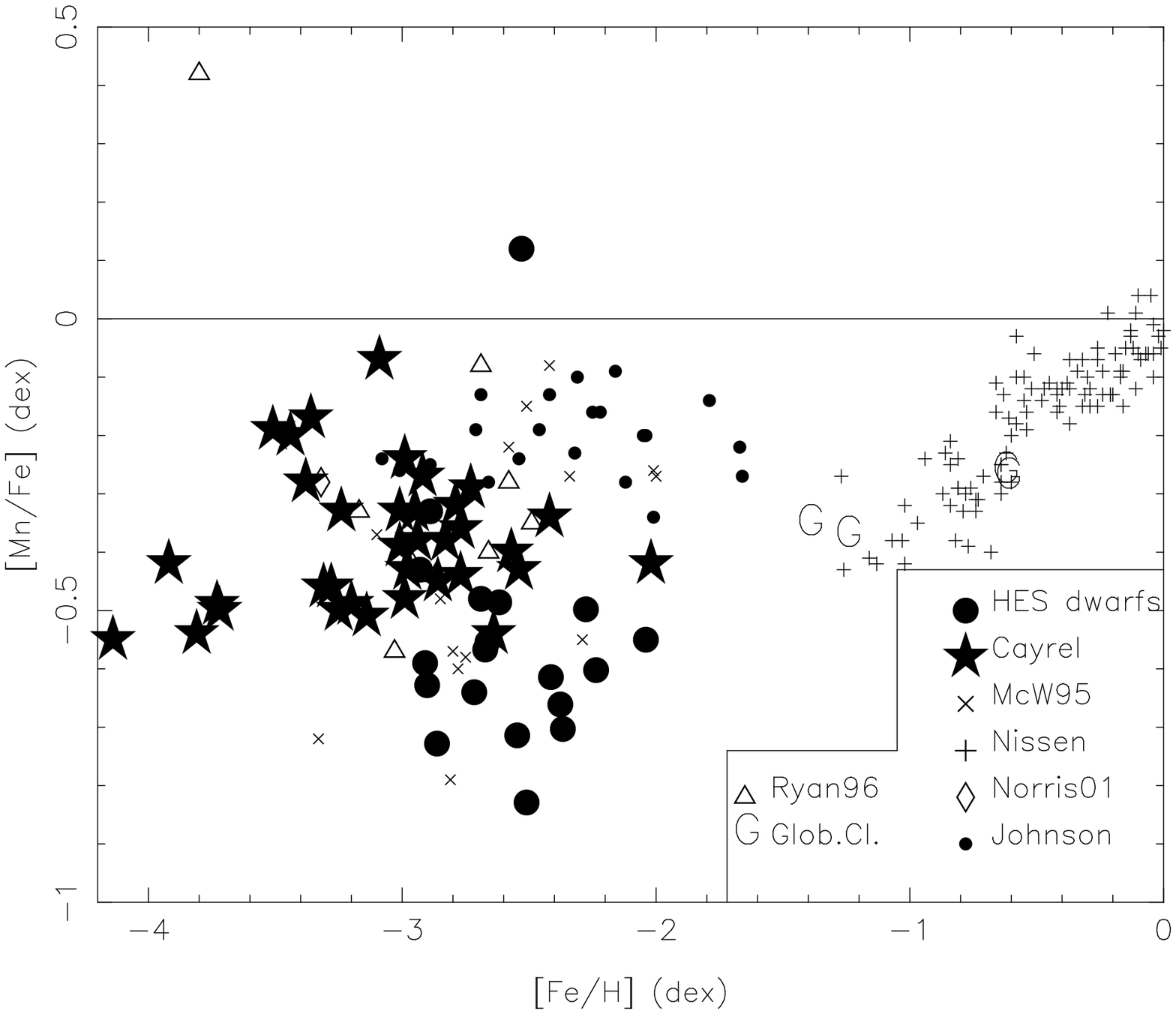}
\caption[]{[Mn/Fe] is shown as a function of metallicity for
a large sample of halo stars compiled from the literature.  The symbol
key is shown on the figure and the full references
are given in the text.  
\label{figure_surveymn}}
\end{figure}

\begin{figure}
\epsscale{1.0}
% Comment out the following line to embed the PS figure into the manuscript
% \plotone{/scr2/jlc/hamburg_survey/hires_summary/programs/surveys_cofe.ps}
\plotone{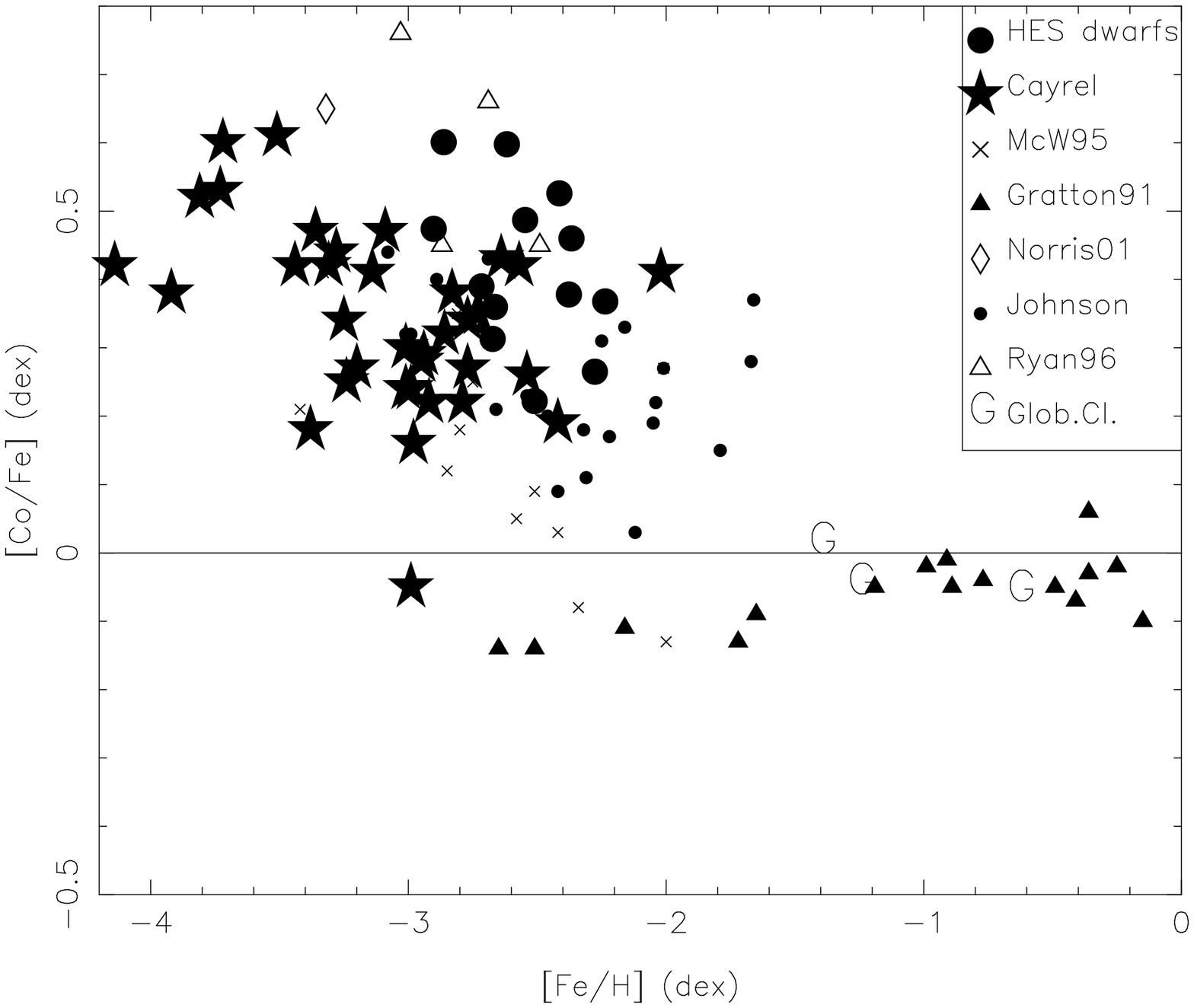}
\caption[]{[Co/Fe] is shown as a function of metallicity for
a large sample of halo stars compiled from the literature. 
The symbol
key is shown on the figure and the full references
are given in the text.  
\label{figure_surveyco}}
\end{figure}

\cite{woosley95} and \cite{umeda02}, among others, present
detailed {\it{ab initio}} calcuations of the yields for low metallicity
Type II SN.  A comparison with the predictions of 
abundance ratios with respect to Fe by the latter group for
SN with masses of 15, 30 and 50 M\subsun\ and explosion energies
ranging from 10$^{51}$ to 10$^{53}$ ergs
is shown in
figure~\ref{figure_nucleo_nomoto}.  These are representative
of  the many cases they
studied, which cover a range from 13 to 50 M\subsun, with a large
range in explosion energy as well. 
These models have particular difficulty
in reproducing the odd-even effect as seen in EMP stars; 
they often overestimate its magnitude, and so underestimate 
one or more of the observed
[Na/Fe],  [Al/Fe], [K/Fe], [Sc/Fe], [Co/Fe] and
[Mn/Fe] ratios in EMP stars. In order to reproduce the observations
for the Fe-peak elements, \cite{umeda04}
introduce additional parameters involving fall back and mixing. 
Some combination of all these parameters does reproduce our data;
SN progenitors at the upper end of the mass range considered are required. 
But why that, and only that, small range of
progenitor mass and  SN explosion parameters occurs
under EMP conditions remains to be explained.

Another important result from our work is in the
suggestion of a plateau 
in the abundance ratios at very low metallicity (i.e. [X/Fe] fixed
as [Fe/H] varies) for most elements (particularly the $\alpha$-elements)
studied here between Mg and Ni except Cr, Co and Mn.  These last three elements
show definite trends between [X/Fe] and [Fe/H].  Another element
which shows such a trend
is Cu (atomic number 29, i.e. another odd element) \citep*[see][and
references therein]{simmerer03}.  Furthermore all of these
trends between [X/Fe] and [Fe/H], both flat and with a definite
slope, are followed by the Galactic 
globular clusters (Table~\ref{table_globcomp}).  We thus
require some metallicity or time dependent effect 
to produce the trends
observed for Cr, Mn, Co and Cu, the latter three 
of which are relatively rare elements
with $N$(Co, Mn or Cu)/$N$(Fe) $< 100$, while maintaining
constant ratios for the abundant $\alpha$-elements.

The bulk of the production of Type II SN is $\alpha$-elements;
the production of Mn, Co and Cu is relatively small and,
at least as modeled by \cite{umeda02} and \cite{umeda04}, 
much more subject
to variation given small changes in the characteristics of
the the progenitor star or the SN explosion.  Either the SN yields 
or the IMF
could be metallicity dependent. In the latter case
the favored mass for the SN progenitors
would evolve monotonically with time, presumably towards smaller masses.
The constancy of
the abundance ratios for the $\alpha$-elements suggests that the IMF
is not varying strongly with time/metallicity, so we suspect that
the production yields of some of the less abundant elements in the region
of the Fe-peak must be varying with metallicity.
Perhaps this is related to 
neutron excess dependent yields for these specific elements
\citep{arnett71,arnett96}, with a larger odd-even 
amplitude for smaller neutron excesses which are characteristic
of metal-poor material that has not been modified by H or He burning.

The most crucial result presented here, in addition to the
abundance ratios themselves, is the small dispersions found
for [X/Fe] for X between Mg and Ni
for the EMP dwarfs in our sample once the C-rich stars
and the dwarf HE2344$-$2800 are eliminated
(see Table~\ref{table_disp}).
These dispersions are consistent with our small estimated observational errors
(up to a factor of $\sim$2 larger in many cases, probably arising from the
neglect of small systematic effects arising within the atomic data,
see \S\ref{section_disp_real})
given in Table~\ref{table_sens}
and hence the intrinsic dispersions may be even smaller.
The only genuine outlier among the dwarfs in our sample
is the star HE2344$-$2800 with abnormally high Cr and Mn,
as originally pointed out by \cite{carretta02}.
The nucleosynthetic processes required to produce these anomalies
is not well understood.

The HES covers a large area on the sky, so different SN presumably
contribute to the chemical inventory of each star.
As the
metallicity decreases and the
number of supernovae contributing to a given parcel of
gas becomes smaller, this uniformity
becomes more challenging to reproduce.  Yet the observations appear
to demand uniformity in relative yield patters for common
elements up to the Fe-peak group.  One would expect that
different SN in the early halo might have had
different mass progenitors and different
mass cuts, eject different masses into the
ISM, which are then diluted and mixed within the ISM in a very
stochastic manner.  Their relative yields and yield patterns were not identical,
as is found in all SNII nucleosynthesis models.
How a diverse class of stars with very different metallicities can exhibit so
homogenous a set of abundance ratios from very variable 
SNII sources is not evident.
One possibility is that even the lowest metallicity stars contain the
average of ejecta from many SNII, thus smoothing out the effect from 
different sources.  A second possibility is that the early SN were much
more uniform in their characteristics.
Invoking a Population III pushes the
issues of homogeneity further back in time, where perhaps the
physics of star formation at very
low metallicity could introduce constraints which
gave rise to a much sharper IMF, a better
defined SN mass, etc. than are characteristic of Type II SN today.   
The simulations of \cite{karlsson01} illustrate
some of the issues involved.  
% Guided by little other than intuition, 
% we suspect
% that many SN may be contributing to the chemical inventory of each star even
% at such low metallicities.
% removed at request of GJW

Among the heavy elements, once the C-rich stars are eliminated,
most of the dwarfs lie within a range Sr/Fe 
of about a factor of 10;
the full range including a few outliers with very low Sr/Fe
is a factor of 80.  Ba and Sr are
reasonably well correlated with each other when both
are detected, but there are five dwarfs in which no line
of Ba~II was detected.  
The production of Sr through the $s$-process 
is discussed at length in \cite{travaglio04}.
Presumably the wide range we see in Sr/Fe  and even wider range
in Ba/Fe reflects the varying contributions of the $r$ and
$s$ neutron capture processes
to the chemical inventory of Sr and Ba in
EMP dwarfs and of mass transfer across binaries, or perhaps
the ratio  of Ba to Sr from the $r$-process alone
varies somewhat at these low metallicities for reasons not yet understood. 

Homogeneity still
largely prevails even at these very low metallicities 
for the more abundant elements Mg through Zn.
But for Sr and Ba, which are much rarer elements, with abundances
a factor of about 1000 below that of Ni,
stochastic effects have become easily visible among these EMP dwarfs.

\subsection{The C-rich Stars}

These stars are main sequence turnoff stars.  They are not blue
stragglers.  They are unevolved, and so any peculiarities must
be primordial, inherent in the ISM when the stars formed, or
accreted from some external source such as a binary companion,
a nearby SN, etc.  Our sample of candidate EMP dwarfs contains
two carbon stars and two stars with C strongly enhanced.
Detailed analyses have already been published for the two C-enhanced
stars \mystar\ \citep{cohen03} and \sarastar\ \citep{lucatello03}.
These stars, and one of the two carbon stars
in our sample, HE~0143$-$0441,
share the property that in addition to the enhancement of the
light elements, the neutron capture elements past the Fe peak are strongly
enhanced.  Very large enhancements of the $s$-process elements
are seen, including detections of lead in all three of these stars.
On the other hand, the fourth EMP carbon star, HE0007$-$1832,
shows no enhancement of the heavy elements and no detection of lead. It has 
Sr/Fe and Ba/Fe ratios consistent with the majority of the non-C-enhanced
stars (see Figures~\ref{figure_basr} and \ref{figure_srpb}).  

The scenario discussed for such stars is similar to that suggested
by \cite{mcclure83} for the 
Ba stars.  It involves mass transfer
in a binary system where the primary star has become an AGB star and 
has transferred mass containing nucleosynthesis debris onto
the secondary star, which is the star we now observe.  This can explain
most of the properties of these stars, and is the origin of our
desire to check the C-enhanced stars for binarity as thoroughly
as possible.  
While it is clear from the work of \cite{mcclure90}, \cite{north00}
and others that most of the Ba stars are binaries, none of the
giant carbon stars with no $s$-process enhancement appear to be
binaries \citep{mcclure97}.  There is thus considerable interest
in whether or not the EMP stars of this type are binaries;
evidence at present is scarce, but see \cite{preston01},
among others.

Detailed models of nucleosynthesis in such systems
have been presented by \cite{gallino98}, \cite{busso99},
\cite{arlandini99} and others.  The yield of $s$-nuclei in an AGB
star depends on the mass of the star and the magnitude of the carbon
pocket.  Thus it should be 
possible to produce a range of enhancements
of the $s$-process elements, from very little to very large,
as is seen in our sample.

\section{Implications for the HES Survey \label{section_hescomp}}

Every effort has been made to find and include the most metal poor dwarfs possible
in choosing the sample from the HES for high resolution observations.
As is indicated in \S\ref{section_abundfe}, there
are no dwarfs in the present sample with [Fe/H] $< -4$ dex, and very few
with [Fe/H] $< -3.5$ dex.   Furthermore
in our follow up efforts thus far no additional turnoff stars with
$\mbox{[Fe/H]}<-3.5$ have yet been identified for future
detailed study.  This rarity of stars below [Fe/H] $\sim -3.5$ dex
must be a genuine characteristic of the general
metal poor star population.  \cite{cayrel03} succeeded in adding two stars
to the six previously known to have [Fe/H] $\le -3.5$ dex, five of which
are studied in detail in \cite{norris01} and the most iron-poor
of all, with [Fe/H] = $-5.3$ dex (found in the HES),
is analyzed by \cite{christlieb04}.  We have  added
two more stars to that select group, HE0218$-$2738 and BS16945-0089,
for a total of 10.

The spectra of EMP candidate dwarfs are very weak lined, and the
assignment of metallicity from  moderate resolution 
spectra is difficult, yet this is crucial to utilization of
the full sample of the HES.  We first consider the accuracy of this
process as it is currently carried out, 
as described in \cite{bee92} and \cite{beers99}.
We define $\Delta$(Fe) as [Fe/H](HK) $-$ [Fe/H](HIRES).  Ideally
$\Delta$(Fe) should be 0 for all stars.
Ignoring the bright calibration stars, 
the mean of $\Delta$(Fe) for the 25 dwarfs in our sample
computed from the values given in
Table~\ref{table_feabund}
is +0.46$\pm$0.31 dex, as can be seen by comparing the 
middle and bottom panels of Figure~\ref{figure_fehist}.  If we exclude
the 3 stars from the HK survey in our sample, then for 22 stars we find
$\Delta$(Fe) = +0.51$\pm$0.23 dex. Excluding the
four C-rich stars in our sample only reduces $\Delta$(Fe) to +0.48 dex.
Our  preliminary impression  at this time is that
this problem is less severe among the HES giants, which appear to have
a mean $\Delta$(Fe) $\approx$ $+0.2$ dex, corresponding
to just the difference in adopted Solar iron abundances,
although we have noticed that among the HES giant 
samples, the C-rich stars systematically tend to have 
abnormally large values of $\Delta$(Fe).

Thus the calibration used by the HES for the approximate abundance
indicator [Fe/H](HK), based on a similar parameter used by the HK Survey,
appears to systematically underestimate the abundance
of EMP candidate dwarfs as measured from our detailed abundance
analyses.  The variation with time over the past 15 years
of concensus choices for the iron abundance of the Sun
can give rise to a maximum  contribution of +0.2 dex to $\Delta$(Fe).
Part of the offset in [Fe/H] scales might also be due to
a selection bias similar to the Malmquist bias.
The metallicity distribution function of halo stars
declines rapidly as [Fe/H] decreases in the metallicity
range of interest.  Uncertainties in [Fe/H](HK) will tend to
scatter intermediate metallicity stars into the low metallicity regime
at a rate which greatly exceeds the converse, thus introducing
a bias towards positive values of $\Delta$(Fe) and producing
a sample of EMP stars selected for high dispersion
spectroscopy whose true metallicity is higher than anticipated.
Monte Carlo trials suggest that for (Gaussian) uncertainties in
[Fe/H](HK) of $\pm$0.3 dex,
an increase of $\sim$0.1 dex in the mean [Fe/H] above that expected 
will result in the HIRES sample. 
This leaves $\sim$0.2 dex presumably arising from calibration 
differences, including differences between the \teff\ scale adopted
here and that of \cite{beers99b}.

Irrespective of the cause of this problem, in spite of our best efforts
thus far, the samples of candidate EMP dwarfs culled from the HES
appear to be contaminated with somewhat higher metallicity,
although still very metal poor, dwarfs even after two stages of vetting
including moderate resolution spectroscoy.

It is also important to note in this context that the counts of EMP
stars within the HK Survey given by \cite{beers99} are based almost
entirely on moderate resolution spectra, not from detailed abundance
analyses from high dispersion spectra.  Since dwarfs dominate
the samples, Beers' statement that the
HK Survey contains 100 known EMP stars may be a substantial overestimate.

The follow up spectra cannot easily detect the weak molecular
features in EMP dwarfs in the \teff\ range considered here, so
the fraction of C-rich stars found among the HIRES sample, which is
$\sim$20\%, is a lower limit to the fraction of such stars in the
full HES sample.  This fraction is best established from the giants,
but there the issue of internal production cannot be so readily
dismissed.

The previously unpublished HES EMP candidates presented 
here have a median $V$ of $\sim15.7$ mag.
Because these stars are all  dwarfs near the TO, they represent
a sample of halo dwarfs with median distance of 2.8 kpc, while
a giant at the same $V$ with luminosity
1 mag below the RGB tip would be at a distance of 30 kpc.
We model the density distribution in the Galactic halo
as $\rho \propto R_{GC}^{-3}$, where $R_{GC}$ is the
galactocentric radius, to calculate the ratio of dwarfs
to giants in the range of magnitude relevant to the HES and to the
HK Surveys by viewing the volumes of the two surveys in
galactocentric coordinates.  At the distances typical
of the stars in the HK Survey (which are considerably brighter), 
the 8 kpc distance to the
Galactic center dominates, and $R_{GC}$ does not drop rapidly
with increasing magnitude within such a survey.  
The fainter giants in the HES
are far above the Galactic plane, hence the rapid decline
in $\rho(R_{GC})$  becomes important, while the
dwarfs are still close enough that for them
$R_{GC} \approx 8$ kpc and hence $\rho$ is approximately constant.   
A rough calculation suggests
that the number density of giants to dwarfs at the main sequence
turnoff at $V=13.5$ is more
than a factor of 10 larger than it is at $V=16$. Therefore it
will be more challenging in the HES than in the HK Survey
to separate the very distant halo giants from dwarfs.

%
%  see program /scr2/jlc/hamburg_survey/dwarf_giant_hes.f

\section{Summary}

We present the chemical abundances in a sample of 17 candidate extremely metal poor 
dwarf stars, including 14
previously unpublished EMP candidates 
from the Hamburg/ESO Survey.  This is combined with
the dwarfs from the Keck Pilot Project and gives a total sample of 28 
EMP candidate dwarfs (22 from the HES) near the main sequence turn off
with \teff\ $\ge 6000$~K. 
 
Dwarfs this hot have very weak atomic and molecular lines,
making the problem of finding EMP dwarfs and analyzing their spectra
difficult.  Only the most abundant elements yield
measurable absorption features.  But dwarfs have the advantage
of being unevolved, with no chance of any interior nuclear
processing beyond burning H to He, and thus no chance
of self-pollution.
We analyze high resolution and high precision
spectra of these stars taken with HIRES at the Keck Observatory.
Our sample has a median [Fe/H] of $-2.7$ dex, extends to $-3.5$ dex,
and is somewhat less metal-poor than was expected from [Fe/H](HK,HES) 
determined from low resolution spectra.  We find,
in agreement with previous surveys \citep*[see, for example][]{bee92},
that there appears to be a sharp drop in the metallicity distribution 
of stars below [Fe/H] $= -3$ dex.
In our follow up efforts thus far no additional turnoff stars with
$\mbox{[Fe/H]}<-3.5$ have yet been identified for future
detailed study.

The abundance ratios [X/Fe] which  we derive 
are similar to those found by earlier
investigations, and are, with the exception of Si, in good
agreement with the very recent results of the First Stars Project
\citep{cayrel03} for a large
sample of EMP giants.  Silicon shows a substantial difference in
behavior between the EMP giants and the dwarfs, which may be due
in part to problems in the analysis, but is not fully understood
at present; Sc/Fe may show a similar, smaller, difference between EMP
dwarfs and giants.  Ignoring Si,
the details of the abundance trends, in particular the existence
of an abundance plateau for the $\alpha$-elements with fixed [X/Fe] 
as [Fe/H] decreases, are better defined than in earlier studies. 
These abundance ratios
are in general in reasonable agreement with predictions 
using the yields of \cite{qian02} and assuming that
only Type II SN contribute to the stellar inventory for these
EMP dwarfs.  Cr, Co and Mn (and Cu, see Simmerer \etal\ 2003), 
show definite
trends of mean [X/Fe] with metallicity.  These behaviors are all
in good agreement with the abundance ratios of Galactic
globular clusters.

The dispersions
of [X/Fe] are surprisingly small, as was also found by the
First Stars Project.   They are still in most cases dominated by
internal errors.  Careful attention to the details of the atomic
physics and the analysis will be required for further improvement.

We interpret these trends in abundance ratios as requiring
a fixed IMF for the SN progenitors combined with
metallicity dependent yields for the low abundance elements
such as Mn, Co and Cu.  The progenitor mass must be biased
towards the high end of the range generally considered for Type II SN
to fit the observations. Even at the low metallicities considered here,
we suspect that many SN contributed to the chemical inventory of
each of the observed stars to reproduce the small dispersions
in [X/Fe] determined by us.

Sr/Fe and Ba/Fe have dispersions significantly larger than
the elements from Mg to Ni.  But by eliminating the C-rich
stars, and thus presumably the bulk of the $s$-process material, we
reduce the dispersions for these elements.

Two dwarfs in the sample are carbon stars, while two
others have significant C enhancements.  All four have \ciso\ $\sim$7.
% This requires convective burning near the H shell.
%  comment by GJW, which I took out 
The C-enhanced stars are a mixed group.  Three of the four have
large enhancements of the heavy neutron capture
elements including lead, implying a strong
$s$-process contribution, presumably from an AGB companion.  The fourth
C-rich star shows no anomalies among the heavy elements
compared to the bulk of the sample.

We examined the available information regarding the binarity
of the sample stars.  We pay particular attention to the C-rich
stars, since their substantial 
$s$-process enrichment supposedly arises via mass transfer from
an AGB star.  To date there are three definite binaries,
two double lined systems and one star with
an orbit \citep{lucatello03}, in our sample.

A sample of EMP giants from the HES will be difficult to assemble,
not just because the stars are faint, but because the ratio
of giants to dwarfs is falling rapidly in the relevant
magnitude regime (13 to 16 mag) due to the rapid drop in 
stellar density in the  halo
as a function of galactocentric radius.
Such a sample, which we are now constructing, 
will consist of giants at typical distances of
30 kpc, and will be very interesting to explore.

\acknowledgements

The entire Keck/HIRES user community owes a huge debt
to Jerry Nelson, Gerry Smith, Steve Vogt, and many other people who have
worked to make the Keck Telescope and HIRES a reality and to
operate and maintain the Keck Observatory.  
We are grateful to the W. M. Keck Foundation for the vision to 
fund the construction of the W. M. Keck Observatory. 
We thank Steve Shectman and the many people who worked to design
and build the Magellan Telescopes. 
This publication makes use of data products from the Two Micron All Sky
Survey, which is a joint project of the University of Massachusetts and
the Infrared Processing and Analysis Center/California Institute
of Technology, funded by the National Aeronautics and Space Administration
and the National Science Foundation. 
JGC and JM are grateful for partial support from  NSF grant
AST-0205951.
N.C. acknowledges financial support 
through a Henri Chretien International Research Grant administered
by the American Astronomical Society, and he is very grateful to
 Gaston Araya
   and JGC for their hospitality in Altadena.
III acknowledges research funding from NASA through 
  Hubble Fellowship grant HST-HF-01151.01-A from the Space 
  Telescope Science Inst., operated by AURA, under NASA 
  contract NAS5-26555.

% REFERENCES

\clearpage

\begin{deluxetable}{lrrr rrrr}
\tablenum{1}
\tablewidth{0pt}
\small
\tablecaption{Sample of New Stars \label{table_newstars}}
\tablehead{\colhead{Star} & \colhead{$V$\tablenotemark{a}} &  
\colhead{RA} & \colhead{Dec} & $E(B-V)$\tablenotemark{b} & \colhead{Obs.Run} &
\colhead{Exp.Time} & \colhead{SNR} \\
\colhead{} & \colhead{(Mag)} & \colhead{J2000} &
\colhead{} & \colhead{(Mag)} & \colhead{} & \colhead{(sec)} & \colhead{} }
\startdata 
Very bright \\
LP0831$-$0700\tablenotemark{c}  & 11.62\tablenotemark{d} &
     3 06 05.4 & $-$22 19 18  &  0.019 & 9/2001 & 900  & $>$100 \\
HK stars \\
BS16545$-$0089 & 14.44\tablenotemark{e} &
   11 24 27.6 & +36 50 28 &  0.019 & 5/2002 & 1800 &  80 \\    
BS16945$-$0054  & 13.67\tablenotemark{e} &
    10 50 57.6  & +34 01 39 & 0.022 & 5/2002 & 2700 & $>$100 \\
HES stars \\
HE0007$-$1832\tablenotemark{f} & 15.46 & 00 09 52.8 & $-$18 16 12 & 0.030 &
     9/2001 & 8400 & 90 \\
HE0028$-$2152 & 15.80 & 00 31 07.7 & $-$21 36 07 & 0.019 &
    9/2002 & 9600 & 100 \\
\enddata
\tablenotetext{a}{$V$ mags from observations at the Las Campanas Swope 1m
telescope unless otherwise noted.}
\tablenotetext{b}{$E(B-V)$ from the map of \cite{schlegel98}.}
\tablenotetext{c}{Found in the NLTT proper motion survey \citep{luyten80},
also known as NLTT 831-70.}
% checked Ryan et al 1991 AJ 102, 303 for correct alternate name, its lp831-70
\tablenotetext{d}{$V$ mag from \cite{ryan89}.}
\tablenotetext{e}{$V$ mag from \cite{anthony00}.}
\tablenotetext{f}{This star has detectable C$_2$ bands.}
\end{deluxetable}

\begin{deluxetable}{lrrr rrrrr rrrrr rrrrr rr}
\tablenum{2}
\tablewidth{0pt}
\small
\tablecaption{\eqw\ for 17 Candidate Dwarf EMP Stars \label{table_eqw}}
\tablehead{\colhead{Line $\lambda$} &
\colhead{Species\tablenotemark{a}} &
\colhead{EP} &
\colhead{Log($gf$)} &
\colhead{BS16545} &
\colhead{BS16945} &
\colhead{LP831$-$70} &
\colhead{HE0007} &
\colhead{HE0028} &
\colhead{HE0054} &
\colhead{HE0102$-$0633} &
\colhead{HE0105$-$2202} &
\colhead{HE0142+0147} &
\colhead{HE0143$-$0441} &
\colhead{HE0458$-$1346} &
\colhead{HE0508$-$1555} &
\colhead{HE1141$-$0610} &
\colhead{HE1320+0038} &
\colhead{HE1346$-$0427} &
\colhead{HE2209$-$1240} &
\colhead{HE2341$-$2029} \\    
\colhead{(\AA)} & \colhead{} & \colhead{(eV)} & \colhead{}
& \colhead{$-$0089} & \colhead{$-$0054} & \colhead{} & \colhead{$-$1832}
& \colhead{$-$2152} & \colhead{$-$2123} \\                  
}
\startdata
4057.52 &    12.0 &   4.34 &  -1.200 &   \nodata &    22.8 &    11.2 &   \nodata &     9.6 &     7.2 &   \nodata &    11.2 &   \nodata &    21.2 &    25.7 &    17.6 &    14.1 &   \nodata &   \nodata &     9.6 &   \nodata  \\
4167.28 &    12.0 &   4.34 &  -1.000 &   \nodata &    31.8 &    11.2 &   \nodata &    17.6 &    10.8 &   \nodata &   \nodata &    22.4 &   \nodata &    32.2 &    24.2 &    28.1 &    16.0 &   \nodata &    16.0 &    17.0  \\
4703.00 &    12.0 &   4.34 &  -0.670 &   \nodata &    55.4 &    22.8 &    24.3 &    37.5 &    23.4 &   \nodata &    36.0 &    35.2 &   \nodata &    46.3 &    46.0 &    40.7 &    27.0 &     6.4 &    38.3 &    35.0  \\
5172.70 &    12.0 &   2.71 &  -0.380 &    69.2 &   158.4 &   111.8 &   102.8 &   124.0 &   124.0 &   104.3 &   124.7 &   128.9 &   163.5 &   166.0 &   132.1 &   136.6 &   112.9 &    69.8 &   123.3 &   116.8  \\
5183.62 &    12.0 &   2.72 &  -0.160 &    80.4 &   175.6 &   130.7 &   130.0 &   140.5 &   140.5 &   117.2 &   129.8 &   148.4 &   187.8 &   183.4 &   148.7 &   149.9 &   127.6 &    86.8 &   145.9 &   134.0  \\
3944.01 &    13.0 &   0.00 &  -0.640 &    14.0 &    75.5 &    35.8 &   \nodata &    49.4 &    49.4 &  34.7 &    34.9 &    57.8 &   \nodata &    82.2 &    54.2 &    56.6 &    36.0 &    14.9 &    69.8 &    41.1  \\
3961.52 &    13.0 &   0.00 &  -0.340 &    19.6 &    76.1 &    62.3 &    63.0 &    62.3 &    62.3 &    42.0 &    50.5 &    65.6 &    77.8 &    91.1 &    67.0 &    60.3 &    58.8 &    21.8 &    77.3 &    51.7  \\
3905.53 &    14.0 &   1.91 &  -1.090 &    36.4 &   125.6 &    79.3 &   \nodata &    95.7 &    95.7 &    82.0 &    99.0 &   103.2 &   239.1 &   141.0 &   101.4 &    84.7 &    79.5 &    48.2 &    96.6 &    96.9  \\
\enddata
\tablenotetext{a}{The species are identified by their atomic number
and ionization stage.  Fe~I is denoted as 26.0, Fe~I as 26.1, etc.}
\tablenotetext{b}{Upper limits to \eqw\ are indicated by negative numbers.}
\end{deluxetable}

\begin{deluxetable}{lrrr rrr}
\tablenum{3}
\tablewidth{0pt}
\small
\tablecaption{Stellar Parameters \label{table_params}}
\tablehead{\colhead{Star} & \colhead{\teff} &   \colhead{\grav} &
\colhead{[Fe/H]\tablenotemark{a}} & \colhead{$v_t$} & \colhead{$v_r$} & 
 \colhead{Notes}  \\
\colhead{} & \colhead{(K)} & \colhead{dex} & \colhead{dex} &
\colhead{(\kms)} & \colhead{(\kms)}  &  \colhead{}  
}
\startdata 
Very bright \\
BD 3$-$740  &  6355 & 4.0 & \nodata & 1.4 & +174.5 & kp \\ 
G139$-$8    &  6200 & 4.5 & \nodata & 1.0 & $-$114.0 &  b,kp \\ 
LP0831$-$0700  & 6270\tablenotemark{h} & 4.5 & $-3.0$ & 1.6 & $-48.2$ & \\ 
% LP0831$-$0700  & 6400 & 4.3 & $-3.0$ & 1.6 & $-48.2$ & \\
HK stars \\
% BS16545$-$0089 & 6500 & 3.8 & $-3.0$ & 1.7 & $-163.5$ \\
BS16545$-$0089 & 6115\tablenotemark{h} & 3.6 & $-3.0$ & 1.7 & $-163.5$ \\
% BS16945$-$0054  & 6575 & 4.2 & $-2.5$ & 1.6 & $-$73.5  \\
BS16945$-$0054  & 6065\tablenotemark{h} & 3.7 & $-2.5$ & 1.6 & $-$73.5  \\
BS17447$-$029 & 6350 & 4.4 & \nodata & 1.9 & $-201.6$ & b,kp \\
\enddata
\tablenotetext{a}{The [Fe/H] in dex of the model from the \cite{kurucz93}
grid of model stellar atmospheres used in the analysis.}
\tablenotetext{b}{Star just below the main sequence turnoff.}
\tablenotetext{c}{\grav\ value from Keck Pilot project slightly adjusted.}
\tablenotetext{d}{This is a rediscovery of CS 22966$-$048, originally
found in the HK Survey.}
\tablenotetext{e}{The HIRES spectra of this star show double lines.}
\tablenotetext{f}{Spectroscopic binary with known orbit, \cite{lucatello03}.}
\tablenotetext{g}{See \cite{cohen03} for a detailed discussion of this 
peculiar star.}
\tablenotetext{h}{\teff\ obtained from $T(exc)$ for Fe~I.}
\tablenotetext{cs}{Carbon star; spectrum shows bands of C$_2$.}
\tablenotetext{kp}{From the Keck Pilot Project}
\end{deluxetable}

\clearpage

Notes to table 3 are repeated here, as they fall off the end of the page
with the table.

a. The [Fe/H] in dex of the model from the \cite{kurucz93}
grid of model stellar atmospheres used in the analysis.

b. Star just below the main sequence turnoff.

c. log(g) value from Keck Pilot project slightly adjusted.

d. This is a rediscovery of CS 22966$-$048, originally
found in the HK Survey.

e. The HIRES spectra of this star are double lined.

f. Spectroscopic binary with known orbit, \cite{lucatello03}.

g. See \cite{cohen03} for a detailed discussion of this 
peculiar star.

h. \teff\ obtained from $T(exc)$ for Fe~I.

cs. Carbon star; spectrum shows bands of C$_2$.

kp. From the Keck Pilot Project

\clearpage

\begin{deluxetable}{l rrr rrr rrr rrr rrr}
\tablenum{4a}
% \tiny
\tabletypesize{\tiny}
\tablewidth{0pt}
\tablecaption{Abundances of the Sample Stars \label{table_abunda}}
\tablehead{\colhead{Species} & 
\multispan{3}{LP0831-070} &
\multispan{3}{BS16545$-$0089} &
\multispan{3}{BS16945$-$0054} &
\multispan{3}{HE0007$-$1832} &
\multispan{3}{HE0028$-$2152} \\
\colhead{} & \colhead{log$\epsilon(X)$} & \colhead{No.} & \colhead{$\sigma$} &
\colhead{log$\epsilon(X)$} & \colhead{No.} & \colhead{$\sigma$} &
\colhead{log$\epsilon(X)$} & \colhead{No.} & \colhead{$\sigma$} &
\colhead{log$\epsilon(X)$} & \colhead{No.} & \colhead{$\sigma$} &
\colhead{log$\epsilon(X)$} & \colhead{No.} & \colhead{$\sigma$}  \\
\colhead{} & \colhead{(dex)} & \colhead{Lines} & \colhead{(dex)} &
\colhead{(dex)} & \colhead{Lines} & \colhead{(dex)} &
\colhead{(dex)} & \colhead{Lines} & \colhead{(dex)} &
\colhead{(dex)} & \colhead{Lines} & \colhead{(dex)} &
\colhead{(dex)} & \colhead{Lines} & \colhead{(dex)} 
}
\startdata
   6.0 &   \nodata & \nodata& \nodata   &   \nodata & \nodata& \nodata   &   \nodata & \nodata& \nodata   &     8.29 &    1 & 
     \nodata   &   \nodata & \nodata& \nodata   \\
   7.0 &   \nodata & \nodata& \nodata   &   \nodata & \nodata& \nodata   &   \nodata & \nodata& \nodata   &     7.00 &    1 & 
    \nodata    &   \nodata & \nodata& \nodata   \\
  12.0 &     5.33 &    5 &    0.23    &     4.38 &    2 &    0.02    &     5.71 &    5 &    0.21    &     5.65 &    5 &    
0.35    &     5.42 &    5 &    0.23    \\
  13.0 &     3.64 &    2 &    0.07    &     2.91 &    2 &    0.09    &     3.87 &    2 &    0.21    &     3.28 &    1 & 
   \nodata  &     3.57 &    2 &    0.12    \\
  14.0 &     4.70 &    1 & \nodata   &     3.88 &    1 & \nodata   &     5.33 &    1 & \nodata   &     S5.06 &    1 &  
   \nodata    &     4.89 &    1 & \nodata   \\
  20.0 &     3.52 &    3 &    0.17    &     3.26 &    3 &    0.16    &     4.27 &    7 &    0.11    &     4.03 &    1 & 
    \nodata    &     4.01 &    7 &    0.14    \\
  21.1 &     0.62 &    3 &    0.07    &    $-$0.29 &    1 & \nodata   &     0.90 &    3 &    0.01    &   \nodata & \nodata &  
    \nodata   &     0.52 &    3 &    0.08    \\
  22.0 & U  2.73 &    3 &    0.18    &   \nodata & \nodata& \nodata   &     3.04 &    3 &    0.07    &   \nodata & \nodata & 
    \nodata    &     2.87 &    4 &    0.08    \\
\enddata
\end{deluxetable}

\begin{deluxetable}{l rrr rrr rrr rrr rrr}
\tablenum{4b}
\tablewidth{0pt}
% \tiny
\tabletypesize{\tiny}
\tablecaption{Abundances of the Sample Stars \label{table_abundb}}
\tablehead{\colhead{Species} & 
\multispan{3}{HE0054$-$2123} &
\multispan{3}{HE0102$-$0633} &
\multispan{3}{HE0105$-$2202} &
\multispan{3}{HE0142+0147} &
\multispan{3}{HE0143$-$0441} \\
\colhead{} & \colhead{log$\epsilon(X)$} & \colhead{No.} & \colhead{$\sigma$} &
\colhead{log$\epsilon(X)$} & \colhead{No.} & \colhead{$\sigma$} &
\colhead{log$\epsilon(X)$} & \colhead{No.} & \colhead{$\sigma$} &
\colhead{log$\epsilon(X)$} & \colhead{No.} & \colhead{$\sigma$} &
\colhead{log$\epsilon(X)$} & \colhead{No.} & \colhead{$\sigma$}  \\
\colhead{} & \colhead{(dex)} & \colhead{Lines} & \colhead{(dex)} &
\colhead{(dex)} & \colhead{Lines} & \colhead{(dex)} &
\colhead{(dex)} & \colhead{Lines} & \colhead{(dex)} &
\colhead{(dex)} & \colhead{Lines} & \colhead{(dex)} &
\colhead{(dex)} & \colhead{Lines} & \colhead{(dex)}
}
\startdata
   6.0 &   \nodata &  \nodata &  \nodata    &   \nodata &  \nodata &  \nodata    &   \nodata &  \nodata &  \nodata    &   \nodata &  \nodata &  \nodata    &     7.89 &    1 &  \nodata    \\
   7.0 &   \nodata &  \nodata &  \nodata    &   \nodata &  \nodata &  \nodata    &   \nodata &  \nodata &  \nodata    &   \nodata &  \nodata &  \nodata    &     5.60 &    1 &  \nodata    \\
  12.0 &     5.36 &    5 &    0.14    &     5.05 &    2 &    0.03    &     5.47 &    4 &    0.29    &     5.53 &    4 &    0.16    &     5.85 &    3 &    0.17    \\
  13.0 &     3.79 &    2 &    0.12    &     3.47 &    2 &  0.12    &     3.61 &    2 &    0.05    &     3.72 &    2 &    0.13    &     4.06 &    1 &  \nodata    \\
  14.0 &     5.11 &    1 &  \nodata    &     4.79 &    1 &  \nodata    &     5.16 &    1 &  \nodata    &     5.16 &    1 &  \nodata    &   S 5.20 &    1 &  \nodata    \\
  20.0 &     4.17 &    7 &    0.12    &     3.89 &    5 &    0.28    &     4.02 &    4 &    0.11    &     4.11 &    6 &    0.13    &     4.60 &    2 &    0.17    \\
  21.1 &     0.92 &    3 &    0.10    &     0.33 &    1 &  \nodata    &     0.75 &    3 &    0.25    &   \nodata &  \nodata &  \nodata    &     1.79 &    1 &  \nodata    \\
  22.0 &     3.05 &    6 &    0.11    &   \nodata &  \nodata &  \nodata    &     3.19 &    3 &    0.16    &     3.02 &    4 &    0.16    &     3.39 &    5 &    0.26    \\
\enddata
\end{deluxetable}

\begin{deluxetable}{l rrr rrr rrr rrr rrr}
\tablenum{4c}
\tablewidth{0pt}
% \tiny
\tabletypesize{\tiny}
\tablecaption{Abundances of the Sample Stars \label{table_abundc}}
\tablehead{\colhead{Species} & 
\multispan{3}{HE0458$-$1346} &
\multispan{3}{HE0508$-$1555} &
\multispan{3}{HE1141$-$0610} &
\multispan{3}{HE1320+0038} &
\multispan{3}{HE1346$-$0427} \\
\colhead{} & \colhead{log$\epsilon(X)$} & \colhead{No.} & \colhead{$\sigma$} &
\colhead{log$\epsilon(X)$} & \colhead{No.} & \colhead{$\sigma$} &
\colhead{log$\epsilon(X)$} & \colhead{No.} & \colhead{$\sigma$} &
\colhead{log$\epsilon(X)$} & \colhead{No.} & \colhead{$\sigma$} &
\colhead{log$\epsilon(X)$} & \colhead{No.} & \colhead{$\sigma$}  \\
\colhead{} & \colhead{(dex)} & \colhead{Lines} & \colhead{(dex)} &
\colhead{(dex)} & \colhead{Lines} & \colhead{(dex)} &
\colhead{(dex)} & \colhead{Lines} & \colhead{(dex)} &
\colhead{(dex)} & \colhead{Lines} & \colhead{(dex)} &
\colhead{(dex)} & \colhead{Lines} & \colhead{(dex)} 
}
\startdata
  12.0 &     5.65 &    5 &    0.31    &     5.70 &    5 &    0.28    &     5.82 &    5 &    0.13    &     5.40 &    4 &    0.19    &     4.54 &    3 &    0.25    \\
  13.0 &     4.06 &    2 &    0.14    &     3.91 &    2 &    0.09    &     4.03 &    2 &    0.19    &     3.70 &    2 &    0.03    &     3.02 &    2 &    0.08    \\
  14.0 &     5.43 &    1 &  \nodata    &     5.23 &    1 &  \nodata    &   5.04 &    1 &  \nodata    &     4.77 &    1 &  \nodata    &     4.17 &    1 &  \nodata    \\
  20.0 &     4.19 &    6 &    0.17    &     4.22 &    7 &    0.11    &     4.33 &    7 &    0.08    &     3.98 &    5 &    0.12    &     3.16 &    3 &    0.19    \\
  21.1 &     0.92 &    3 &    0.05    &     1.02 &    3 &    0.03    &     1.11 &    3 &    0.12    &     0.48 &    2 &    0.02    &    $-$0.33 &    1 &  \nodata    \\
  22.0 &     3.06 &    6 &    0.18    &     3.23 &    5 &    0.14    &     3.16 &    5 &    0.17    & U  2.98 &    1 &  \nodata    &   \nodata &  \nodata &  \nodata    \\
  22.1 &     3.00 &   13 &    0.11    &     3.08 &   15 &    0.10    &     3.14 &   16 &    0.09    &     2.61 &   12 &    0.05    &     2.02 &    9 &    0.12    \\
  23.0 &   \nodata &  \nodata &  \nodata    &   \nodata &  \nodata &  \nodata    & U  2.60 &    1 &  \nodata    &   \nodata &  \nodata &  \nodata    &   \nodata &  \nodata &  \nodata    \\
\enddata
\end{deluxetable}

\clearpage

\begin{deluxetable}{l rrr rrr r}
\tablenum{4d}
\tablewidth{0pt}
\small
\tablecaption{Abundances of the Sample Stars \label{table_abundd}}
\tablehead{\colhead{Species} & 
\multispan{3}{HE2209$-$1240} &
\multispan{3}{HE2341$-$2029} & \colhead{Sun\tablenotemark{a}} \\
\colhead{} & \colhead{log$\epsilon(X)$} & \colhead{No.} & \colhead{$\sigma$} &
\colhead{log$\epsilon(X)$} & \colhead{No.} & \colhead{$\sigma$} &
\colhead{log$\epsilon(X)$}  \\
\colhead{} & \colhead{(dex)} & \colhead{Lines} & \colhead{(dex)} &
\colhead{(dex)} & \colhead{Lines} & \colhead{(dex)} & \colhead{(dex)}  
}
\startdata
   6.0 &    \nodata &  \nodata &  \nodata &   \nodata &  \nodata &  \nodata & 8.39 \\
   7.0 &    \nodata &  \nodata &  \nodata &   \nodata &  \nodata &  \nodata & 7.80 \\
  12.0 &     5.43 &    5 &    0.20 &        5.56 &    4 &    0.22 &      7.54 \\
  13.0 &     3.91 &    2 &    0.14 &        3.79 &    2 &    0.11 &      6.47 \\
  14.0 &     4.95 &    1 &  \nodata &       5.28 &    1 &  \nodata &      7.55 \\
  20.0 &     3.98 &    6 &    0.08 &        4.24 &    6 &    0.09 &      6.36 \\
  21.1 &     0.50 &    3 &    0.08 &        1.02 &    2 &    0.03 &      3.10 \\
  22.0 &   U 2.93 &    2 &    0.04 &        3.15 &    3 &    0.21 &      4.99 \\
\enddata
\tablenotetext{a}{See discussion  of the adopted Solar abundances 
in \S\ref{at_data}.}
\end{deluxetable}

\begin{deluxetable}{lrrc | lrrc}
\tablenum{5}
\tablewidth{0pt}
\small
\tablecaption{Fe Abundances -- Preliminary Estimates vs. Detailed Analyses 
\label{table_feabund}}
\tablehead{\colhead{Star}  &  \colhead{[Fe/H]} & \colhead{[Fe/H]} &  
\colhead{Mod. Res.} &
\colhead{Star}  &  \colhead{[Fe/H]} & \colhead{[Fe/H]} &  
\colhead{Mod. Res.} \\
\colhead{} & \colhead{(HK)} & \colhead{(Keck)} & \colhead{Spec\tablenotemark{a}} &
\colhead{} & \colhead{(HK)} & \colhead{(Keck)} & 
\colhead{Spec\tablenotemark{a}} \\
\colhead{} & \colhead{(dex)} & \colhead{(dex)} & 
\colhead{} &
\colhead{} & \colhead{(dex)} & \colhead{(dex)} & \colhead{}
}
\startdata 
Very bright \\
BD 3$-$740  & \nodata   &  $-$2.69 & \nodata & G139$-$8    & \nodata  &  $-$2.04 & \nodata \\  
LP0831$-$0700  & \nodata  &  $-$2.86 & \nodata  \\
HK stars \\
BS16545$-$0089 &  $\le-3.0$\tablenotemark{b}    & $-3.55$ & \nodata &
BS16945$-$0054  &   $-$3.00\tablenotemark{b}  & $-2.28$ & \nodata \\
BS17447$-$029 &   $-$3.15  & $-$2.91 & KPP \\
HES stars \\
HE0007$-$1832 &   $-3.00$ & $-2.65$  & ESO & HE0024$-$2523 &   $-$3.08 & $-$2.72 & KPP \\
\enddata
\tablenotetext{a}{Moderate resolution spectra from ESO 3.6m by
Christlieb, P200 by Cohen \& Ramirez, and from 
Magellan 6.5m (LCO) by McWilliam
or from Keck Pilot Project (KPP).}
\tablenotetext{b}{[Fe/H](HK) from \cite{anthony00}; believed to be on same
system as values of [Fe/H](HK) for the HES stars.}
\end{deluxetable}

\begin{deluxetable}{l ccccc}
\tablenum{6}
\tablewidth{0pt}
\small
\tablecaption{Abundance Changes for Small Changes in Stellar 
Parameters\tablenotemark{a}
\label{table_sens}}
\tablehead{\colhead{Species} & \colhead{${\Delta}$log$\epsilon(X)$} &
\colhead{${\Delta}$log$\epsilon(X)$} &
\colhead{$\Delta$log$\epsilon(X)$} &
\colhead{$\Delta$log$\epsilon(X)$} & 
\colhead{[X/Fe]\tablenotemark{b}} \\
\colhead{} & \colhead{(\teff$+250$K)} & \colhead{(\grav$+0.5$ dex)} &
\colhead{(Model[Fe/H])$+0.5$ dex)} & \colhead{($v_t+0.2$~\kms)} & 
\colhead{1 Star} \\
\colhead{} & \colhead{(dex)} & \colhead{(dex)} & \colhead{(dex)} &
\colhead{(dex)} & \colhead{(1$\sigma$)(dex)} 
}
\startdata
Mg~I & $-$0.17 &   0.11 &   0.13 &   0.03 & 0.06 \\
Al~I & $-$0.22 &   0.07 &   0.14 &   0.06 & 0.08 \\
Si~I & $-$0.29 &   0.20 &   0.17 &   0.07 & 0.12 \\
Ca~I & $-$0.17 &   0.04 &   0.13 &   0.02 & 0.05 \\
Sc~II & $-$0.13 &  $-$0.17 &   0.07 &   0.01 & 0.06 \\
Ti~I & $-$0.22 &   0.00 &   0.10 &   0.00 & 0.03 \\
Ti~II & $-$0.10 &  $-$0.17 &   0.08 &   0.03 & 0.05 \\
Cr~I & $-$0.24 &   0.01 &   0.11 &   0.04 & 0.06 \\
\enddata
\tablenotetext{a}{Computed from the line list of HE0458$-$1346.}
\tablenotetext{b}{1$\sigma$ uncertainty in [X/Fe] for a single (typical) star.
Fe~II is used for the ionized species. The adopted uncertainties
in stellar parameters ($\pm100$~K in \teff, $\pm0.1$ dex in \grav,
$\pm0.2$~\kms, etc) and in
\eqw\ are discussed in \S\ref{section_anal}.
Potential systematic errors are not included.}
\tablenotetext{c}{Uncertainty in log$\epsilon$(Fe) 
of a single (typical) star for the adopted uncertainties
in stellar parameters ($\pm100$~K in \teff, $\pm0.1$ dex in \grav,
$\pm0.2$~\kms, etc) and in
\eqw; see \S\ref{section_anal}.  Potential systematic errors are
not included.}
\end{deluxetable}

\begin{deluxetable}{l rrr}
\tablenum{7}
\tablewidth{0pt}
\small
\tablecaption{Slopes of Mean Abundance Ratios [X/Fe] Versus [Fe/H] for the 28 
Candidate EMP Dwarfs \label{table_slopes}}
\tablehead{\colhead{Species} & 
\multicolumn{3}{c}{No C-rich Stars}  \\
\colhead{} & \colhead{Slope\tablenotemark{a}} & \colhead{No. of} & 
\colhead{$\sigma$\tablenotemark{b}} \\
\colhead{} & \colhead{(dex/dex)} & \colhead{Stars} & \colhead{(dex)}
}
\startdata
\mg &  $+0.03\pm0.06$ & 24 & 0.20 \\
\al &  $-0.23\pm0.06$ & 24 & 0.12 \\
\si & $+0.03\pm0.06$  & 19 & 0.14 \\
\ca & $-0.13\pm0.05$ & 24 & 0.13 \\
\scme\tablenotemark{c} & $-0.01\pm0.05$ & 23 & 0.16 \\
\ti\tablenotemark{c} & $+0.00\pm0.05$ & 24 & 0.11 \\
\crme &  $+0.02\pm0.05$ & 24 & 0.15\tablenotemark{d} \\
\mn & $-0.07\pm0.10$ & 19 & 0.19\tablenotemark{d} \\
\co & $-0.23\pm0.15$ & 13 & 0.11 \\
\nime &  $-0.20\pm0.07$ & 15 & 0.09 \\
\enddata
\tablenotetext{a}{Slope of least squares linear fit to
relationship between [X/Fe] and [Fe/H].}
\tablenotetext{b}{The 1$\sigma$ rms dispersion of [X/Fe] about the 
linear fit, excluding
the non-detections and the C-rich stars.}
\tablenotetext{c}{Fe~II is used in determining the abundance ratio instead of Fe~I.}
\tablenotetext{d}{A significantly lower $\sigma$ would result if 
the outlier HE2344$-$2800 were omitted.}
\end{deluxetable}

\begin{deluxetable}{l rrr rrr rrr}
\tablenum{8}
\tablewidth{0pt}
\small
\tablecaption{Mean Abundance Ratios [X/Fe] and Their 
Dispersions for the 28 Candidate EMP Dwarfs\label{table_disp}}
\tablehead{\colhead{Species} & \multicolumn{3}{c}{All Stars} &
\multicolumn{3}{c}{No C-enh. Stars} & \multicolumn{3}{c}{Special Cases} \\
\colhead{} & \colhead{$<$[X/Fe]$>$} & \colhead{$\sigma$} & \colhead{Num.} &
\colhead{$<$[X/Fe]$>$} & \colhead{$\sigma$} & \colhead{Num.} & 
\colhead{$<$[X/Fe]$>$} & \colhead{$\sigma$} & \colhead{Num.} \\
\colhead{} & \colhead{(dex)} & \colhead{(dex)} & \colhead{Stars} & \colhead{(dex)} &
\colhead{(dex)} & \colhead{Stars} & 
\colhead{(dex)} & \colhead{(dex)} & \colhead{Stars} 
}
\startdata
Mg/Fe & +0.42 & 0.22 & 28 & +0.39 & 0.21 & 24 & 
    +0.56\tablenotemark{b} & 0.12 & 20 \\
Mg/Fe\tablenotemark{h} & +0.56 & 0.18 & 19  &  +0.52 & 0.21 & 19 &
  +0.56\tablenotemark{i} & 0.15 & 16  \\
Al/Fe\tablenotemark{a} & $-$0.10 & 0.19 & 28 & $-$0.09 & 0.16 & 24 &
    \nodata &  \nodata &  \nodata \\
Si/Fe & +0.14 & 0.17 & 23 & $+0.04$ & 0.14 & 19 & 
    \nodata &  \nodata &  \nodata \\
% recalcuated sigma(Si/Fe) for all stars with synthese for C-enhanced stars    
% Si increased 0.05 dex, gf -1.09 instead of -1.04    
Ca/Fe & +0.33 & 0.13 & 28 & +0.31 & 0.12 & 24 & 
    \nodata &  \nodata &  \nodata \\
Sc/Fe\tablenotemark{c} &
       +0.28 & 0.21 & 26 & +0.24 & 0.16 & 23 &  
        \nodata &  \nodata &  \nodata \\
Ti/Fe\tablenotemark{c} &       
       +0.36 & 0.12 & 28 & +0.36 & 0.12 & 24 & 
       \nodata &  \nodata &  \nodata \\
Cr/Fe & $-$0.23 & 0.13 & 27 & $-$0.21 & 0.13 & 24 & 
      $-$0.23\tablenotemark{d} & 0.09 & 23 \\
\enddata
\tablenotetext{a}{A constant non-LTE correction is assumed for Al.}
\tablenotetext{b}{The four stars with Solar Mg/Fe have been excluded.}
\tablenotetext{c}{Fe~II is used in determining the abundance ratio instead of Fe~I.}
\tablenotetext{d}{HE2344$-$2800 is excluded.}
\tablenotetext{e}{Three stars (LP-831$-$070, G138$-$9 and HE0242$-$0732)
with very low Sr/Fe omitted.}
\tablenotetext{f}{Three stars (LP-831$-$070, G138$-$9 and HE0130$-$2303)
with very low Ba/Fe and five stars with no detected Ba lines omitted.}
\tablenotetext{g}{Only detected in some of the C-enhanced stars.}
\tablenotetext{h}{The stars for which only the Mg triplet was detected 
are excluded.}
\tablenotetext{i}{The outlier G139$-$8 is also excluded.}
\end{deluxetable}

\begin{deluxetable}{lcr}
\tablenum{9}
\tablewidth{0pt}
\small
\tablecaption{Line-By-Line Values of Mg Abundances\tablenotemark{a} 
\label{table_mg}}
\tablehead{\colhead{Wavelength} & 
\colhead{$\Delta$[log$\epsilon$(Mg)]\tablenotemark{b}} &
\colhead{$\sigma$\tablenotemark{c}} \\
\colhead{(\AA)} & \colhead{(dex)} & \colhead{(dex)}
 }
\startdata
4057.5 & +0.07 & 0.04 \\
4167.3 & +0.07 & 0.02 \\
4703.0 & +0.12 & 0.03 \\
5172.7 & $-0.13$ & 0.03 \\
5183.6 & $-0.12$ & 0.03 \\
\enddata
\tablenotetext{a}{The average of the deviation of
log$\epsilon$(Mg) inferred from each Mg~I line used here with respect to
the mean abundance in the star for a set of 7 stars chosen as described
in the text.}
\tablenotetext{b}{For each of the five Mg~I lines used, this is
$<$\{log$\epsilon$(Mg,line)$-<$log$\epsilon$(Mg,star)$>$\}$>$, where the
difference is averaged over 7 stars.}
\tablenotetext{c}{This is the 1$\sigma$ uncertainty of the mean.}
\end{deluxetable}

\begin{deluxetable}{l rrrrr}
\tablenum{10}
\tablewidth{0pt}
\small
\tablecaption{Comparison of Mean Abundance Ratios 
and Dispersions With Those of \cite{cayrel03}
\label{table_cayrel}}
\tablehead{\colhead{Species} & \colhead{Mean [X/Fe]} & 
\colhead{Mean [X/Fe]\tablenotemark{a}}
& \colhead{$\Delta$\tablenotemark{a}} & \colhead{$\sigma$} & 
\colhead{$\sigma$} \\
\colhead{} & \colhead{EMP Dwarfs\tablenotemark{b}} & 
\colhead{Cayrel Giants\tablenotemark{c}} & \colhead{(Dwarfs-Giants)}
& \colhead{EMP Dwarfs} &  \colhead{Cayrel Giants} \\
\colhead{} & \colhead{(dex)} & \colhead{(dex)} & \colhead{(dex)}
& \colhead{(dex)} & \colhead{(dex)}
}
\startdata
Mg/Fe & +0.39 & +0.28 & +0.11 & 0.13\tablenotemark{d} & 0.13 \\ 
Al/Fe & $-0.09$ & $-$0.08\tablenotemark{e} & $-$0.01 & 0.16 & 0.18 \\ 
Si/Fe & $+0.04$ & +0.45 & $-0.41$ & 0.14 & 0.15 \\
Ca/Fe & +0.31 & +0.34 & $-0.03$ & 0.12 & 0.10 \\
Sc/Fe & +0.24 & +0.08 & +0.16 & 0.16 & 0.11 \\
Ti/Fe & +0.36 & +0.23 & +0.13 & 0.12 & 0.09 \\
Cr/Fe & $-0.21$ & $-0.35$ & +0.14 & 0.09\tablenotemark{f} & 0.05 \\
Mn/Fe & $-0.55$ & $-0.44$ & $-0.11$ & 0.13\tablenotemark{f} & 0.12 \\
Co/Fe & +0.42 & +0.27 & +0.15 & 0.12 & 0.13 \\ 
Ni/Fe & $-0.02$ & $-0.04$ & +0.02 & 0.12 & 0.11 \\
\enddata
\tablenotetext{a}{\cite{cayrel03} adopt log$\epsilon$(Fe) = 7.50 dex,
0.05 dex higher than we do.  No correction for this 
difference has been applied.}
\tablenotetext{b}{The C-rich stars in our sample are excluded.}
\tablenotetext{c}{Calculated from the fits given in 
\cite{cayrel03} at [Fe/H] = $-3.0$ dex.}
\tablenotetext{d}{We use the value from 
Table~\ref{table_sens} ignoring the C-rich stars and the
four stars with Solar Mg/Fe.}
\tablenotetext{e}{We adjust the value from \cite{cayrel03} to the
slightly smaller
non-LTE correction for Al~I adopted here.}
\tablenotetext{f}{We ignore the discrepant star HE2344$-$2800 as well as
the C-rich stars.}
\end{deluxetable}

\begin{deluxetable}{l rrrrr rr}
\tablenum{11}
\tablewidth{0pt}
\small
\tablecaption{Comparison With Galactic Globular Clusters
\label{table_globcomp}}
\tablehead{ \colhead{} &
\multicolumn{2}{c}{EMP Dwarfs} & \colhead{NGC~6397\tablenotemark{i}}  &
\colhead{M3\tablenotemark{a}}  &
\colhead{M5\tablenotemark{b}} & 
\colhead{M71\tablenotemark{c}} & \colhead{47 Tuc\tablenotemark{h}} \\
\colhead{Species} & \colhead{$<$[X/Fe]$>$} & 
\colhead{$\sigma$[X/Fe]\tablenotemark{d}} &
\colhead{$<$[X/Fe]$>$} &  \colhead{$<$[X/Fe]$>$} &
\colhead{$<$[X/Fe]$>$} &   \colhead{$<$[X/Fe]$>$} &
\colhead{$<$[X/Fe]$>$}  \\
\colhead{} & \colhead{(dex)} & \colhead{(dex)} &
\colhead{(dex)} & \colhead{(dex)} &  \colhead{(dex)} & \colhead{(dex)} &
\colhead{(dex)} 
}
\startdata
\fe & $-2.66$ & \nodata & $-2.02$  & $-1.45$ &  $-1.30$ & $-0.68$ & $-0.67$ \\ 
\mg & 0.39 & 0.21 & 0.04 & 0.22 &  0.29 &  0.35 & 0.40 \\
\al & $-0.09$ &  0.16 & \nodata & 0.51\tablenotemark{e} & 0.45\tablenotemark{f} & 
  +0.22 & +0.27  \\
\si & $+0.04$ & 0.14 &  \nodata & 0.30  & 0.21  & 0.27 & 0.30  \\
\ca & +0.31 &  0.12 & 0.24 & 0.23 & 0.33  & 0.36  & 0.20 \\
\scme & +0.24 &  0.16 &  0.27 & $-$0.11  & 0.05  & 0.03 & 0.13 \\
\ti & +0.36 &  0.12 & 0.47 & 0.16 &  0.16  & 0.23 & 0.32 \\
\crme & $-0.23$\tablenotemark{g}
 & 0.09 & $-0.12$ & 0.00  & $-$0.13  & $-0.07$ & +0.11 \\
\enddata
\tablenotetext{a}{Data from \cite{sneden04} and from \cite{cohen04a}.}
\tablenotetext{b}{Data from \cite{ramirez03}.}
\tablenotetext{c}{Data from \cite{ramirez02} as updated in
\cite{ramirez03}.}
\tablenotetext{d}{1$\sigma$ dispersion about the mean, C-rich stars are excluded.} 
\tablenotetext{e}{[Al/Fe] from \cite{sneden04}.}
\tablenotetext{f}{[Al/Fe] from \cite{ivans01}.}
\tablenotetext{g}{We ignore the discrepant star HE2344$-$2800 as well as
the C-rich stars.}
\tablenotetext{h}{Data from \cite{carretta04}.}
\tablenotetext{i}{Data from \cite{thevenin01}.}
\end{deluxetable}

\clearpage

\clearpage

\clearpage

\clearpage

\clearpage

\end{document}